\documentclass[12pt]{article}
\usepackage{float}
\usepackage{graphicx}
\usepackage{psfig,epsfig}%,supertabular,theorem,cite,amsmath,amssymb}

\newcommand{\ips}   {\ensuremath{\mathrm{ps^{-1}}}}

\newcommand{\Bs}    {\ifmmode {B^0_s}  \else {$B^0_s$} \fi}
\newcommand{\Bsbar} {\ifmmode {\bar{B}^0_s}  \else {$\bar{B}^0_s$} \fi}
\newcommand{\Bd}    {\ifmmode {B^0_d}  \else {$B^0_d$} \fi}
\newcommand{\Bdbar} {\ifmmode {\bar{B}^0_d}  \else {$\bar{B}^0_d$} \fi}
\newcommand{\Bsmix} {\ifmmode {B^0_s - \bar{B}^0_s}  \else {$B^0_s$ - $\bar{B}^0_s$} \fi}
\newcommand{\Bdmix} {\ifmmode {B^0_d - \bar{B}^0_d}  \else {$B^0_d$ - $\bar{B}^0_d$} \fi}

\newcommand{\rb} {\bar{\rho}}
\newcommand{\eb} {\bar{\eta}}
\newcommand{\dmd}{\Delta m_{B_{d}}}  %{\Delta m_d}
\newcommand{\dms}{\Delta m_{B_{s}}}  %{\Delta m_s}
\newcommand{\ek} {\epsilon_K}
\newcommand{\dckm} {\delta_{\rm{ckm}}}
\newcommand{\fsbd}{f^2_{B_d}\hat{B}_{B_d}}
\newcommand{\fsbs}{f^2_{B_s}\hat{B}_{B_s}}

\newcommand{\eq}[1]{Eq.~(\ref{#1})}
\newcommand{\fig}[1]{Figure~(\ref{#1})}

\newcommand{\nn}{\noindent}

\newcommand{\fbbd}{f^2_{B_d}{B}_{B_d}}

\newcommand{\fbd}{f_{B_d}}

\def\be{\begin{equation}}
\def\ee{\end{equation}}
\def\bea{\begin{eqnarray}}
\def\eea{\end{eqnarray}}
\def\beas{\begin{eqnarray*}}
\def\eeas{\end{eqnarray*}}
\def\barr{\begin{array}}
\def\earr{\end{array}}
\def\nn{\nonumber}

\def\mrm{\mathrm}
\def\lt{\left}
\def\rt{\right}

\def\rly#1{\mathrel{\raise.3ex\hbox{$#1$\kern-.75em\lower1ex\hbox{$\sim$}}}}

\def\amp {{\cal{A}}} 
\def\samp{\sigma_{\cal{A}}}
\def\bit{\begin{itemize}}
\def\eit{\end{itemize}}

\begin{document}

%$~$

\vspace{2cm}

\begin{center}

\begin{flushright}
FTPI-MINN-06/10\\
UMN-TH-2438/06
\end{flushright}

\vspace{1cm}

{\Large{\bf{
Impact of \boldmath{$\dms$} on the determination of the unitary triangle
and bounds on physics beyond the standard model}}}

\vspace{2cm}

{\bf L. Velasco-Sevilla
\footnote{liliana@physics.umn.edu}
}\\

{\it William I. Fine Theoretical Physics Institute\\
University of Minnesota}

\vspace{1cm}

{\bf N. Leonardo
\footnote{neonard@mit.edu}
}\\

{\it Department of Phsyics \\
Massachusetts Institute of Technology}

\vspace{3cm}
\today

\thispagestyle{empty}

{\bf Abstract}

\end{center}
We study the impact on the unitarity triangle of the \Bsmix flavor oscillation data analyses being carried out at the Tevatron Run II. 
In the current version of this work we consider the first direct experimental bound on the mass difference $\dms$ recently reported by D\O\ along with the first measurement achieved by CDF, $\dms=(17.33\pm 0.4) \ {\rm{ps^{-1}}}$.
In particular we show how these measurements further constrain the compatibility between the experimental value of $\sin 2\beta^{\;\rm{exp.}}=(0.591,0.783)$ at $99\%$ C.L. and the value obtained by the fit of the CKM parameters considering just CP conserving experimental observables: $\sin 2\beta^{\;\rm{CP-conserv.}}=(0.742,0.856)$ at $99\%$ C.L. We also obtain bounds on certain classes of processes beyond the SM.

\newpage
%
%%%%%%%%%%%%%%%%%%%%%%%%%%%%%%%%%%%%%%%%%%%%%%%%%%%%%%%%%%%%%%%%%%%%%%%%%%%%%%%%%%%%%%
%
\section{Introduction}
%
%%%%%%%%%%%%%%%%%%%%%%%%%%%%%%%%%%%%%%%%%%%%%%%%%%%%%%%%%%%%%%%%%%%%%%%%%%%%%%%%%%%%%%
%

We are at an exciting era of $b$-flavoured hadron physics, as the ongoing experimental efforts in this area can be used to test to a better accuracy the CKM matrix and hence the Standard Model (SM) giving at the same time bounds on processes beyond the SM (BSM).

Various high precision $B$ physics measurements have been made possible 
notably by the $B$ factories operating at the $\Upsilon(4S)$ resonance. 
Mixing in the $\Bd$ system, with the light $B$ mesons being there copiously produced,  has been measured accurately. 

The situation is different for the heavier neutral $B$ meson system, $\Bs$, 
which has been intensively studied by the Tevatron experiments, CDF and D\O. For the last few years they have been accumulating large $\Bs$ data samples, developing and optimizing complex data analysis techniques which finally lead to the measurement of mixing in this system. A task that has been rather challenging for almost two decades.

All combined experimental data prior to Tevatron Run II have given only a lower \emph{bound} on the mass difference that characterizes flavour mixing in the $\Bsmix$ system, namely $\dms >14.5~\ips$. 

The first $\Bs$ mixing results from both of Tevatron collaborations after Run II, namely $\dms >16.6~\ips$ with a combined sensitivity of $20.0~\ips$, were presented last year
~\cite{eps05-panic05-conf} and translated already into sizable contributions to the world average. We have studied its effect on the determination of the unitary triangle (UT) in the previous version of this work.

More recently after having extended their $\dms$ analyses to 1~fb$^{-1}$ of data, both collaborations at Tevatron have reported results of unprecedented relevance. D\O\  has established a first direct experimental two-sided bound $\dms\in[17,21]~\ips$ at 90\%~CL~\cite{d0-prl06}. CDF presented shortly afterward a first and precise \emph{measurement} $\dms = 17.31^{+0.34}_{-0.19}~\ips$~\cite{cdf-prl06a}.

How then these novel results translate not only in the determination of the CKM parameters and their compatibility with each experimental measurement, but also in constraining models BSM?  This is the motivation of the present work.

The CKM fitter group~\cite{ckmfitter:wp} and the UT fit group~\cite{utfit:wp} have been developing {state-of-the-art} analyses of the CKM parameters which are also regularly updated with new experimental information. However we perform this analysis  to study the impact of $\dms$ on the determination of the unitary triangle (UT) parameters in the SM and also to have a code compatible with such reference analyses in order to test models BSM.
 
To this end we first update the UT fits of the SM comparing the bound (corresponding to the combined oscillation data before the CDF and D\O\ results at Run II had become available) and the current measurement. Then we begin our tests of models BSM within the Minimal Flavour Violation (MFV) scenario and briefly mention consequences for an example of a horizontal symmetry.

During the last and the present decades many groups have been developing analyses in order to test the unitary triangle and CP violation using various approaches, including the Bayesian method~\cite{ckm:fit:bay}, the Range-fit (Rfit) method~\cite{ckm:fit:fre}, based on a frequentist approach, a Gaussian method~\cite{ckm:fit:gau}, and the scan method~\cite{ckm:fit:sca}. These approaches treat in somewhat different ways the available information on the experimental and the theoretical uncertainties. We use the Bayesian approach to construct a global inference function for the Wolfenstein parameters $\rb$ and $\eb$, from which a {\it posterior} probability density function (pdf) for all the fitted (within the analysis) and related parameters to the unitary triangle and the constraints used to determine it can be derived.

The analysis of the UT and CP violation tests the SM, hence giving it a more robust success or, if physics BSM is present already at sizable levels, detecting inconsistencies with the various unitary tests. 
In what it has been called the {\it Classic analysis} of the UT \cite{Bona:2005eu}, the CP conserving parameters -- $|V_{ub}|/|V_{cb}|$, $\dmd$ and $\dmd/\dms$ -- and  CP violating ones -- $\ek$ and $\sin 2\beta$ -- are taken into account. The ratio $|V_{ub}|/|V_{cb}|$ is not expected to have a big contribution from beyond tree-level processes of semi-leptonic decays from which $|V_{ub}|$ and $|V_{cb}|$ are extracted. The CP asymmetry in $B \to J/\psi K_S$ could give deviations from measuring not just the SM unitary angle $\beta$ but $\beta+\theta_{BSM}$. Over the past years the SM prediction of this quantity using the other UT parameters was consistent with its observed value. But an important result of the present analysis (as well as other recent analyses) is that there is an incompatibility with the input experimental value $\sin 2\beta^{\;\rm{exp.}}=(0.623,0.751)$ at $95\%$ confidence level (CL) and the output value of the fit including just the CP conserving parameters: $\sin 2\beta^{\;\rm{CP-conserv.}}=(0.760,0.836)$ at 95\%~CL. Now they are just compatible at 99\%~CL.{\footnote{Using the bound~\eq{eqn:dms:f05} they were compatible at 95\%~CL output since the value of this case for the CP conserving processes is $\sin 2\beta=0.793^{-0.021}_{+0.024}$.}}   This is an important hint for BSM processes but of course a better measurement of $\dms$ and a better determination of the parameters $|V_{ub}|$ and $|V_{cb}|$ will be needed in order to shed light on this apparent incompatibility.

We just consider $\Delta F=2$ flavour changing processes, leaving out of our analysis $\Delta F=1$ processes, which despite being relevant in some cases the present experimental bounds are not at the level of precision \cite{Laplace:2002ik} as most quantities entering in the {\it Classic analysis}.

Mixing in the kaon and in the $B_{d,s}$ neutral meson systems, from which $\ek$, $\dmd$ and $\dms$ are respectively obtained,  can be particularly sensitive to physics BSM.  The most general way to identify the deviation of $\ek$, $\dmd$ and $\dms$ from their SM prediction is by measuring the deviation from unity of the ratios
\bea
\label{eq:ratio_bsm_sm}
{\rm{Im}}  \{ \langle K^0| {\mathcal{H}}^{\rm{Total}}_{\rm{eff}} |\bar{K}^0 \rangle \}
&/& 
{\rm{Im}}  \{ \langle K^0| {\mathcal{H}}^{\rm{SM}}_{\rm{eff}} |\bar{K}^0 \rangle\} \quad {\rm{and}}\nn \\
\langle B^0_{d,s}|{\mathcal{H}}^{\rm{Total}}_{\rm{eff}} |\bar{B}^0_{d,s} \rangle 
&/&
\langle B_{d,s}|{\mathcal{H}}^{\rm{SM}}_{\rm{eff}} |\bar{B}_{d,s}\rangle,\nn
\eea
respectively, where ${\mathcal{H}}^{\rm{Total}}_{\rm{eff}}$ is the effective Hamiltonian that takes into account all the processes, both in the SM and BSM, of the relevant $\Delta F=2$ ($K^0$-$\bar{K}^0$ and $B^0_{d,s}$-$\bar{B}^0_{d,s}$) flavour changing neutral current processes.

However, we have to bear in mind that there are two main restrictions on measuring these ratios. The first one is that there are still big uncertainties in some quantities entering in the computation of the effective SM Hamiltonian matrix elements, chiefly $B_K$ and $\fbbd$, which obscure differentiation of these and the effects of processes BSM. The second one is that in general these ratios are related in different extensions of models BSM  and just for the minimal flavour violation (MFV) scenario of the Minimal Supersymmetric Standard Model (MSSM) the form of the effective Hamiltonians and the QCD next to leading order (NLO) determination of its parameters has been computed \cite{Krauss:1998mt}.

For the first of the aforementioned restrictions not much can be done currently but for the second one we can test models for which we can extract meaningful information from the current way in which the UT analysis it is performed. Since we want to test what is the impact of the new results on $\dms$, we explore the cases of physics BSM for which the ratio $\dmd/\dms$ is equal to the case of the SM. 

The MFV violation scenario, which  essentially requires that all flavour and CP violating interactions are linked to the known Yukawa couplings, is a well known example of this condition. 
The MFV scenario can be easily implemented in the MSSM, as it has been done extensively since the pioneering work of~\cite{Bertolini:1990if}, but using an effective operator approach~\cite{D'Ambrosio:2002ex} it can also be implemented for other cases. With the current bound we can see that the MFV supersymmetric scenario appears plausible with the current $\dms$ value.

Given the new experimental information we thus study the possibility of having significant contributions of processes BSM in the determination of the CKM elements, and we show that, under certain constraints, somewhat large effects of processes beyond the SM are still possible. We mention briefly the consequences of this analysis for models of fermion masses with an $SU(3)$ horizontal (flavour) symmetry.

%
%%%%%%%%%%%%%%%%%%%%%%%%%%%%%%%%%%%%%%%%%%%%%%%%%%%%%%%%%%%%%%%%%%%%%%%%%%%%%%%%%%%%%%
%
\section{Constraints of the CKM matrix}
%
%%%%%%%%%%%%%%%%%%%%%%%%%%%%%%%%%%%%%%%%%%%%%%%%%%%%%%%%%%%%%%%%%%%%%%%%%%%%%%%%%%%%%%
%
Here we just quickly state the formulas used in the SM fit of the CKM matrix, $V$, and remark the processes from which they are extracted in order to point out which constraints we use to put bounds on processes beyond the SM. The inputs of the fits are presented in Tables (\ref{tbl:fixedp})-(\ref{tbl:fittedp}) of Appendix \ref{sec:app_exp_inp}. We recall here that $V$ can be expressed in terms of the four Wolfenstein parameters: $A$, $\lambda$ $\rho$ and $\eta$. All elements of $V$ can be obtained as an expansion in $\lambda$ up to the desired $n$-th order, using the unitary condition from the basic definitions
\bea
|V_{us}|=\lambda,\quad  |V_{cb}|=A\lambda^2,\quad |V_{ub}|=A\lambda^3 (\rho-i\eta),
\eea
which are valid to all orders in $\lambda$. The parameters $\lambda$ and $A$ are obtained from {\it tree-level} kaon and bottom semi-leptonic decays which have large branching ratios making their determination independent, to an excellent approximation, of any processes BSM. The re-scaled parameters $\rb=c \rho$ and $\eb=c \eta$, with $c\equiv 1-\lambda^2/2$, are the coordinates of the apex of the unitary triangle (UT), with sides $R_b$ and $R_t$ defined by
\bea
R^2_b=\rb^2+\eb^2,\quad \quad
R^2_t=(1-\rb)^2+\eb^2
\label{eq:Rb_Rt}
\eea
and angles $\alpha$, $\beta$ and $\gamma$ defined by
\bea
\sin\beta=\frac{\eb}{R_t},\quad \sin\gamma=\frac{\eb}{R_b}, \quad \alpha=\pi-\beta-\gamma.
\label{eq:abg_def}
\eea
In the {\it Classic fit} of the SM CKM matrix, $\rb$ and $\eb$ are extracted from \\

\begin{tabular}{lcl}
(i) CP conserving processes: &  \& & (ii) CP conserving processes: \\
  &  \\
$\frac{|V_{ub}|}{|V_{cb}|}$, $\dmd$ and $\frac{\dmd}{\dms}$
& &
$\ek$ and $\sin 2\beta$.\\
 & &  
\end{tabular}

\noindent In {(i)}, $|V_{ub}|$ and $|V_{cb}|$ are extracted from semi-leptonic decays of $B$ mesons.  The parameters $\dmd$ and $\dms$ are extracted from measurements of $B^0_{d,s}$-$\bar{B}^0_{d,s}$ oscillations, which are $\Delta B=2$ flavour changing neutral current processes (FCNC). From {(ii)}, the asymmetry parameter $|\ek|$ is induced by $K^0$-$\bar{K}^0$ transitions, which are $\Delta S=2$ FCNC processes. 
Finally the measurement of $\sin 2\beta$ is dominated by the channel $\Bd \to J/\psi K_S$ and complemented by other decay modes. 

The parameters $|\epsilon_K|$, $|V_{ub}/V_{cb}|$, $\dmd$,  $\dms$ and $\sin 2\beta$ are expressed directly or indirectly in terms of $\bar{\rho}$ and $\bar{\eta}$, and  a set of theoretical and experimental parameters $x$, some of which have been well measured and some have still large uncertainties.  Thus, in the fits the parameters $x$ which have been well measured are kept fixed and those with dominant uncertainties are varied and fitted.

\vspace*{0.5cm}

\noindent{\large{$|V_{ub}|/|V_{cb}|$}}
\vspace*{0.5cm}

The CKM matrix elements $|V_{ub}|$ and $|V_{cb}|$ are measured in exclusive and inclusive semi-leptonic $B$ decays. 
In terms of the  re-scaled Wolfenstein parameters, $|V_{ub}|/|V_{cb}|$ is expressed as:
\be
\label{VubVcbtheory}
\frac{|V_{ub}|}{|V_{cb}|}=\frac{\lambda}{c}\sqrt{\bar{\rho}^2 +\bar{\eta}^2}\,.
\ee
\vspace*{0.5cm}

\noindent{\large{{${\dmd}$ and ${\dms}$ }}}
\vspace*{0.5cm}

Within the SM, the mass differences $\Delta m_{d,s}$ of the oscillating systems $B^0_{d,s}$-$\bar{B}^0_{d,s}$ are very well approximated by the  relevant electroweak transition (``box") diagrams, described by the Inami-Lim functions~\cite{Inami:1980fz},  which are dominated by top-quark exchange. 
$\Delta m_{B_{d,s}}$ are proportional to the moduli of the matrix elements of 
the effective Hamiltonian for the $\Delta B=2$ transitions,  
\be
\label{eq:dmbds_sm}
\Delta m_{B_{q}}=
\frac{G^2_F}{6\pi^2} m_{B_q}f^2_{B_q}{B}_{B_q}  \eta_{B} S(x_t)|V_{tq}^*V^{\phantom{*}}_{tb}|^2
\propto
|\langle  B_{q}|{\mathcal{H}}^{\rm{SM}}_{\rm{eff}} |\bar{B}_{q} \rangle |,
\quad (q=d,s)
\ee
where $G_F$ is the Fermi constant, $\eta_{B}$ is a QCD correction factor calculated in NLO, $m_{B_q}$ and $m_W$ are the $B_q$ meson and $W$ boson masses respectively.

The dominant uncertainties in~\eq{eq:dmbds_sm} come from the evaluation of the hadronic quantities: the $B$ meson decay constants $f_{B_q}$ and the bag parameters $B_{B_q}$, which parameterize the value of the hadronic matrix element.
The Inami-Lim function, $S(x_t)$, describes the $|\Delta B|=2$ transition amplitude in the absence of strong interaction, where the mass of the top quark enters via $x_t\equiv \frac{m^2_t}{M^2_{\mrm{W}}}$.
{\footnote{\label{fn:sxt}
$S(x_t)=x_t\left[\frac{1}{4}+\frac{9}{4}\frac{1}{1-x_t}-\frac{3}{2}\frac{1}{(1-x_t)^2} \right]-\frac{3}{2}\left[\frac{x_t}{1-x_t} \right]^3 \ln x_t.$}}

In terms of the Wolfenstein parameters, \eq{eq:dmbds_sm} above for the $\Bd$ case can be expressed as
\be 
\label{dmdtheory}
\dmd=C_{\Delta_m} A^2\lambda^6 [(1-\bar{\rho})^2 + \bar{\eta}^2] m_{B_d} f^2_{B_d}B_{B_d} \eta_{B} S(x_t),
\ee
where we have defined the constant $C_{\Delta m}\equiv\frac{G^2_F M^2_W}{6\pi^2}$. The parameters with dominant uncertainties in~\eq{dmdtheory} are $f^2_{B_d}{B}_{B_d}$, $A$ and $\lambda$;  among these, $A$ and $\lambda$ are varied parameters of the fit.

The size of side $|V_{td}|/(\lambda|V_{cb}|)$ of the unitary triangle can be 
obtained from the ratio of $\dmd$ and $\dms$, 
\be
\label{M12dM12srat}
\frac{\dmd}{\dms}
= \frac{m_{B_d} \, |V_{td}|^2}{m_{B_s} \, \xi^2 \, |V_{ts}|^2}  \qquad \rm{with} \quad 
\xi=\frac{f_{B_s}\sqrt{B_{B_s}}}{f_{B_d}\sqrt{B_{B_d}}},
\ee
which is expected to be less dependent on the absolute values of 
the non-perturbative quantities 
$f_B$ and $B_B$, and hence it can be characterized by $\xi$, 
an SU(3) breaking correction whose value is obtained from lattice QCD calculations 
with better precision than the individual matrix element quantities themselves. 
In the fit we use $\xi$ and $f_{B_s}\sqrt{B_{B_s}}$ instead of $f_{B_d}\sqrt{B_{B_d}}$, which makes the constraint $\dmd$ more effective \cite{Bona:2005vz}.
The constraint we use from $\dms$ is expressed, from~\eq{M12dM12srat}, as
\be
\label{dmstheory}
\dms=\dmd\frac{m_{B_s}}{m_{B_d}}\xi^2 \frac{c^2}{\lambda^2}\frac{1}{(1-\bar{\rho})^2+\bar{\eta}^2},
\ee
where $\dmd$  is here taken as an experimental input. The parameters with dominant uncertainties in~\eq{dmstheory} 
are $~\xi, ~A, ~\lambda$, which are varied parameters of the fit.

\vspace*{0.5cm}

\noindent{\large{{$|\epsilon_K|$}}}
\vspace*{0.5cm}

The parameter $\epsilon_K$ expresses the measurement of indirect CP violation
in the neutral kaon system. In terms of the Wolfenstein parameters it is given by
\bea
\label{eq:ek_sm}
|\epsilon_K|                &=& C_{\epsilon} B_K {\rm{Im}} ~{A_{\rm{SM}}(K)}
\propto  \langle K^0| {\mathcal{H}}^{\rm{SM}}_{\rm{eff}} |\bar{K}^0 \rangle \,, \quad \rm{with}\nn\\ 
{\rm{Im}} ~{A_{\rm{SM}}(K)} &=&
A^2 \lambda^6 \bar{\eta} [-\eta_{cc}x_c + A^2 \lambda^4\left(1-\bar{\rho}\right)\eta_{tt}S(x_t) +\eta_{ct} S(x_c,x_t)]\,,\nn\\
C_{\epsilon}                &=&\frac{G^2_F f^2_K m_K m^2_W}{6\sqrt{2}\pi^2\Delta m_{K}}\,.
\eea
The short distance QCD corrections are codified in the coefficients $\eta_{cc}$, $\eta_{tt}$ and $\eta_{ct}$, and are functions of the charm and top quark masses and of the QCD scale parameter $\Lambda _{QCD}$; these parameters have been calculated in NLO.
The Inami-Lim functions, %\cite{Inami:1981fz}, 
which describe the $|\Delta S|=2$ transition amplitude in the absence of strong interactions, 
are given by $S(x_t)$, as in the $\dmd$ case  $^{\ref{fn:sxt}}$, and by $S(x_c,x_t)$.
{\footnote{
\label{eq:Sxcxt}
$S(x_c,x_t)=-x_c \ln x_c + x_c \left[ \frac{x_t^2-8x_t +4}{4(1-x_t)^2}\ln x_t +\frac{3}{4}\frac{x_t}{x_t-1} \right], \quad x_q\equiv \frac{m^2_q}{M^2_{\mrm{W}}}.$}}

The parameters with dominant uncertainties are  $B_K$, $\eta_{cc}$, $\eta_{ct}$, $m_c$ and $m_t$ which are hence varied in the fit. 
\vspace*{0.5cm}

\noindent{\large$\sin 2\beta$}
\vspace*{0.5cm}

A direct determination of the angles of the unitary triangle can be achieved via measurements of CP asymmetries in various $B$ decays. These angles are defined in reference to the normalized unitary triangle:
\bea
\alpha={\mathrm{arg}}\left[\frac{-V_{td}V^*_{tb}}{V_{ud}V^*_{ub}} \right],\quad
\beta ={\mathrm{arg}}\left[\frac{-V_{cd}V^*_{cb}}{V_{td}V^*_{tb}} \right],\quad
\gamma={\mathrm{arg}}\left[\frac{-V_{ud}V^*_{ub}}{V_{cd}V^*_{cb}} \right].
\eea

The value of $\sin 2\beta$ is measured in $b \to c \bar{c} s$ transitions, such as $\Bd \to J/\psi K^0$ decays.
The UT angles can be expressed directly in terms of the re-scaled Wolfenstein parameters; 
in particular we have 
\bea
\sin 2\beta&=&\frac{2\bar{\eta}(1-\bar{\rho})}{\bar{\eta}^2+(1-\bar{\rho})^2}\label{sin2bWolp} \,.
\eea
%
%%%%%%%%%%%%%%%%%%%%%%%%%%%%%%%%%%%%%%%%%%%%%%%%%%%%%%%%%%%%%%%%%%%%%%%%%%%%%%%%%%%%%%
%
\section{How to account for indications of physics beyond the SM\label{sec:howtoa_BSM}}
%
%%%%%%%%%%%%%%%%%%%%%%%%%%%%%%%%%%%%%%%%%%%%%%%%%%%%%%%%%%%%%%%%%%%%%%%%%%%%%%%%%%%%%%
%
The ratio $|V_{ub}|/|V_{cb}|$ is not expected to have a big contribution from beyond tree-level processes of semi-leptonic decays from which $|V_{ub}|$ and $|V_{cb}|$ are extracted. 
Mixing in the kaon and $B_{d,s}$ systems however 
can be particularly sensitive to physics beyond the SM (BSM). 
Their deviation from the SM can be expressed,  
without loss of generality, 
by
\bea
r_{B_{d,s}}e^{2\theta_{B_{d,s}}}&=&%\left(
\frac{\langle B_{d,s}|{\mathcal{H}}^{\rm{Total}}_{\rm{eff}} |\bar{B}_{d,s}\rangle}{\langle  B_{d,s}|{\mathcal{H}}^{\rm{SM}}_{\rm{eff}} |\bar{B}_{d,s} \rangle} %\right) 
\,,\quad \nn\\%\theta_{d,s}=\frac{1}{2}{\rm{arg}}\{ r_{B_{d,s}}\}\nn\\
r_{\ek}        &=&\frac{ {\rm{Im}}  \{ \langle K^0| {\mathcal{H}}^{\rm{Total}}_{\rm{eff}} |\bar{K}^0 \rangle \}  } {{\rm{Im}}   \{ \langle K^0| {\mathcal{H}}^{\rm{SM}}_{\rm{eff}} |\bar{K}^0 \rangle\}  }  \;. 
\label{eq:gen_ek_dms}
\eea

New CP phases may arise from the interference between mixing and decay amplitudes in the $B_{d,s}$ systems.
 Thus CP violating asymmetries would involve not only $\alpha$, $\beta$ and $\gamma$ but additionally $\theta_{B_d}$ and $\theta_{B_s}$, defined by 
\bea
\sin 2\beta=\sin(2\beta^{\rm{SM}}+2\theta_{B_d}), \quad \alpha=\alpha^{\rm{SM}}-\theta_{B_d}
\label{eq:bsm_cont_b_a}
\eea
and similarly for the $\Bs$ system whose phase information may be obtained from measurements of CP asymmetries in $\Bs\to J/\psi \phi$ which however are not yet available. For the kaon system, $\Delta m_K$ is not considered due to the lack of control in the QCD long distance effects in the $K^0$-$\bar{K}^0$ system and hence a possible additional phase BSM in this sector is neglected. In general the BSM contributions to the amplitudes of the processes responsible for  $\dms$ and $\ek$ are no longer proportional to the CKM matrix elements as in the SM,
\bea
\langle  B_{q}|{\mathcal{H}}^{\rm{SM}}_{\rm{eff}} |\bar{B}_{q} \rangle &\propto& (V_{tq}V^*_{tb})^2,\quad (q=d,s) \,, \nn\\
\langle K^0| {\mathcal{H}}^{\rm{SM}}_{\rm{eff}} |\bar{K}^0 \rangle &\propto& (V_{cd}V^*_{cs})^2, (V_{td}V^*_{ts})^2, (V_{cd}V^*_{cs})(V_{td}V^*_{ts}) \,, \label{eq:sm_effham}
\eea
but at best there will be only additional contributions with the same phase.

This is the point that can be used if we want to extract useful information of processes BSM from the analysis of the unitary triangle. Meaningful bounds on such processes can be extracted from $B^0_{d,s}$-$\bar{B}^0_{d,s}$ and $K^0$-$\bar{K}^0$ as long as the corresponding effective Hamiltonians are still proportional to the combination of CKM elements as given in~\eq{eq:sm_effham}, and such that there is only a change in the contribution to the box diagrams of these $\Delta F=2$ transitions. This amounts to a replacement of
$\eta_2 S(x_t)$ in $\ek$, as appears in~\eq{eq:ek_sm}, and a replacement of $\eta_{B} S(x_t)$ in $\dmd$ and $\dms$, as appears in~\eq{eq:dmbds_sm}, by new functions describing the contribution of particles BSM. These can be generically written as 
\bea
\dmd &=& \frac{G^2_F M^2_W}{2\pi^2} m_{B_d} \fsbd |V^*_{td}V_{tb}|^2\eta_B[S(x_t)+\delta S_{\dmd}] \,, \quad {\rm{and}}\nn\\
|\ek| &=& C_{\epsilon}B_K
A^2 \lambda^6 \bar{\eta} \left[-\eta_{cc}x_c +\eta_{ct} S(x_c,x_t)% \right.\nn \\
%& &\left.
+ A^2\lambda^4(1-\bar{\rho}) \eta_{tt}[S(x_t)+\delta S_{\ek}  ]  \right] \,,
\eea
respectively, by normalizing the QCD correction factors of the processes BSM  with respect to those of the SM.

The set of theories which satisfy these conditions is quite restrictive, but it is important to note that the bounds put on them are at the same level of precision at which the SM unitary triangle parameters can be currently determined. On the other hand we cannot test by these means new exciting theories for which there are contributions to box and penguin diagrams that are not proportional to the CKM matrix elements as the SM top contribution, or there are complex phases beyond the phases of the CKM or new local operators (contributing to the relevant amplitudes) and hence introducing additional non-perturbative factors $B_i$ and box and penguin diagrams.

%
%%%%%%%%%%%%%%%%%%%%%%%%%%%%%%%%%%%%%%%%%%%%%%%%%%%%%%%%%%%%%%%%%%%%%%%%%%%%%%%%
\section{Statistical analysis of the unitary triangle}
%
%%%%%%%%%%%%%%%%%%%%%%%%%%%%%%%%%%%%%%%%%%%%%%%%%%%%%%%%%%%%%%%%%%%%%%%%%%%%%%%%
%
%
%%%%%%%%%%%%%%%%%%%%%%%%%%%%%%%%%%%%%%%%%%%%%%%%%%%%%%%%%%%%%%%%%%%%%%%%%%%%%%%%
\subsection{Likelihood technique}
We briefly introduce here the general lines of the inference method we adopt,  
mostly for completeness and for reference in later sections. 
The general approach is well documented in~\cite{Ciuchini:2000de} where further generic clarifications may be found. 

The likelihood technique for the analysis of the unitary triangle in the SM has been widely studied. 
The Bayesian approach has been investigated by the UT Fitter group~\cite{utfit:wp}. 
We have chosen to follow such an approach in implementing our code, while 
placing emphasis on the importance of the increasing knowledge of $\dms$.

 In the Bayesian approach, the combined probability distribution for $\rb$ and $\eb$ is identified with the likelihood
\bea
{\cal{L}} (\rb, \eb)
&\propto&
\int 
\prod_{j=1,M} f(\hat{c_j} | c_j(\rb, \eb, \{x_i\})) 
\times 
\prod_{i=1,N} f_i(x_i) ~dx_i
~\times  f_0(\rb, \eb), 
\label{eqn:ckm:pdf-re}
\eea
where $f(\hat{c_j} | c_j(\rb, \eb, \{x_i\}))$ is the conditional probability density function (pdf) of the constraints  $c_j \in \{|V_{ub}|/|V_{cb}|$, $\dmd$, $\dmd/\dms$, $\ek$, $\sin 2\beta \}$, given their dependence as functions of the CKM parameters, $\rb$, $\eb$, $A$, $\lambda$ and $x_i$. 
Besides the constraints themselves, there are two classes of parameters involved: {(i)} {\it fitted}, for which pdfs $f_i(x_i)$ are constructed (e.g. the top mass); 
and {\it (ii)} {\it fixed}, which are taken as constant (e.g. the $W$ mass).

The {\it prior} probability density functions $f(\hat{c}_j)$ for the constraints are considered to be Gaussian distributions, possibly convoluted with flat distributions. 
The exception was $\dms$, prior to its first measurement, for which a more complete set of experimental information was used. 

More specifically, the pdf used in this case, in order to be consistent with the Bayesian approach, is implemented via the likelihood ratio, $R$~\eq{eqn:ckm:dmsprob}, after accessing the amplitude point $(\amp,\samp)$ associated to the frequency value obtained by evaluating the r.h.s. of  \eq{dmstheory}. 
The method is described in further detail in a following section. 

For  $f_i(x_i)$ we consider Gaussian and in some cases flat probabilities. A Gaussian pdf is chosen when the uncertainty is dominated by statistical effects, or there are many contributions to the systematics error, so that the central limit theorem applies. 
Otherwise, a flat (uniform) distribution is used for the uncertainty.
When both Gaussian and flat uncertainty components are available for a parameter, the resulting pdf is obtained by convoluting the two distributions. I.e., for an observable parameter $x$ of true value $\bar{x}$, with Gaussian and uniform uncertainty components, $\sigma_g$, $\sigma_u$, one has for the parameter and its pdf, $f(x)$, 
\begin{eqnarray*}
x    = \bar{x} + x_g + x_u \quad \rightarrow \quad
f(x) = \delta(x-\bar{x}) \otimes \rm{Gaus}(x|\sigma_g) \otimes \rm{Unif}(x|\sigma_f).
\end{eqnarray*}

The integration of \eq{eqn:ckm:pdf-re} can be performed using Monte Carlo methods, 
followed by its normalization.  
Hence we can calculate the resulting pdfs for parameters involved in the fit. 

The expression of~\eq{eqn:ckm:pdf-re} shows explicitly that whereas \emph{a priori}  all values of $\rb$ and $\eb$ are equally likely by assumption, \emph{i.e.} $f_0(\rb,\eb)=const.$, \emph{a posteriori}  the probability clusters in a region of maximal likelihood.

The probability regions in the $(\rb$,$\eb)$ plane are constructed from the pdf obtained in \eq{eqn:ckm:pdf-re}. These are called \emph{highest posterior density} regions, 
and are defined such that ${\cal{L}} (\rb, \eb)$ is higher everywhere inside the region than outside, 

$
P_w := \{ 
z=(\rb,\eb) ~:~ \int_{P_w} {\cal{L}}(z) dz = w; 
~{\cal{L}}(z') < min_{P_w}{\cal{L}}(z), \forall_{z' \notin P_w} 
\}
$.

The one dimensional pdfs are obtained similarly. For example, the $\rb$ pdf is obtained as ${\cal{L}} (\rb) \propto \int {\cal{L}} (\rb, \eb) d\eb$, from which its expected value can be calculated together with the corresponding highest posterior density intervals. 

A similar procedure can be in principle used in order to obtain the pdf for other desired parameters. Technically, one may also use the probability function for transformed variables; \emph{i.e.}, that for ${\bf{u}} ({\bf{x}})$ one has 
$f({\bf{u}}) = f({\bf{x}}) |\partial {\bf{x}}/\partial {\bf{u}}|$,
where the last factor denotes the Jacobian. In this way, the pdf for a parameter $x$ can be obtained from 
\bea
{\cal{L}} (x)
&\propto&
\int {\cal{L}} (x,  \eb) ~d\eb \propto
\int {\cal{L}} (\rb, \eb) ~ \lt|\frac{d\rb}{dx}\rt| ~d\eb,
\label{eq:lik_parx}
\eea
where ${\cal{L}} (\rb, \eb)$ has been computed in \eq{eqn:ckm:pdf-re} above.

Besides the probability distribution in the $(\rb,\eb)$ plane, we are most interested in obtaining the \emph{posterior} probability distribution for the $\dms$ parameter.

%
%%%%% %%%%%%%%%%%%%%%%%%%%%%%%%%%%%%%%%%%%%%%%%%%%%%%%%%%%%%%%%%%%%%%%%%%%%%%%%%%%%%%%%
%%%%%%%%%%%%%%%%%%%%%%%%%%%%%%%%%%%%%%%%%%%%%%%%%%%%%%%%%%%%%%%%%%%%%%%%%
%
%
%%%%%%%%%%%%%%%%%%%%%%%%%%%%%%%%%%%%%%%%%%%%%%%%%%%%%%%%%%%%%%%%%%%%%%%%%
%
%
%%%%%%%%%%%%%%%%%%%%%%%%%%%%%%%%%%%%%%%%%%%%%%%%%%%%%%%%%%%%%%%%%%%%%%%%%%%%%%%%%%%%%%
%
\section{Using the available ${\mathbf{B_s}}$ flavour oscillation results}
%
%%%%%%%%%%%%%%%%%%%%%%%%%%%%%%%%%%%%%%%%%%%%%%%%%%%%%%%%%%%%%%%%%%%%%%%%%%%%%%%%%%%%%%
\subsection{Experiments -- short overview}

The study of flavour oscillations in the $B^0_s$--$\bar{B}^0_s$ system has been the subject of many past experimental analyses, performed at ALEPH, CDF, DELPHI, OPAL and SLD. 
A wide range of data analysis techniques has been employed, which have been developed for taking advantage of detector capabilities and the characteristics of the collected data samples.  These include from fully exclusive, where the $B_s$ decays are completely reconstructed, to fully inclusive methods, which aim at identifying the mesons decay vertices; along with several $b$-flavour tagging techniques.   
The most sensitive among these analyses were those based on semi-exclusive or semi-inclusive lepton samples, where the $D_s$, resulting from the semi-leptonic $B_s$ decay, was either inclusively or exclusively reconstructed. 

In general, the more inclusive analyses benefit from considerably larger number of candidates, while the more exclusive ones take advantage of the more precise information about the $B$ decay candidates which is available. 
In practice, therefore, the resulting sensitivity to $B$ oscillations is dependent on a trade-off between the quantity and the quality of the events forming the samples.  Furthermore, the relative weight of the latter increases quickly when probing higher oscillation frequencies.
Accordingly, the significance of an oscillation signal at relatively lower frequencies benefits more readily from high statistics samples, provided the events are characterized by sufficiently adequate resolutions. This has to some extent been the case until rather recently.
The situation becomes different for a signal of a higher frequency, as is known to be the case of the $\Bs$ system --- a measurement of $\dms$ requires samples with very good effective resolutions together with adequately large yields.

A new generation of $\Bs$ mixing analyses is being undertaken by the CDF and D\O\ collaborations, which are accumulating large samples of $\Bs$ meson decays produced in the $p\bar{p}$ collisions of the Tevatron at Fermilab. 
Both collaborations reported first preliminary results during the year of 2005~\cite{eps05-panic05-conf}.

Samples of partially reconstructed semi-leptonic $B_s \to D_s l \nu$ decays have been collected by both experiments, whereas CDF has also gathered substantial samples of   fully reconstructed $B_s \to D_s (\pi\pi) \pi$ decays. The $D_s$ meson is reconstructed in several exclusive channels.

The partially reconstructed decay samples have originally provided the predominant contributions to the combined sensitivity, in view of the considerably larger yields. Nevertheless, as the sizes of the fully reconstructed decay samples increased, these have started to provide unprecedented effective resolution and large sensitivities at higher probe frequencies.

The preliminary results reported last year have contributed already significantly to the world average~\cite{hfag:webpage}, increasing the combined sensitivity to $20~\ips$ and further pushing the lower exclusion limit. 

Tevatron projections based on the those preliminary analyses have demonstrated the potential in Run II for covering the SM favored region for $\dms$.  
Indeed this has already taken place most recently as, after using optimized analysis techniques and increased datasets, sensitivities close to 30~$\ips$ have been reached. And indeed evidence of oscillation signal within the SM favored region 
has been established. 

Following the Tevatron the excitement of the study of the $\Bs$ system 
will be transferred to the LHC experiments: the LHCb, as well as the general purpose detectors ATLAS and CMS. 
A confirmation of the $\dms$ measurement, with an increased precision, is readily expected. 
Additional analyses will be pursued, which will provide crucial information 
for instance on the phase of the mixing amplitude and on CP violating processes.

\subsection{Measurements -- amplitude scan}

The study of oscillations is based on the analysis of the proper decay time $t$ of flavor tagged $\Bs$ candidates. 

It involves the measurement of the $B$ meson decay distance and momentum, as well as a determination of whether the particle-antiparticle system decayed with the same or the opposite flavor with which it was originally produced. 

The model used to describe the proper time distribution involves the exponential decay dictated by the system's lifetime $\tau$, modulated by an oscillating term describing the probability for the $B_s$ to have mixed,
\begin{equation}
P(t) \sim \frac{1}{\tau} \exp^{-\frac{t}{\tau}} 
\left(1 - \amp \, \mathrm{cos} \dms t \right)
\;.
\label{eqn:t-pdf}
\end{equation}
This expression needs to be adapted to take into account detector resolutions, reconstruction effects, and imperfections of the $b$-flavour tagging methods used.

The parameter $\amp$ has been introduced in~\ref{eqn:t-pdf} in an \emph{ad-hoc} fashion  for accessing the amplitude of the oscillations using the so-called amplitude method~\cite{Moser:1996xf}. Instead of extracting from the data the quantity of interest -- $\dms$ -- directly, in this method the likelihood is maximized as a function of $\amp$, while $\dms$ is a fixed parameter of the fit. 
The procedure is repeated for many different probe values of $\dms$, and the result is  thus conveyed as a set of measured $\amp(\dms)$ and $\samp(\dms)$ in a spectrum of probed oscillation frequencies. 
The value of $\amp$ should fluctuate around zero unless the true oscillation frequency of the system is being probed, for which case $\amp$ should be consistent with unity, within errors.

A given frequency value is excluded at 95\%~CL, following a one-sided Gaussian test,   if the corresponding amplitude and its uncertainty satisfy $A + 1.645 \samp<1$; 
the lower exclusion \emph{limit} of an experiment is defined as the value below which all frequencies satisfy the latter condition. 
The \emph{sensitivity} of an analysis is defined as the largest frequency for which $\,1.645 \samp<1$ holds; \emph{i.e.} it is the largest excluded frequency if 
$\amp$ were exactly zero as it is expected to be the case in the absence of a signal, removing the dependence on fluctuations of the central value $\amp$ which affects limits. 

The amplitude method was originally proposed~\cite{Moser:1996xf} as a convenient approach to set limits, as well as for combining such exclusion regions obtained with different analyses. It has provided a consistent fitting procedure through which 
all experiments have commonly expressed their results. 
The combination of amplitude results is standardly done most straightforwardly as averages of Gaussian measurements. 
This is expected to hold in general to good approximation; more detailed combinations could be performed for instance by combining the likelihood profiles for each $\amp$ measurement, which however would not be feasible as it would be based on detailed information which is no longer available for previous measurements. 

In the averages performed by HFAG~\cite{hfag:webpage} the measurements are adjusted on the basis of common physics input values, and possible statistical correlations are taken into account. 

Prior to the inclusion of the Tevatron Run II results, the world-average exclusion limit was~\cite{hfag:webpage} 
\bea
{\dms} > 14.5~\ips \qquad (95\%~\rm{CL}) \,, 
\label{eqn:dms:prerunii}
\eea
with a sensitivity of 18.2~$\ips$. 

With the inclusion of the Tevatron results reported in the past year (2005)~\cite{eps05-panic05-conf},the limit was increased to
\bea
{\dms} > 16.6~\ips \qquad (95\%~\rm{CL}) \,, 
\label{eqn:dms:f05}
\eea
with a combined sensitivity of 20.0~$\ips$. 

More recently, after the analysis of the first fb$^{-1}$ of Tevatron data, both collaborations have reported their corresponding results. 
D\O\ has published a first direct double-sided experimental bound~\cite{d0-prl06} 
placing $\dms$ in the interval 
\bea
17 < {\dms} < 21 ~\ips \qquad  (90\%~\rm{CL}) \,; 
\eea
the measured sample sensitivity is 14.1~$\ips$.  
CDF has published a first direct, precise measurement~\cite{cdf-prl06a},  
\bea
{\dms} &=& 17.31^{+0.33}_{-0.18} (stat.) \pm 0.07 (syst.)~\ips \,,
\label{eqn:dms:cdf06}
\eea 
having been extracted from data samples whose measured sensitivity is 25.8~$\ips$; 
the measurement has already a 2\% level precision.  The probability for the observed signatures being due to background fluctuations were evaluated by the collaborations  to be of the order of 5\% (D\O\ ) and 0.2\% (CDF). 
This is still below the threshold commonly set for attaining the
status of \emph{observation} ($5\sigma$); a (potential) confirmation of the measurement with an even larger significance is being awaited.%

\subsection{Likelihood implementation}

In the fit to the CKM parameters we use the complete information about $\dms$ which is provided by the full amplitude scan. 

\begin{figure}[H]
~\quad
\begin{minipage}[t]{6.8cm}
\begin{center}
\includegraphics[width=1\textwidth]{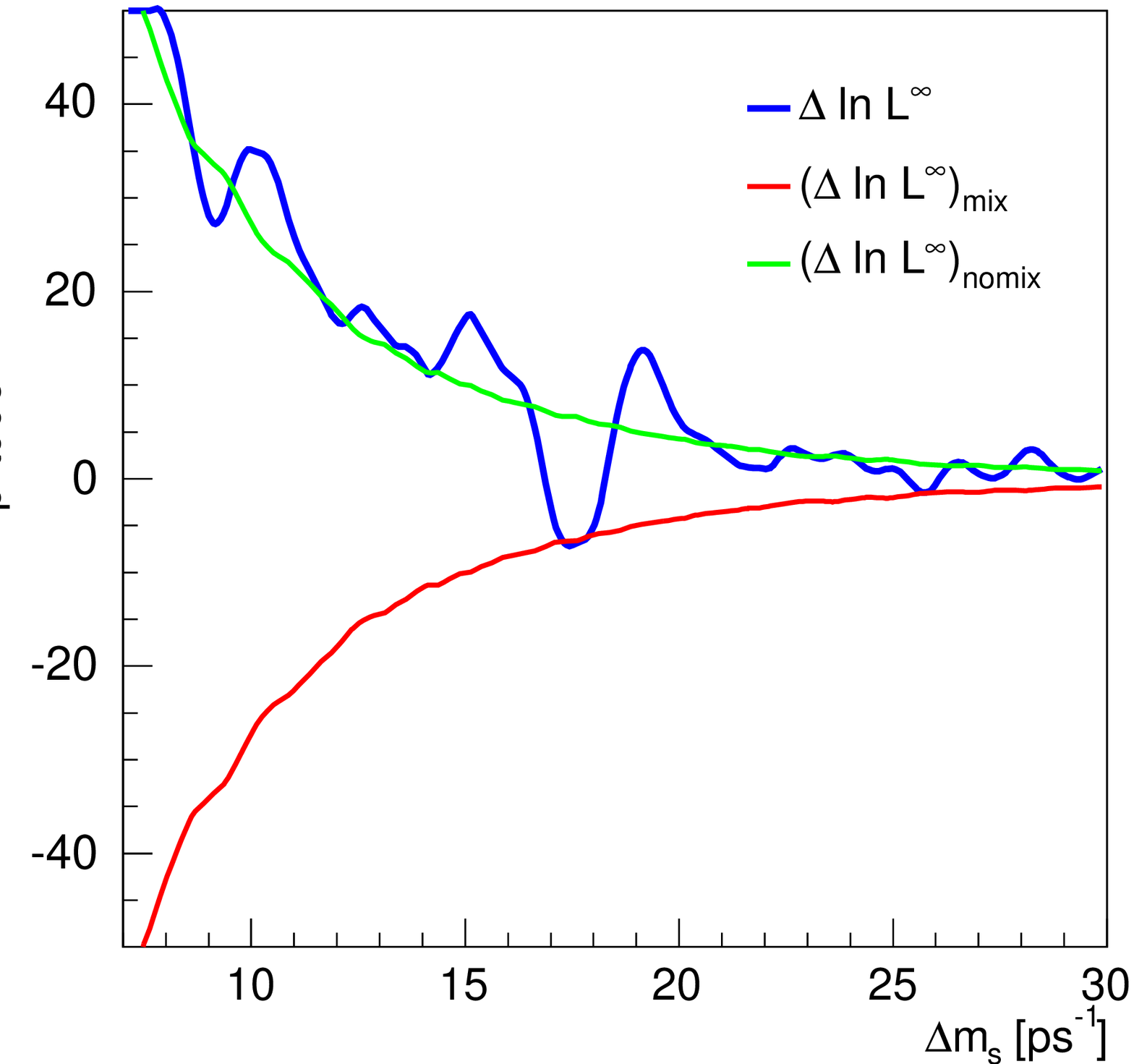}
\vspace*{-0.3cm}
(i)
\end{center}
\end{minipage}
~
\begin{minipage}[t]{6.8cm}
\begin{center}
\includegraphics[width=1\textwidth]{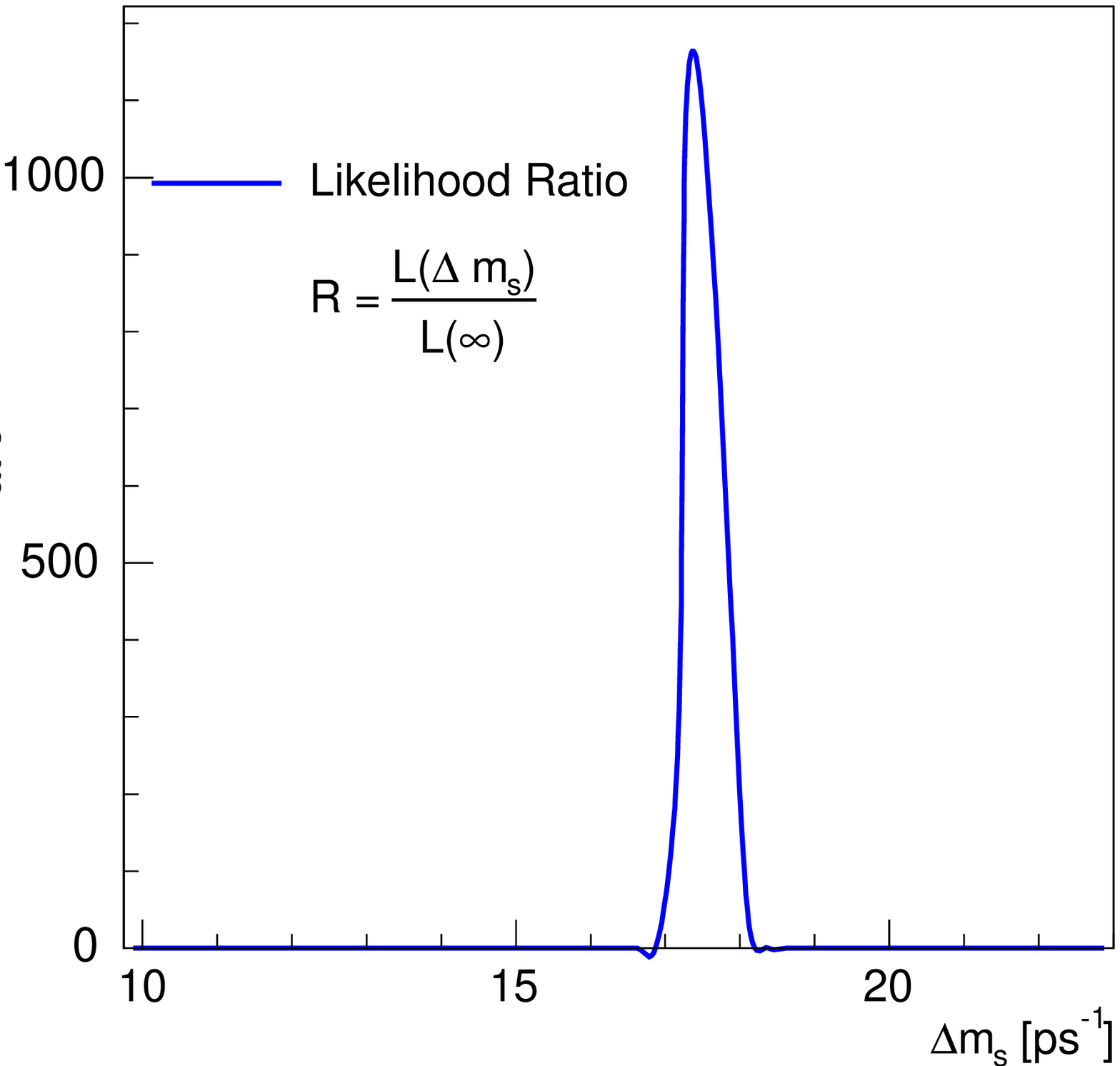}
(ii)
\end{center}
\end{minipage}
\hfill
\vspace*{-0.3cm}
\caption{\small Likelihood for $\dms$ obtained from the combined amplitude measurements, including Tevatron's Run II.}
\label{fig:lhood-dms}
\end{figure}

The measured values of the amplitude and its uncertainty, $\amp$ and $\samp$, may be used to derive~\cite{Moser:1996xf}, in the Gaussian approximation, the log-likelihood function, $\Delta \mathrm{ln} {\cal{L}}^{\infty}(\dms)$, referenced to its value for an infinite oscillation frequency
\beas
\Delta \mathrm{ln} {\cal{L}}^{\infty}(\dms)
&=&
\ln {\cal{L}}(\infty) - \ln {\cal{L}}(\dms) 
=
\lt( \frac{1}{2}-\amp \rt) \frac{1}{\samp^2},\\
\Delta \mathrm{ln} {\cal{L}}^{\infty}(\dms)_{\rm{mix}}
&=&
- \frac{1}{2} \frac{1}{\samp^2},\quad
\Delta \mathrm{ln} {\cal{L}}^{\infty}(\dms)_{\rm{nomix}}
\ \, =\ \,
+ \frac{1}{2} \frac{1}{\samp^2} \,.
\eeas
The last two relations give the expected average log-likelihood value for the cases when $\dms$ corresponds to the true (\emph{mixing} case) or is far from (\emph{no-mixing} case) oscillation frequency of the system, characterized respectively by unity and null expected amplitude values. These are shown in Figure~(\ref{fig:lhood-dms} i).
The log-likelihood difference, according to the central limit theorem of likelihood theory is $\chi^2$-distributed: $\Delta \ln {\cal{L}} = \frac{1}{2} \chi^2$. We translate therefore the amplitude scan into the likelihood ratio 
\be
R(\dms) 
= e^{-\Delta ln {\cal{L}}^{\infty}(\dms)}
= \frac{{\cal{L}}(\dms)}{{\cal{L}}(\infty)}
= e^{-\frac{\frac{1}{2}-\amp(\dms)}{\samp^2(\dms)}},
\label{eqn:ckm:dmsprob}
\ee
through which the constraint for $\dms$ is implemented in the fit. This is represented in Figure~(\ref{fig:lhood-dms}-ii).

We re-iterate that the exponent in \eq{eqn:ckm:dmsprob} corresponds to the $\chi^2$, or log-likelihood, difference between the cases where an oscillation signal is present and absent, for which the true amplitude value is $1$ and $0$, respectively. 
Note that hypotheses for $\dms$ associated with larger and more precisely measured $\amp$-values in the scan contribute a larger weight in the fit.

\subsection*{Extending the amplitude spectrum}
The amplitude fits are usually performed for $\dms$ values lower than a given threshold, beyond which the fit behavior may become unstable. On the other hand, in the framework of the CKM fit, it is in principle desirable to have $R$ defined for all positive frequency values, which in turn demands for a continuation of the amplitude spectrum beyond what is measured. 

The extrapolation of the value of $\samp$ may be achieved through an analytical, Fourier-based description~\cite{Moser:1996xf} of the measured significance curve,  
\beas
\samp \propto {e}^{\frac{1}{2}\sigma_t^2 \dms^2}\,,
\eeas
where $\sigma_t$ denotes the uncertainty in the measured proper decay time
(this expression would in fact need to be properly averaged over the samples signal uncertainty distributions).

An extrapolation of $\amp$ itself as such is not possible, however it is here sufficient to note that its expected values lie in the vicinity of either zero or unity as already mentioned. Therefore, for the unmeasured part of the spectrum, the exponent in ~\eq{eqn:ckm:dmsprob} becomes small, and quickly approaches zero. We take this asymptotic limit as the criterion for extending $R$ beyond the experimentally probed frequencies. 
\begin{figure}[H]
%\hfill
\begin{minipage}[t]{5.7cm}
\begin{center}
\vspace*{-5.4cm}
\includegraphics[width=1\textwidth]{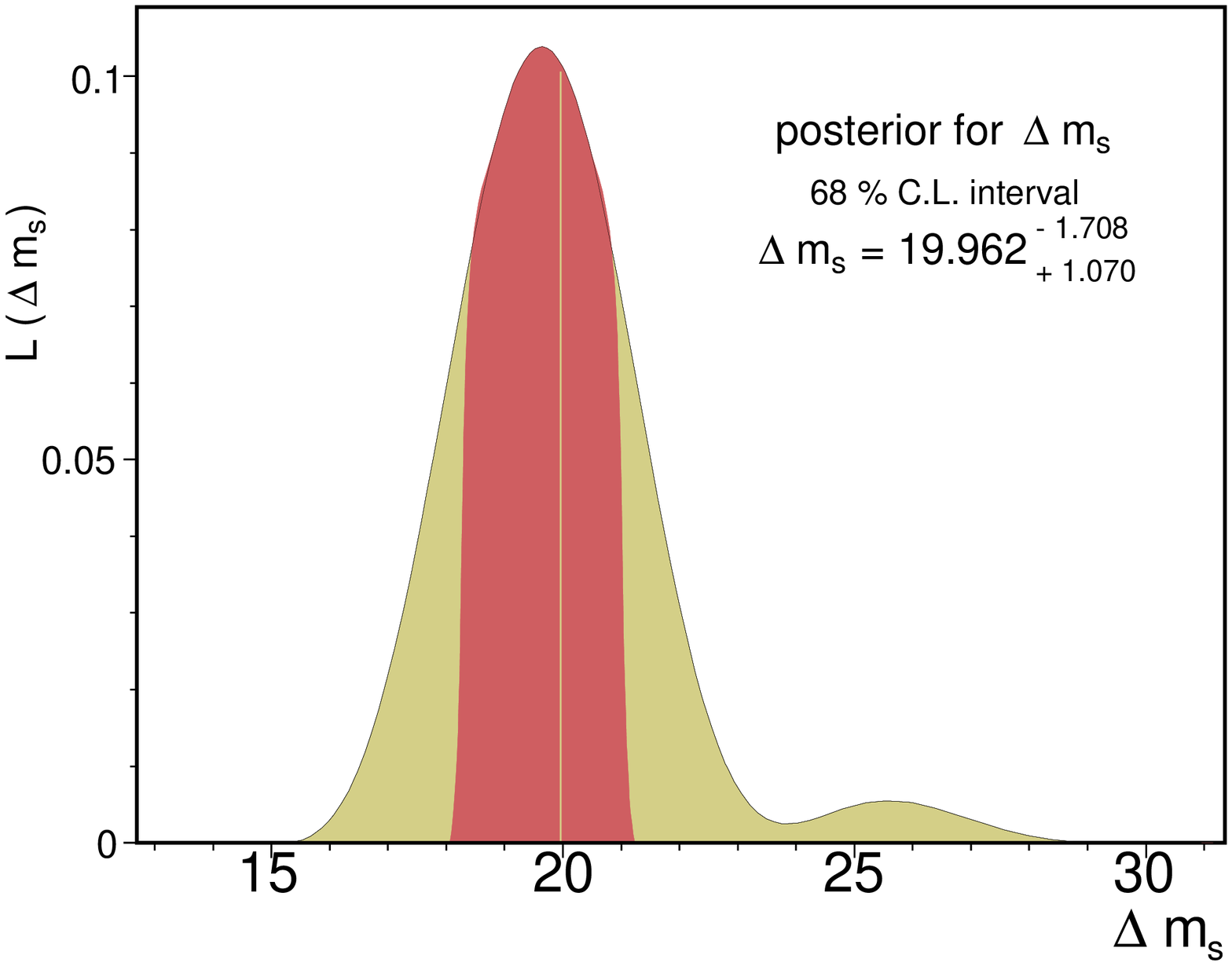}%\caption{\small (i)}
(i)
\end{center}
\end{minipage}
\hfill
\begin{minipage}[t]{5.7cm}
\includegraphics[width=1\textwidth]{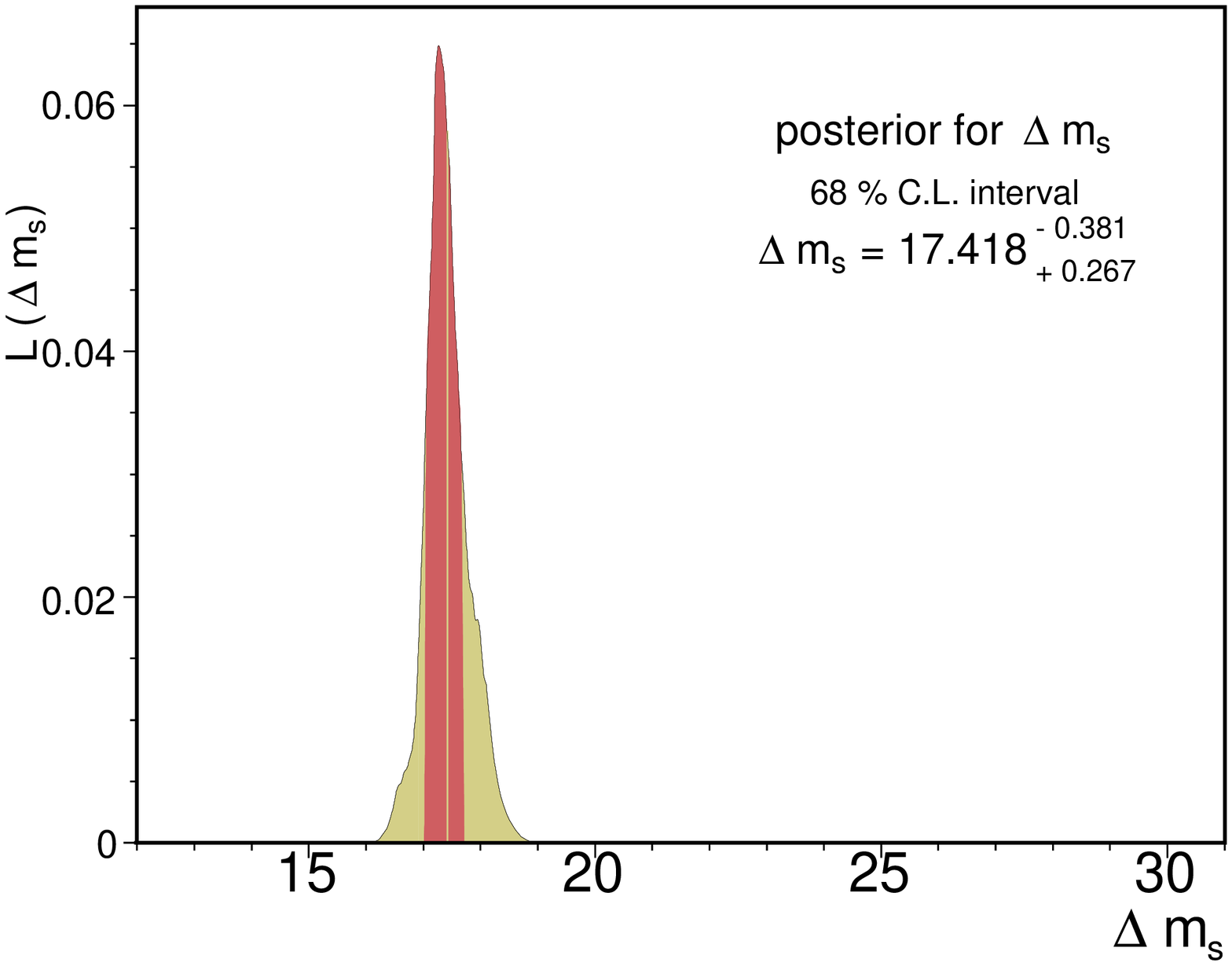}%\caption{\small (ii)}
(ii)
\end{minipage}
\begin{minipage}[t]{5.7cm}
\begin{center}
\includegraphics[width=1\textwidth]{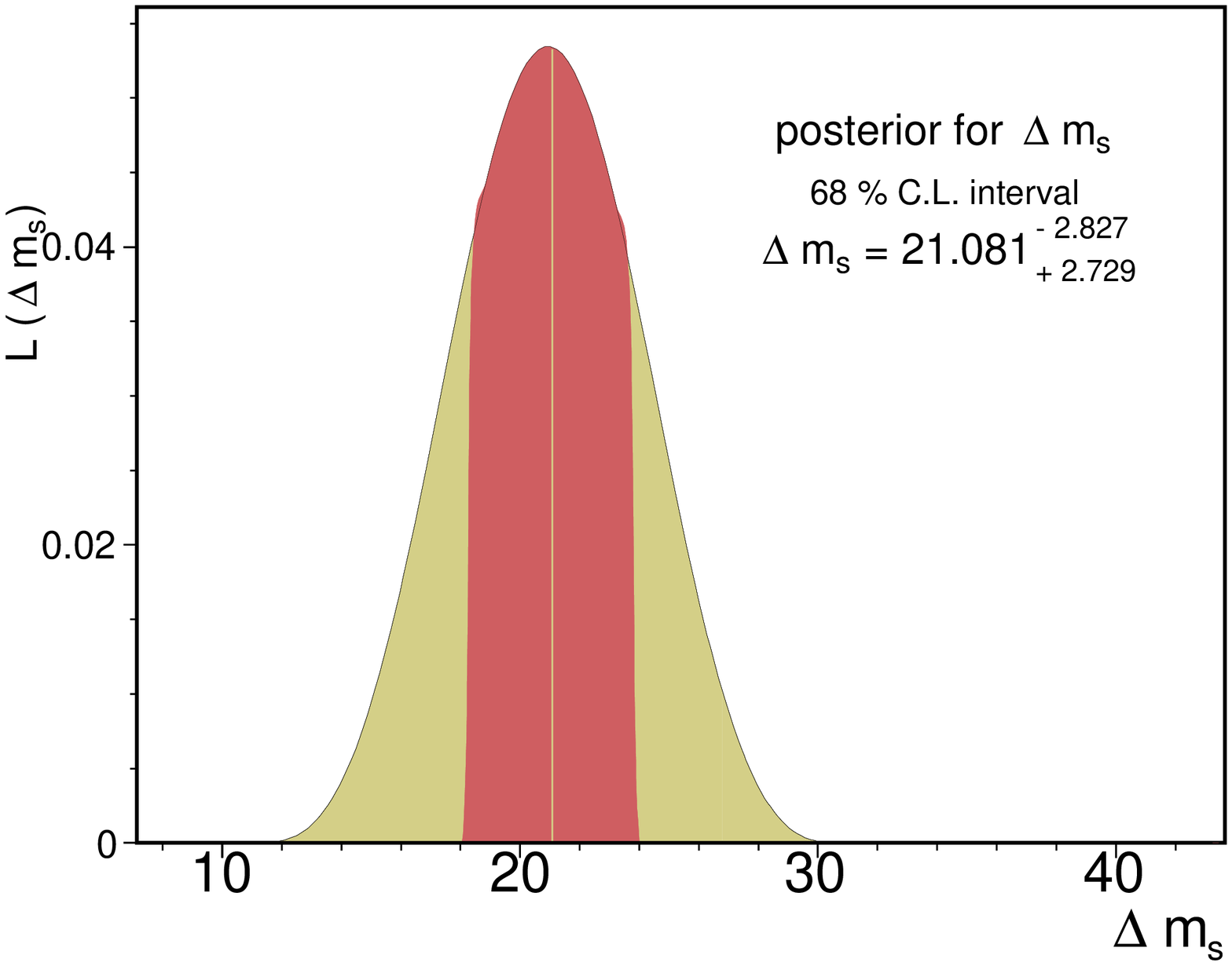}%\caption{\small $\dms$}
(iii)
\end{center}
\end{minipage}
\hfill
\vspace*{-0.2cm}
\caption{\small Probability density function of $\dms$ as a result of the unitary triangle fit using (i) the data corresponding to $\dms>16.6$ ps $^{-1}$ bound, and (ii) the data corresponding to $(\dms=17.33\pm 0.4)$ ps $^{-1}$; (iii) shows the output of $\dms$ without using its constraint in the fit.}
\label{fig:dms}
\end{figure}

\clearpage
%
%%%%%%%%%%%%%%%%%%%%%%%%%%%%%%%%%%%%%%%%%%%%%%%%%%%%%%%%%%%%%%%%%%%%%%%%%%%%%%%%%%%%%%%%%%%%%%%%%%%%%%%%%%%%%%%%%%%%%%%%%%%%%%%%%%%%%%%%%%%%%%%%%%%%%%%%%%%%%%%%%%%%%%%%%%%                                       
%
\section{Tests of the SM in the UT analysis}
%
%
%%%%%%%%%%%%%%%%%%%%%%%%%%%%%%%%%%%%%%%%%%%%%%%%%%%%%%%%%%%%%%%%%%%%%%%%%%%%%%%%%%%%%%%%%%%%%%%%%%%%%%%%%%%%%%%%%%%%%%%%%%%%%%%%%%%%%%%%%%%%%%%%%%%%%%%%%%%%%%%%%%%%%%%%%%%
%
%%%%%%%%%%%%%%%%%%%%%%%%%%%%%%%%%%%%%%%%%%%%%%%%%%%%%%%%%%%%%%%%%%%%%%%%%%%%%%%%%%%%%%
\subsection{Classic fit with {\emph bound} and {\emph first measurement} of ${\mathbf{\dms}}$} 
%
%%%%%%%%%%%%%%%%%%%%%%%%%%%%%%%%%%%%%%%%%%%%%%%%%%%%%%%%%%%%%%%%%%%%%%%%%%%%%%%%%%%%%
%
\begin{figure}[ht]
%\vspace*{-0.5.7cm}
\begin{center}
\includegraphics[width=10cm]{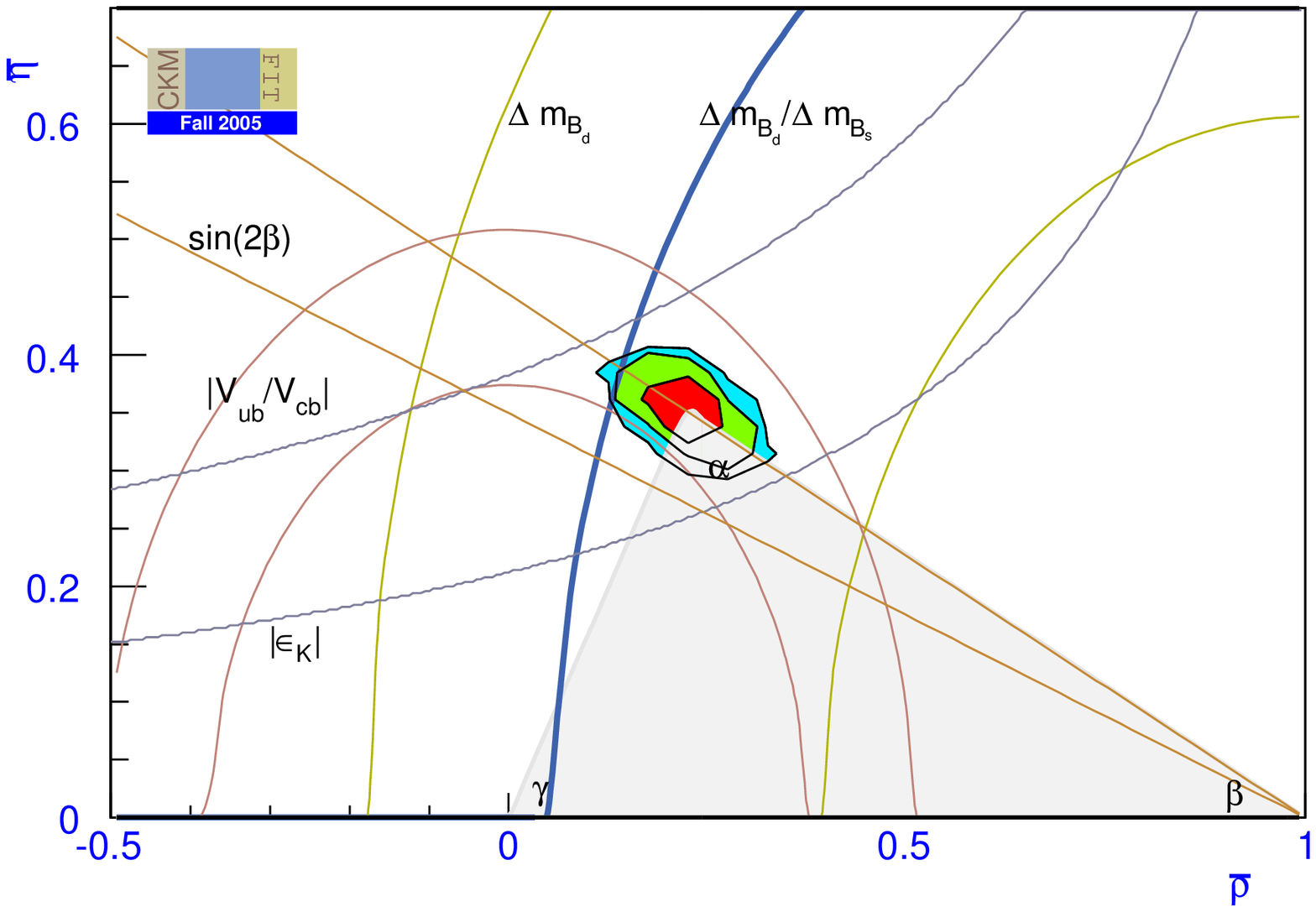}(i)
\includegraphics[width=10cm]{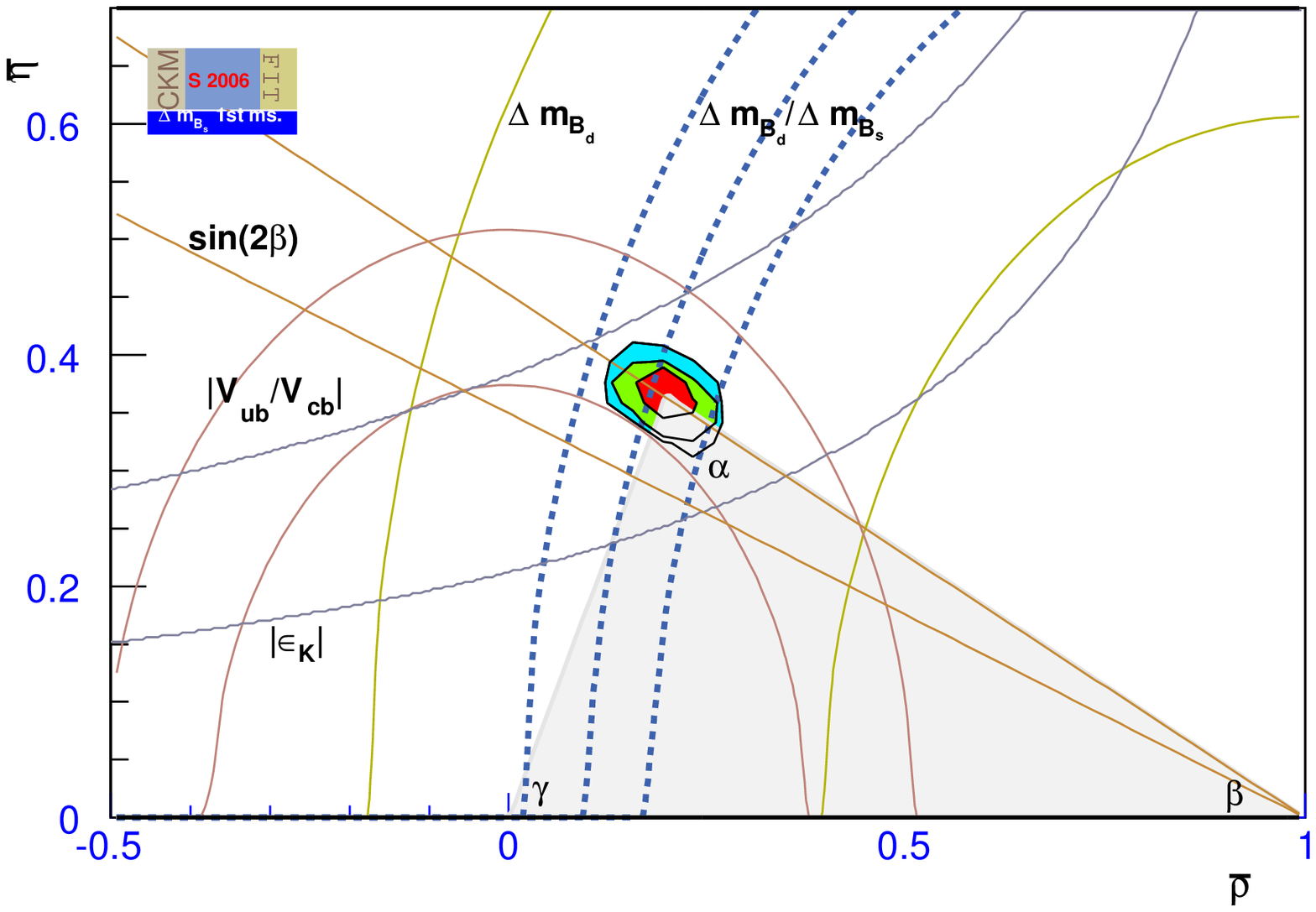}(ii)
\vspace*{-0.4cm}
\caption{\small The first (i) of these plots represents the $(\rb,\eb)$ plane with the bound results and the second one (ii) with the data of the first measurement on $\dms$.} 
\label{fig:rb_eb_2d_r0}
\end{center}
\end{figure}
Although $\Delta F=1$ processes such as the penguin transitions occurring in the charmless decays $B\to \pi\pi,~ \rho\pi,~ \rho\rho$, which determine $\alpha$, and the CP asymmetries in semi-leptonic $B$ decays ($B\to l^- X$) are important, we do not consider these constraints here. 
The experimental constraint on $\alpha$ through the mentioned decays is currently sufficient even without necessarily assuming the SM, but the relation $\alpha=\pi-\beta-\gamma$ is still used for this bound where $\gamma$ here comes from the decay amplitudes and $\beta$ from the $\Bdmix$ mixing, so this $\Delta F=2$ process imposes already a good constraint on $\alpha$.

 The CP asymmetry in semi-leptonic decays, $A_{SL}$, is a crucial constraint of the UT analysis because once $r_{B_d}$ and $\theta_{B_d}$ are defined through~\eq{eq:gen_ek_dms}, the generalization of $A_{SL}$ to account for BSM processes depends on  $r_{B_d}$ and $\theta_{B_d}$ \cite{Bona:2005eu} at the NLO in the penguin term in the  $\Delta F=1$ amplitude. However the present experimental bounds are not precise enough to put constraints on $\rb$ and $\eb$ \cite{Laplace:2002ik}.

The two main fits that we have performed include the constraints $|V_{ub}|/|V_{cb}|$, $\dmd$, $\dmd/\dms$, $\ek$ and $\sin 2\beta$. 
The first one (i) uses the bound ~\eq{eqn:dms:f05}, and the second one (ii) uses the first measurement~\eq{eqn:dms:cdf06} by CDF.{\footnote{In the original version of this work we used the available information from the ${B}^0_{d,s}$-$\bar{B}^0_{d,s}$ oscillation experiments for the summer conferences of last year ($\dms>14.5~\ips$) and other fit using the data after the summer and fall conferences ($\dms>16.6 ~\ips$), averaged by Heavy Flavor Averaging Group (HFAG), using the information by the  CDF, D\O\ , ALEPH, DELPHI and OPAL collaborations.}

In~\fig{fig:rb_eb_2d_r0}(i) we present the 68\%, 95\% and 99\%~CL of the two dimensional pdf for $\rb$ and $\eb$ with the values of Tables (\ref{tbl:fixedp}) and (\ref{tbl:fittedp}) using the bound~\eq{eqn:dms:f05} and in~\fig{fig:rb_eb_2d_r0}(ii) using the current measurement \eq{eqn:dms:cdf06}.    

As we can see in the $(\rb,\eb)$ plane, the radius of the circle describing the constraint $\dmd/\dms$ has been reduced and with it the overlap region of all the constraints, specially for $|V_{ub}|/|V_{cb}|$ and $\sin 2\beta$. As a consequence the central value of $\rb$ has been increased and the central value of $\eb$ has been slightly reduced. In \fig{fig:rb_eb_r0a} and \fig{fig:rb_eb_r0b} we present the pdfs of $\rb$ and $\eb$ for these cases. As we can see by comparing these figures, the one dimensional (1d) pdf of $\rb$ has been shifted to the right while the pdf of $\eb$ has been left practically unchanged. In \fig{fig:dms} we present the $\dms$ 1d projection using the (i) bound,  %and (ii) the fit using data of the $\dms$ 
(ii) the measurement, and (iii) the fit output of $\dms$ without using this constraint. % in the fit.

As expected with the current measurement of $\dms$, the overlap region 
within the %between is 
allowed 68\%~CL can determine with a smaller uncertainty the values of $\rb$ and $\eb$. The last bound of $\dms$ was closer to the region of maximal likelihood of the other constraints, and the relative uncertainties of $\rb$ and $\eb$ were $15\%$ and $9\%$ respectively, in contrast to the current $11\%$ and $4\%$.

We also compare how well the parameters defining a unitary triangle, $x=$ $R_b$, $R_t$, $\alpha$, $\beta$ and $\gamma$, are fitted in each of the cases (i) and (ii), by a $\chi^2$ relative to the experimental value,  
\bea
%\chi^2_{x}=\left|\frac{\chi_{\hat{x}}-\chi_{x}}{\sigma_{\hat{x}}}\right|^2,
\chi^2_{x}=\left|\frac{{\hat{x}}-{x}}{\sigma_{\hat{x}}}\right|^2,
\eea
where $\hat{x}$ is computed as given in \eq{eq:Rb_Rt} with the values of $\rb$ and $\eb$ returned by the fit, as shown in Figures (\ref{fig:rb_eb_r0a}, \ref{fig:rb_eb_r0b}), and $x$ represents the values as shown in Figures (\ref{fig:Rb_Rt_r0a}) and (\ref{fig:Rb_Rt_r0b}). We find that the fit including the measurement of $\dms$ (ii) is fitted slightly worse to the constraints on unitary than the bound (i) fit:
\bea
\chi^2_{R_{t\,(i)}}      &=& 2.3 \times 10^{-6}, \quad \
\chi^2_{R_{t\,(ii)}}      = 5.9 \times 10^{-5}, \quad \nn\\
\chi^2_{R_{b\,(i)}}      &=& 0.008   \quad \quad \quad \quad
\chi^2_{R_{b\,(ii)}}      = 0.018\nn\\
\chi^2_{\sin 2\beta_{\,(i)}}  &=& 1.1 \times 10^{-3} ,\ 
\quad \chi^2_{\sin 2\beta_{\,(ii)}} = 1.4 \times 10^{-2}\nn\\
\chi^2_{\sin\gamma_{\,(i)}}  &=& 1.2 \times 10^{-6} , \quad 
\chi^2_{\sin\gamma_{\,(ii)}} = 5.8  \times 10^{-4}.
\eea
Since $\alpha$ is determined through the constraint $\alpha=\pi-\beta-\gamma$, 
the difference in $\sin\alpha$ %$\delta \sin\alpha$ 
has no meaning for this kind of comparison.
\begin{figure}[ht]
%\hfill
\vspace*{-0.2cm}
\begin{minipage}[t]{5.7cm}
{\centering
\includegraphics[width=1\textwidth]{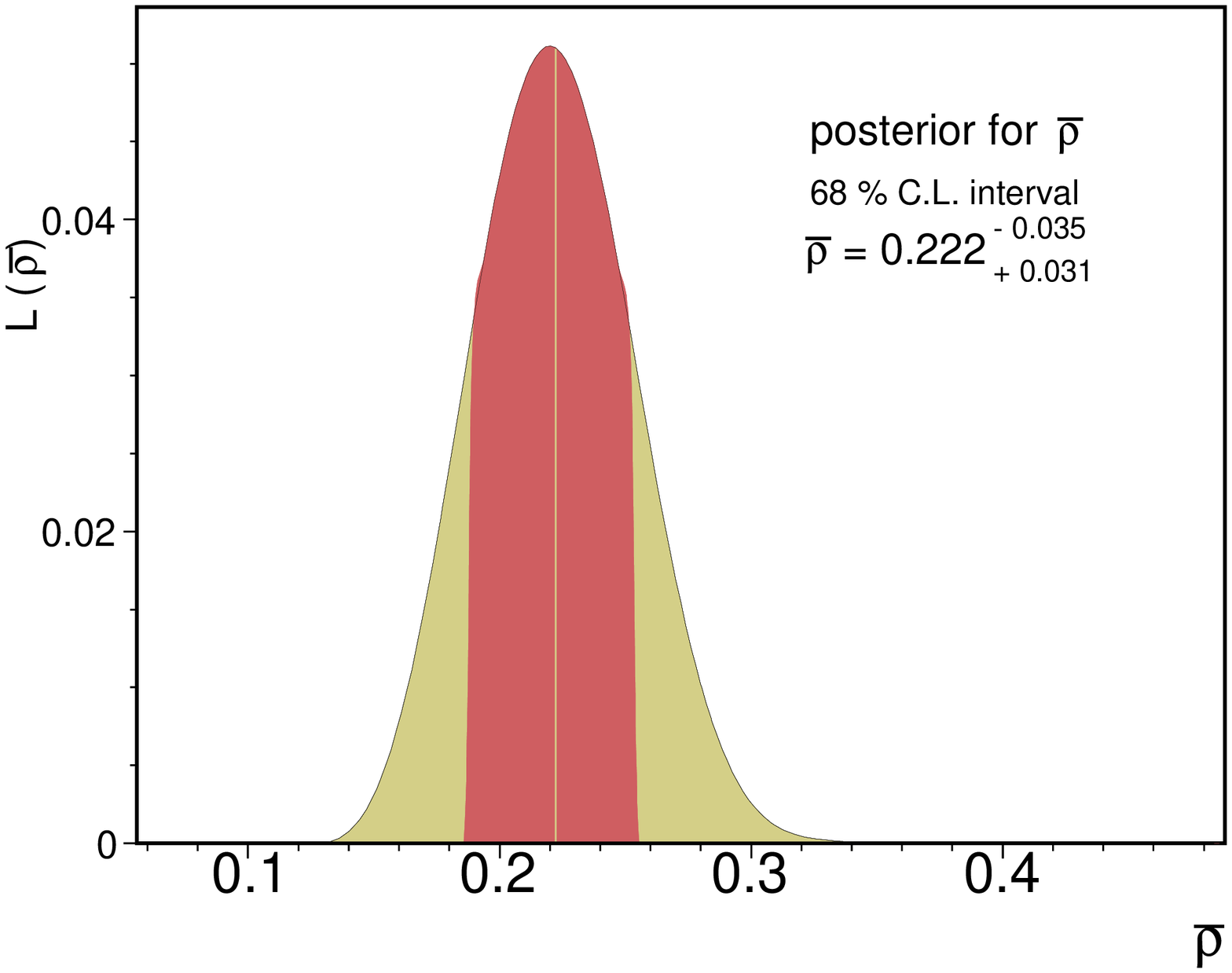}%\caption[$\rb$]{$\rb$} \label{fig:rb_r0a}
}
\end{minipage}
\hfill
\begin{minipage}[t]{5.7cm}
{\centering
\includegraphics[width=1\textwidth]{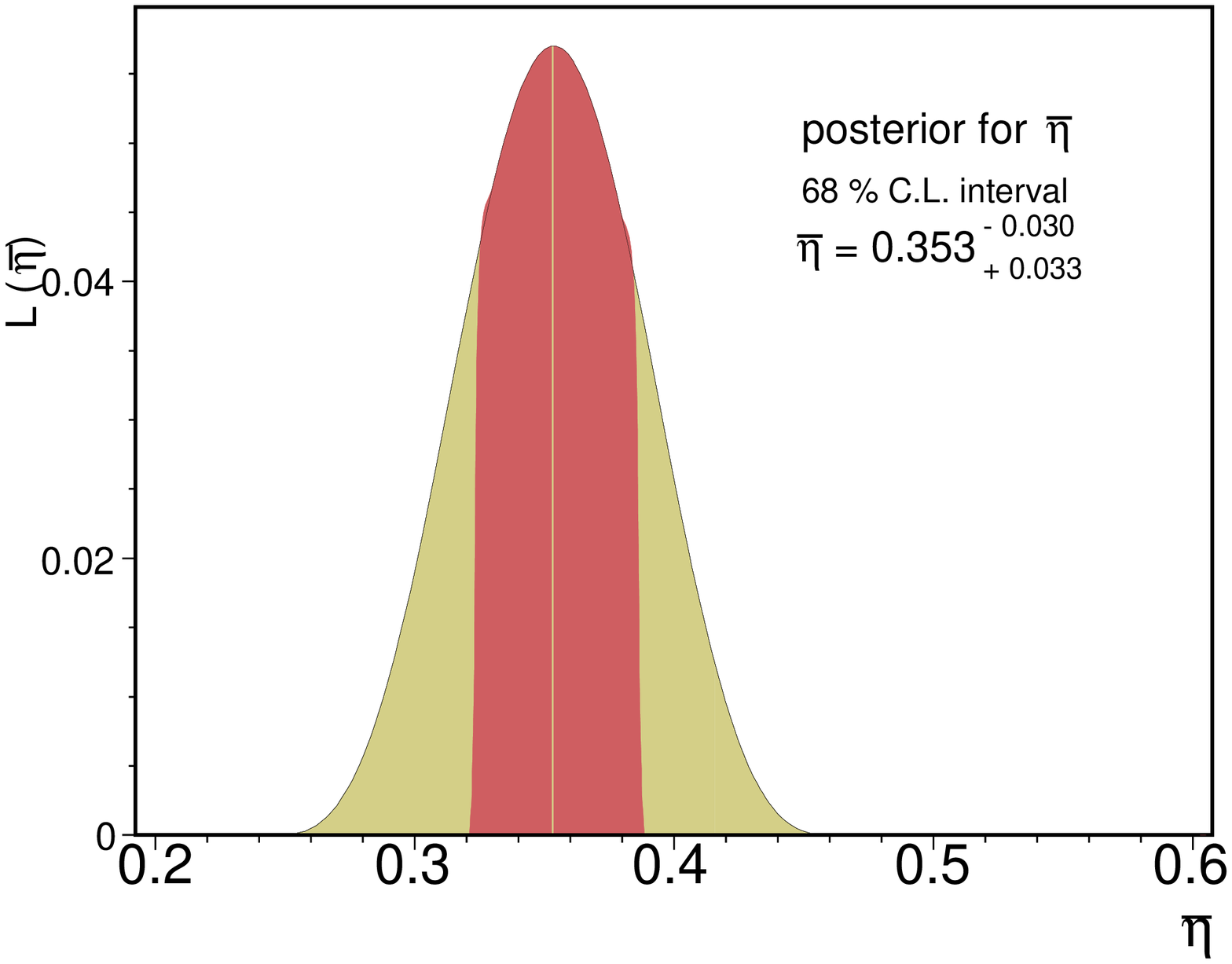}%\caption[$\eb$]{$\eb$} \label{fig:eb_r0a}
}
\end{minipage}
\hfill
\vspace*{-0.8cm}
\caption{\small $\rb$ and $\eb$ one dimensional projections using the bound results~\eq{eqn:dms:f05}.}
\label{fig:rb_eb_r0a}
\end{figure}
\begin{figure}[ht]
%\hfill
\vspace*{-0.6cm}
\begin{minipage}[t]{5.7cm}
{\centering
\includegraphics[width=1\textwidth]{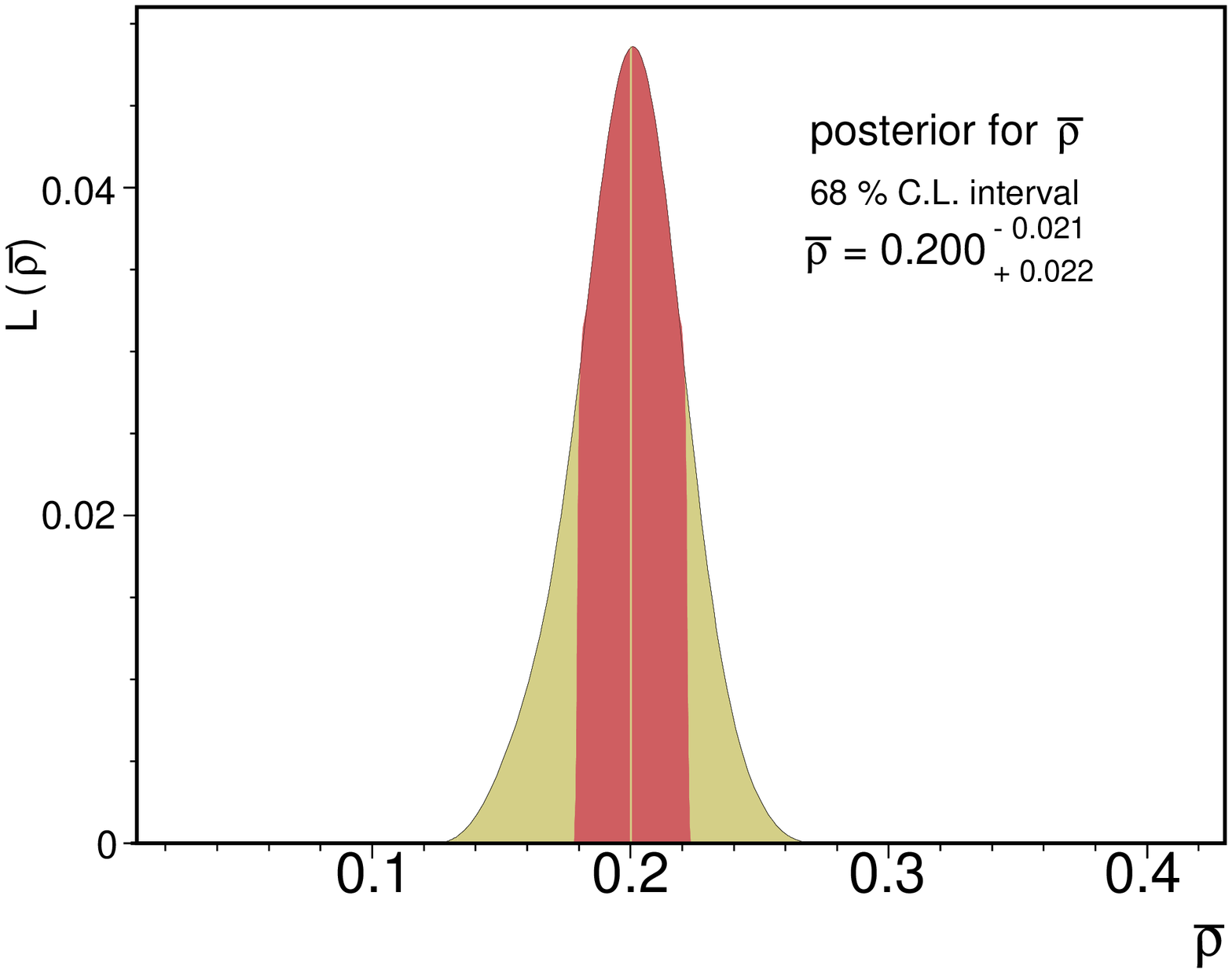}%\caption[$\rb$]{$\rb$} \label{fig:rb_r0a}
}
\end{minipage}
\hfill
\begin{minipage}[t]{5.7cm}
{\centering
\includegraphics[width=1\textwidth]{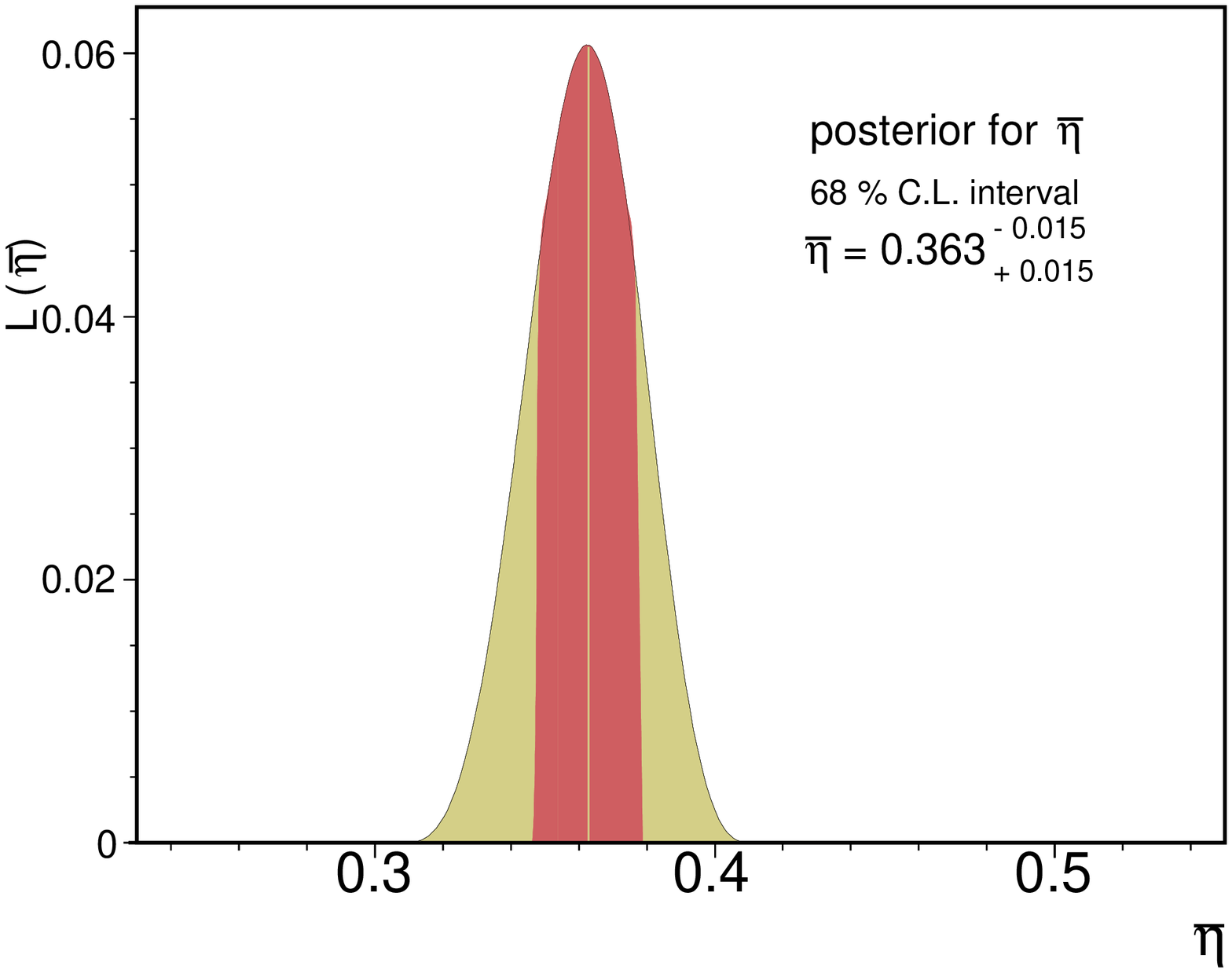}%\caption[$\eb$]{$\eb$} \label{fig:eb_r0a}
}
\end{minipage}
\hfill
\vspace*{-0.8cm}
\caption{\small $\rb$ and $\eb$ one dimensional projections using the results of the fit with the $\dms$ measurement~\eq{eqn:dms:cdf06}.}
\label{fig:rb_eb_r0b}
\end{figure}
\begin{figure}[ht]
%\hfill
\vspace*{-0.6cm}
\begin{minipage}[t]{5.7cm}
\centering{
\includegraphics[width=1\textwidth]{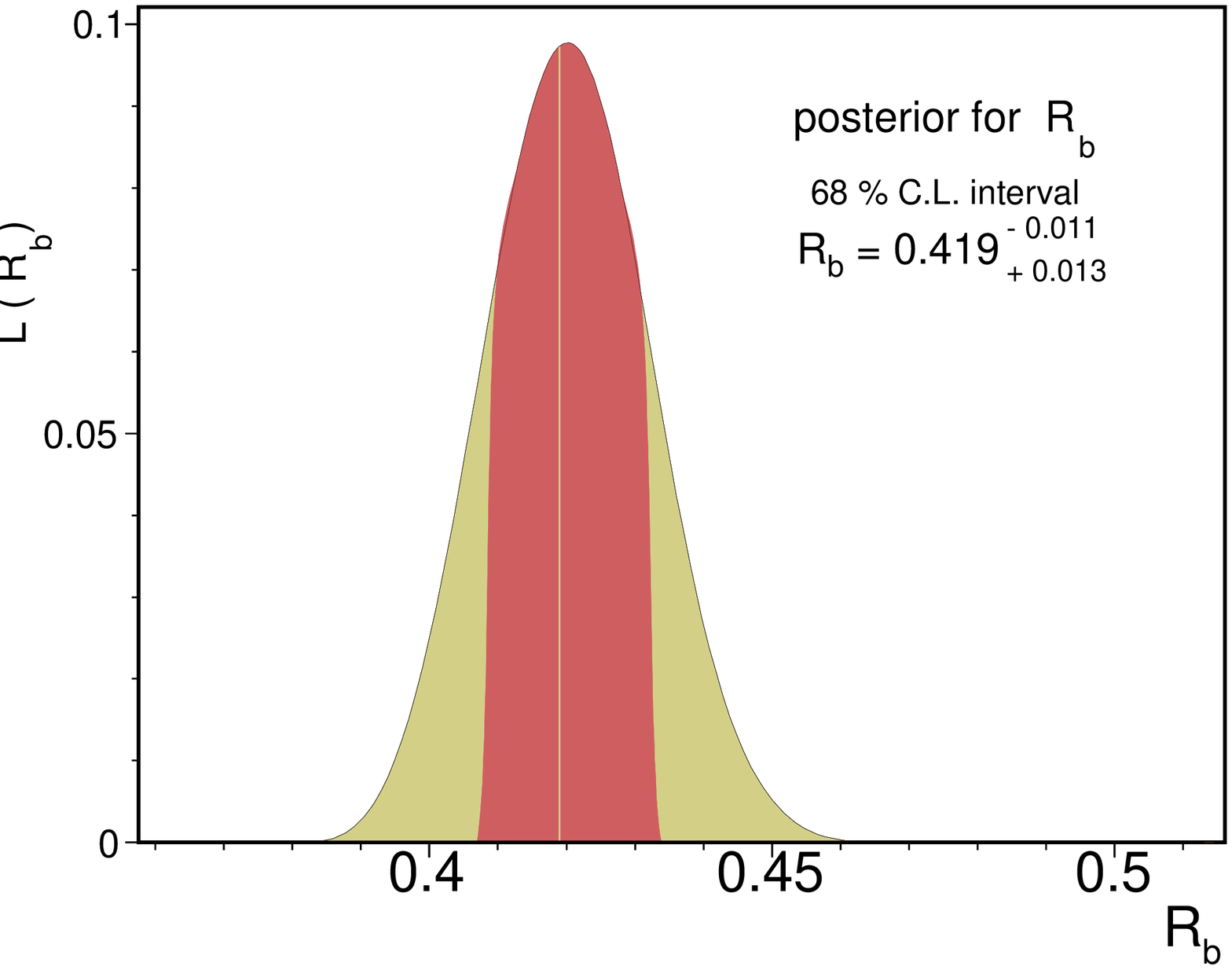}
}
\end{minipage}
\hfill
\begin{minipage}[t]{5.7cm}
\centering{
\includegraphics[width=1\textwidth]{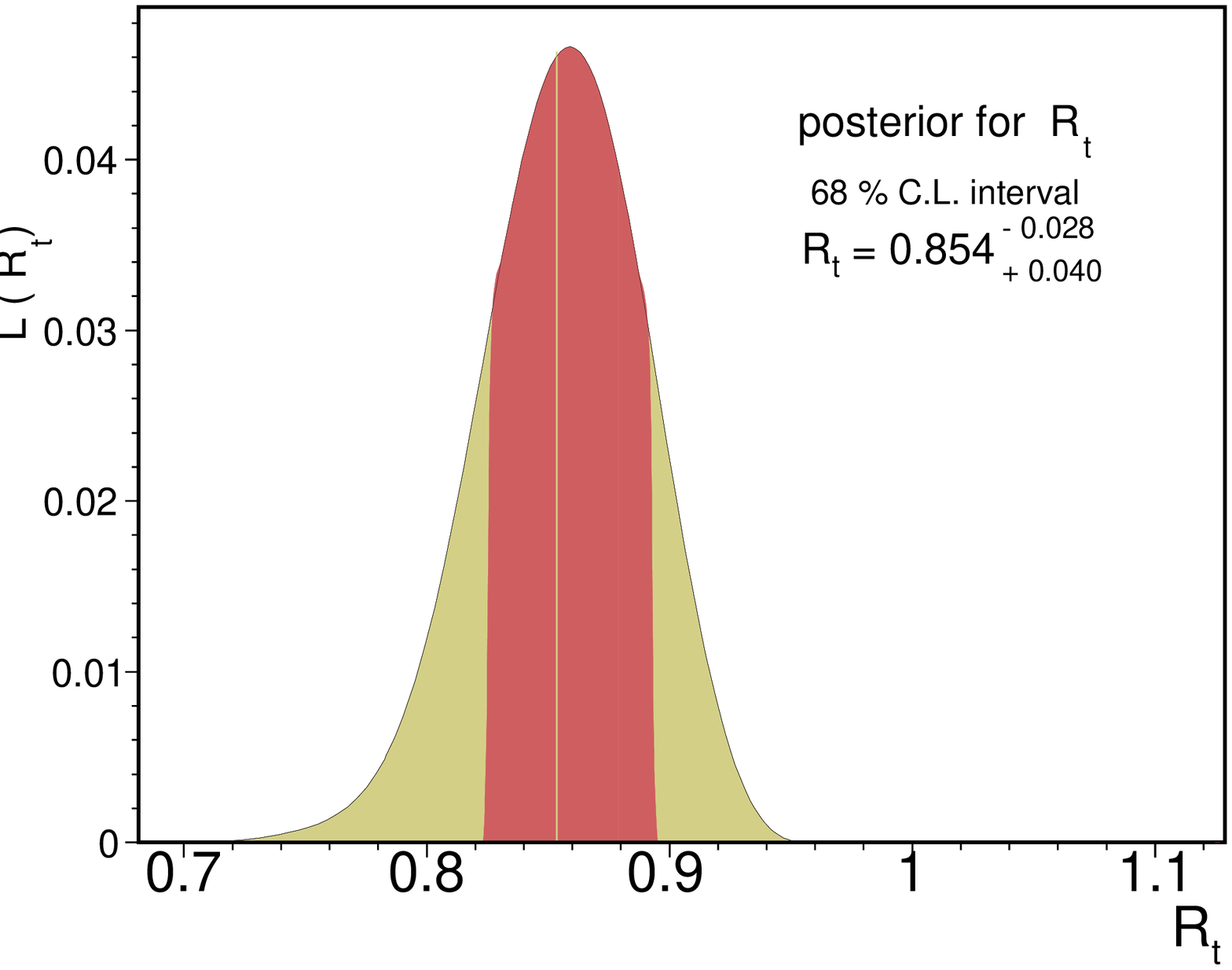}
}
\end{minipage}
\hfill
\vspace*{-0.9cm}
\caption{\small $R_b$ and $R_t$ one dimensional projections using the bound results.}
\label{fig:Rb_Rt_r0a}
\end{figure}
\begin{figure}[ht]
%\hfill
\vspace*{-0.6cm}
\begin{minipage}[t]{5.7cm}
\centering{
\includegraphics[width=1\textwidth]{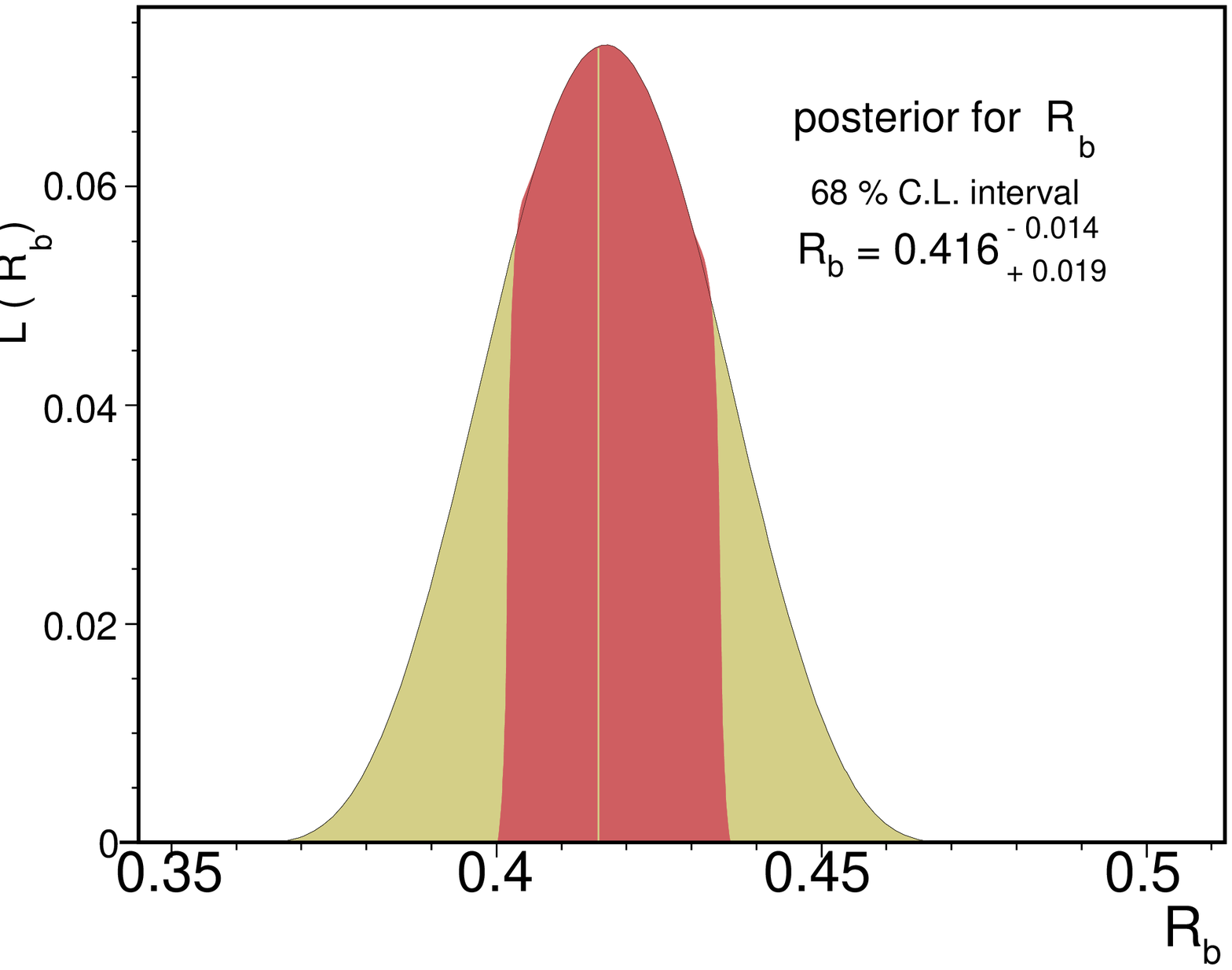}
}
\end{minipage}
\hfill
\begin{minipage}[t]{5.7cm}
\centering{
\includegraphics[width=1\textwidth]{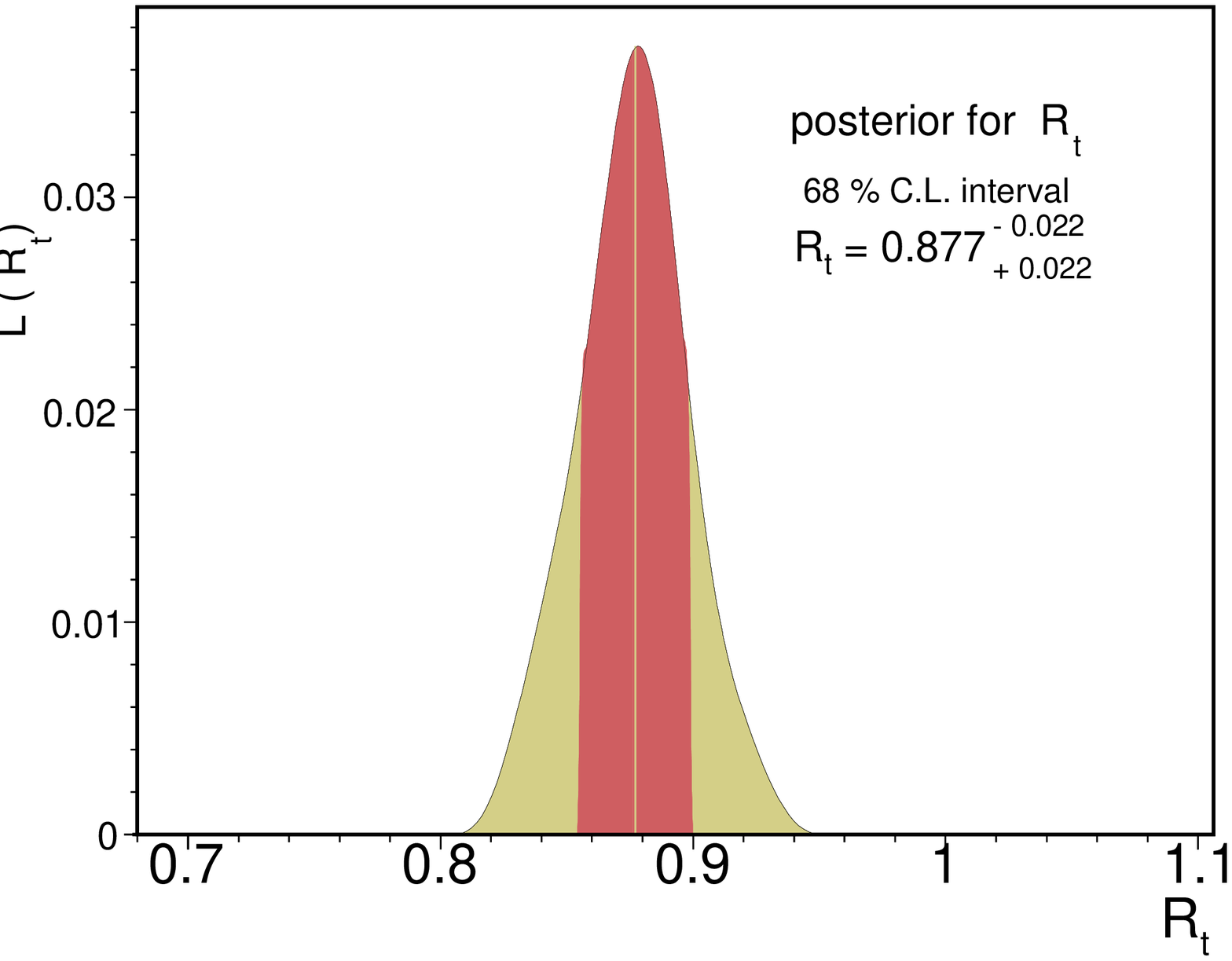}
}
\end{minipage}
\hfill
\vspace*{-0.9cm}
\caption{\small $R_b$ and $R_t$ one dimensional projections using the data of the $\dms$ measurement.}
\label{fig:Rb_Rt_r0b}
\end{figure}
\begin{figure}[ht]
%\hfill
\begin{minipage}[t]{5.7cm}
{\centering
\includegraphics[width=1\textwidth]{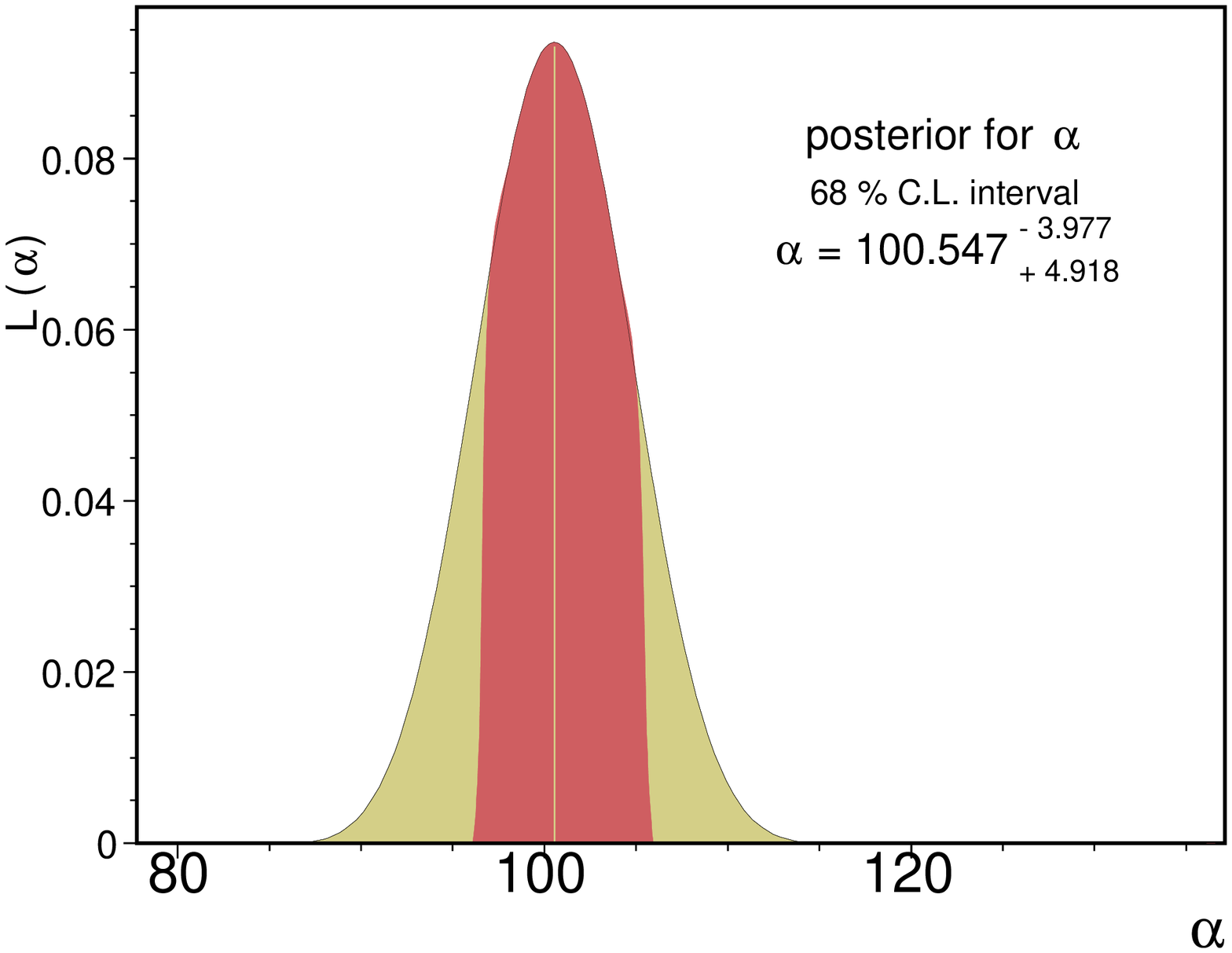}
}
\end{minipage}
\hfill
\begin{minipage}[t]{5.7cm}
{\centering
\includegraphics[width=1\textwidth]{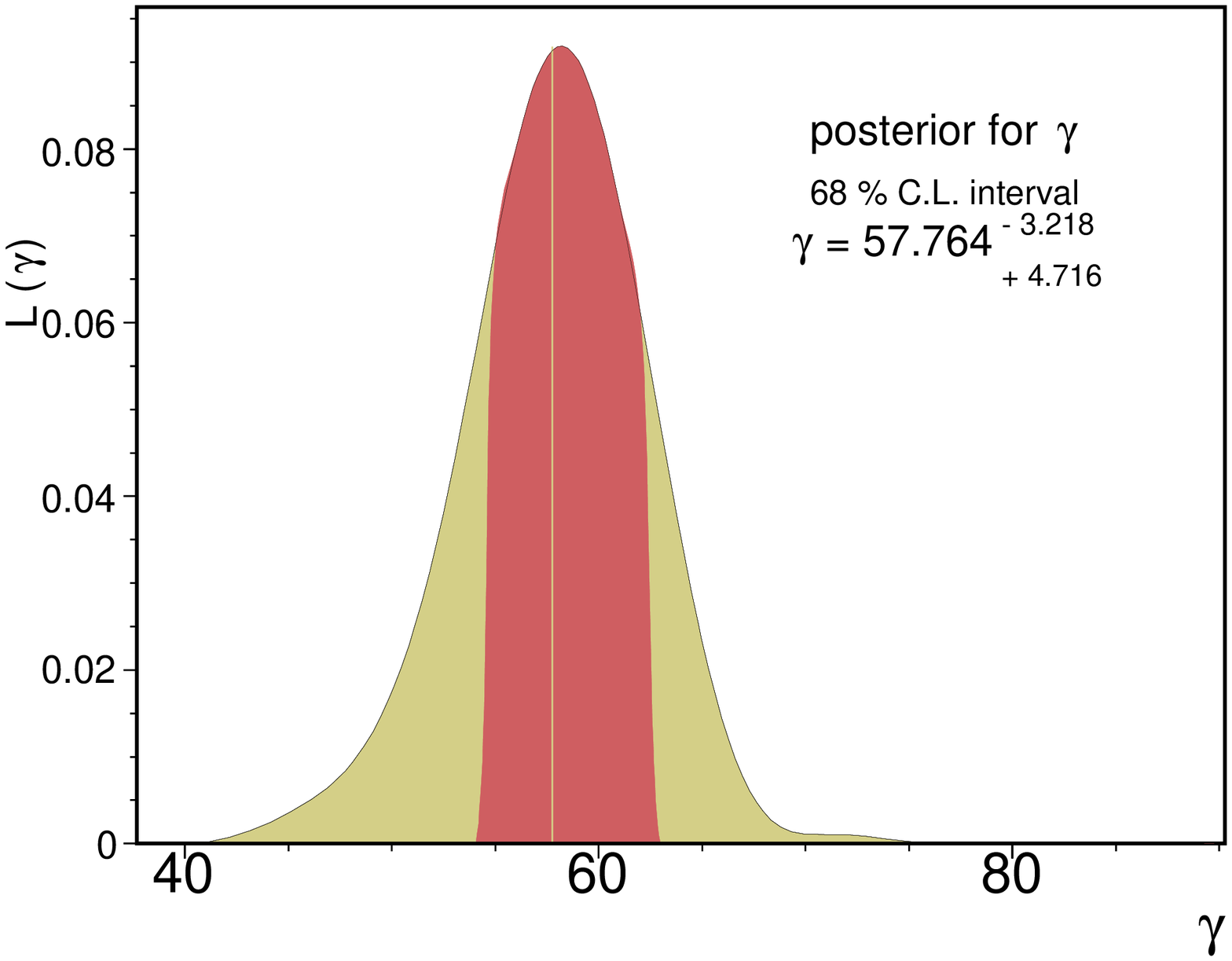}
}
\end{minipage}
\hfill
\vspace*{-0.7cm}
\caption{\small $\alpha$ and $\gamma$ one dimensional projections using the bound results.}\label{fig:alpha_gamma_r0a}
\end{figure}
\begin{figure}[ht]
%\hfill
\begin{minipage}[t]{5.7cm}
{\centering
\includegraphics[width=1\textwidth]{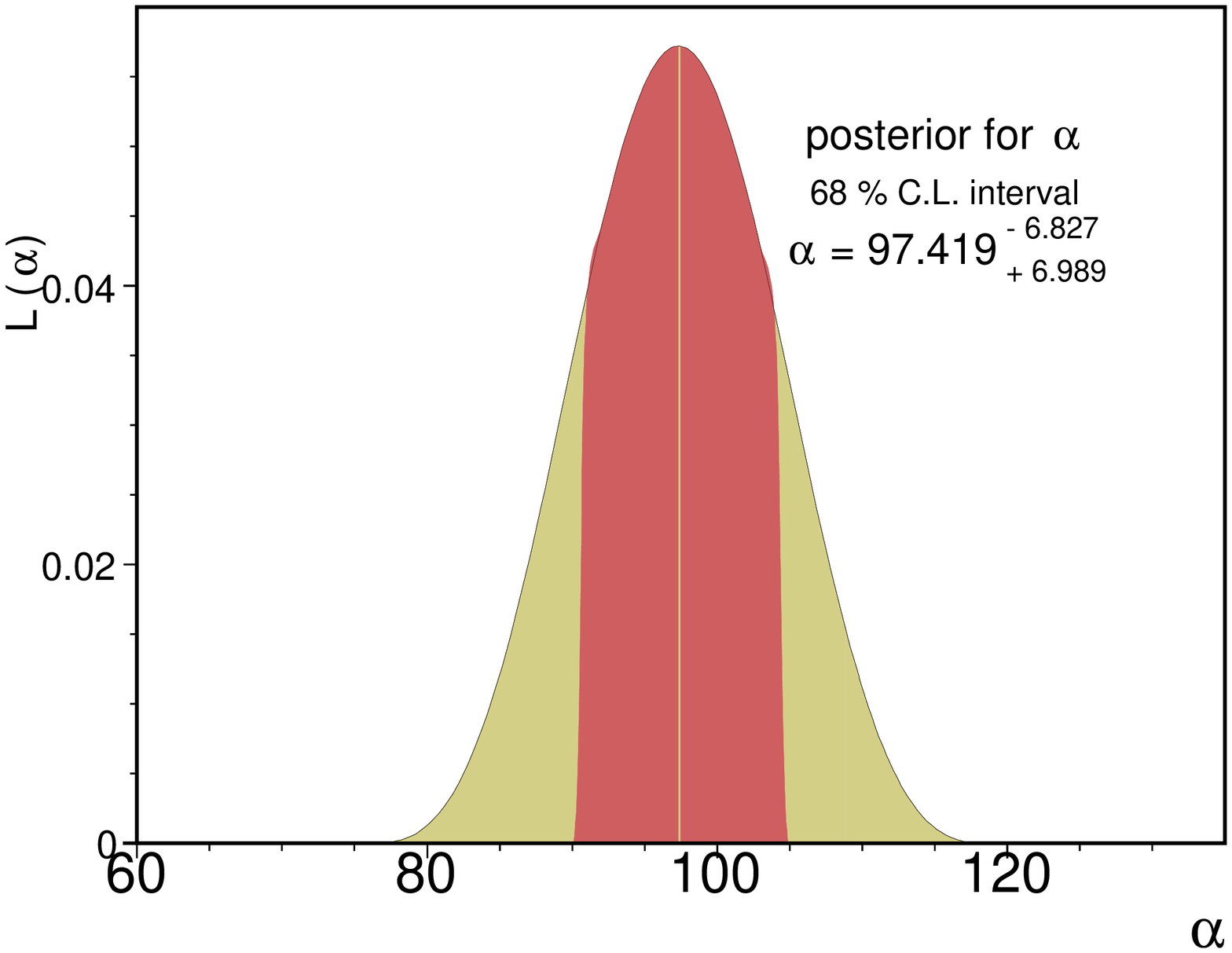}
}
\end{minipage}
\hfill
\begin{minipage}[t]{5.7cm}
{\centering
\includegraphics[width=1\textwidth]{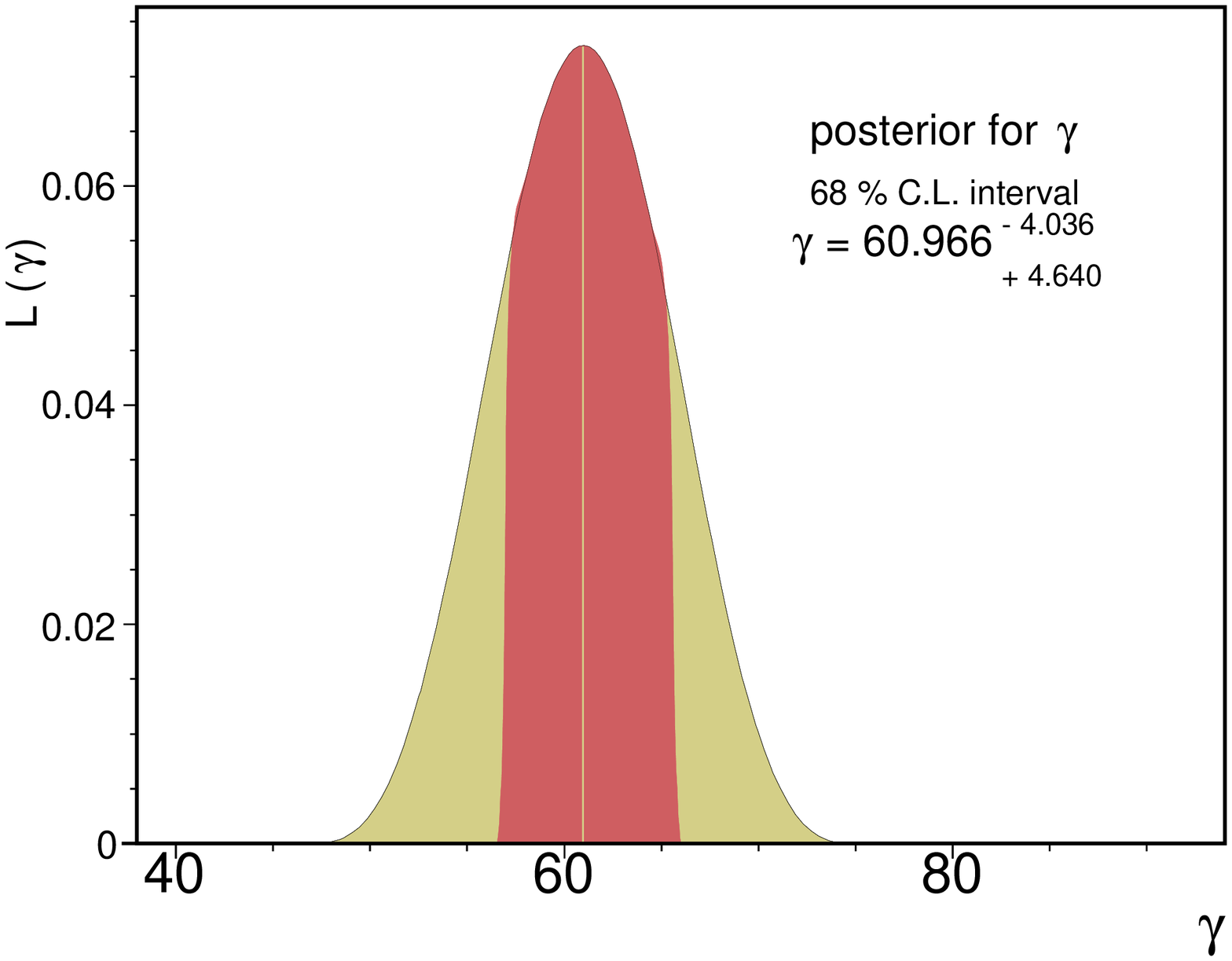}
}
\end{minipage}
\hfill
\vspace*{-0.7cm}
\caption{\small $\alpha$ and $\gamma$ one dimensional projections using the results of the fit with the $\dms$ measurement.}\label{fig:alpha_gamma_r0b}
\end{figure}
\begin{figure}[ht]
%\hfill
\begin{minipage}[t]{5.7cm}
\begin{center}
\includegraphics[width=1\textwidth]{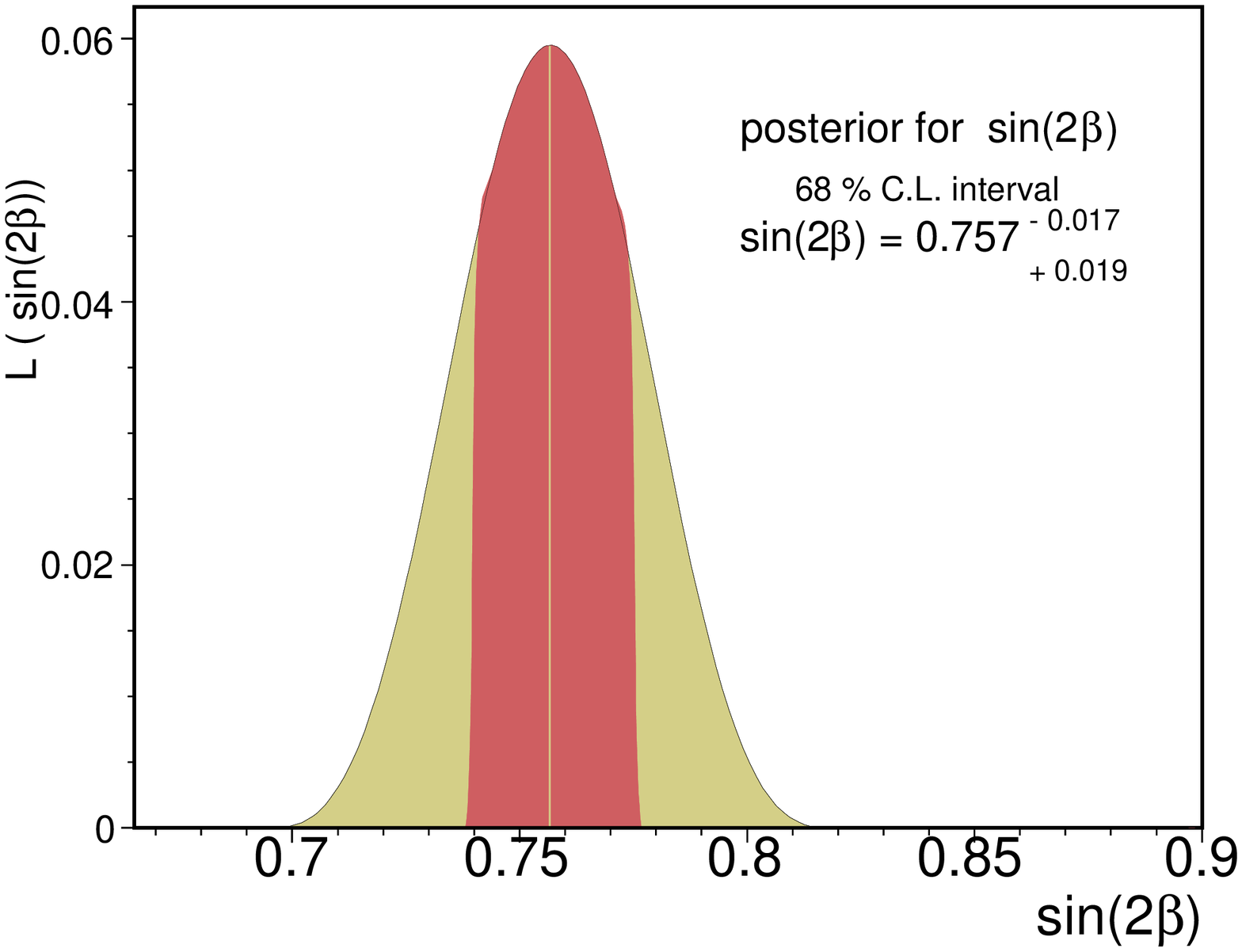}
(i)
\end{center}
\end{minipage}
\hfill
\begin{minipage}[t]{5.7cm}
\begin{center}
\includegraphics[width=1\textwidth]{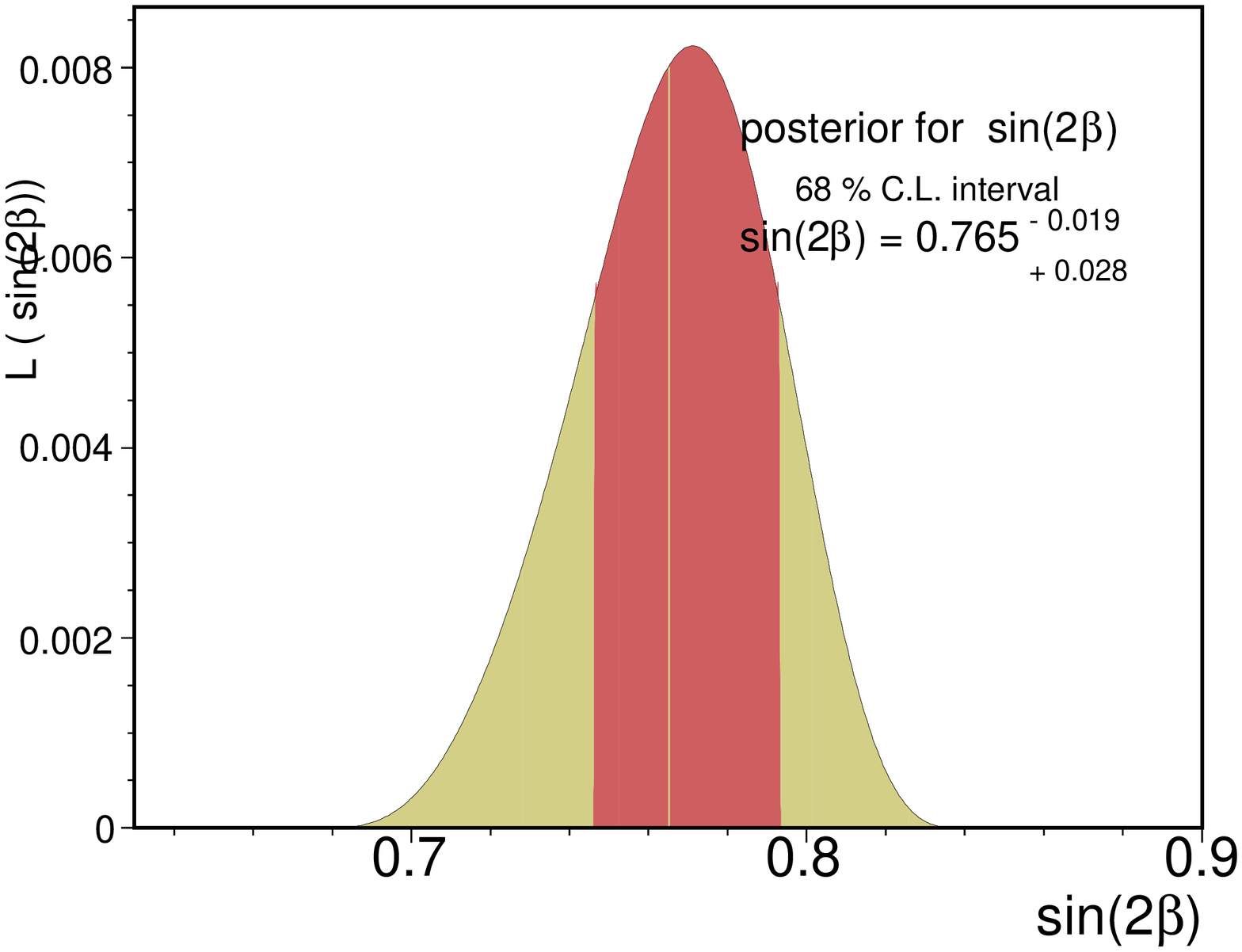}
(ii)
\end{center}
\end{minipage}
\hfill
\vspace*{-0.3cm}
\caption
{$\sin 2\beta$ one dimensional projections using the (i) bound and (ii) the fit using the data of the fit with the $\dms$ measurement.}
\label{fig:sin2beta_r0}
\end{figure}
\clearpage

%%%%%%%%%%%%%%%%%%%%%%%%%%%%%%%%%%%%%%%%%%%%%%%%%%%%%%%%%%%%%%%%%%%%%%%%%%%%%%%%%%%%%%
%
\subsection{Indirect evidence of CP violation}
%
%%%%%%%%%%%%%%%%%%%%%%%%%%%%%%%%%%%%%%%%%%%%%%%%%%%%%%%%%%%%%%%%%%%%%%%%%%%%%%%%%%%%%%
%
The analysis of the unitary triangle (UT) allows for the comparison of the fit using only the CP conserving processes, $|V_{ub}|/|V_{cb}|$, $\dmd$ and $\dms$, with the current experimental values of the CP violation parameters $\ek$ and $\sin 2\beta$. This is a way to check how consistent is the fit of the parameters that are  sensitive just to the sides of the UT with respect to the measurements that are sensitive to the angles, which it is an indirect measurement of the amount of CP violation.
 The $(\rb,\eb)$ plane of this fit is shown in \fig{fig:rb_eb_2d_r1b}.
\begin{figure}[ht]
%\vspace*{-0.5.7cm}
\begin{center}
\includegraphics[width=10cm]{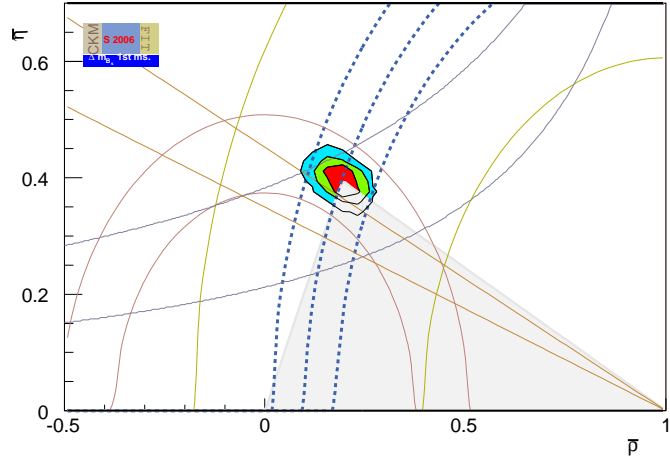}
\vspace*{-0.5cm}
\caption{\small $(\rb,\eb)$ plane with the results of the fit using the $\dms$ measurement and  using only the CP conserving constraints $|V_{ub}|/|V_{cb}|$, $\dmd$ and $\dmd/\dms$. In dotted lines we have plotted the CP violating constraints: $\sin 2\beta$ and $\ek$ to compare them with the pdf of this fit.} 
\end{center}
\label{fig:rb_eb_2d_r1b}
\end{figure}
We also compare the fitted 1d pdfs of the parameters $|\ek|$ and $\sin 2\beta$ to those given for the case when these parameters have been included in the fit as constraints.
\begin{figure}[ht]
%\hfill
\begin{minipage}[t]{5.7cm}
\begin{center}
\includegraphics[width=1\textwidth]{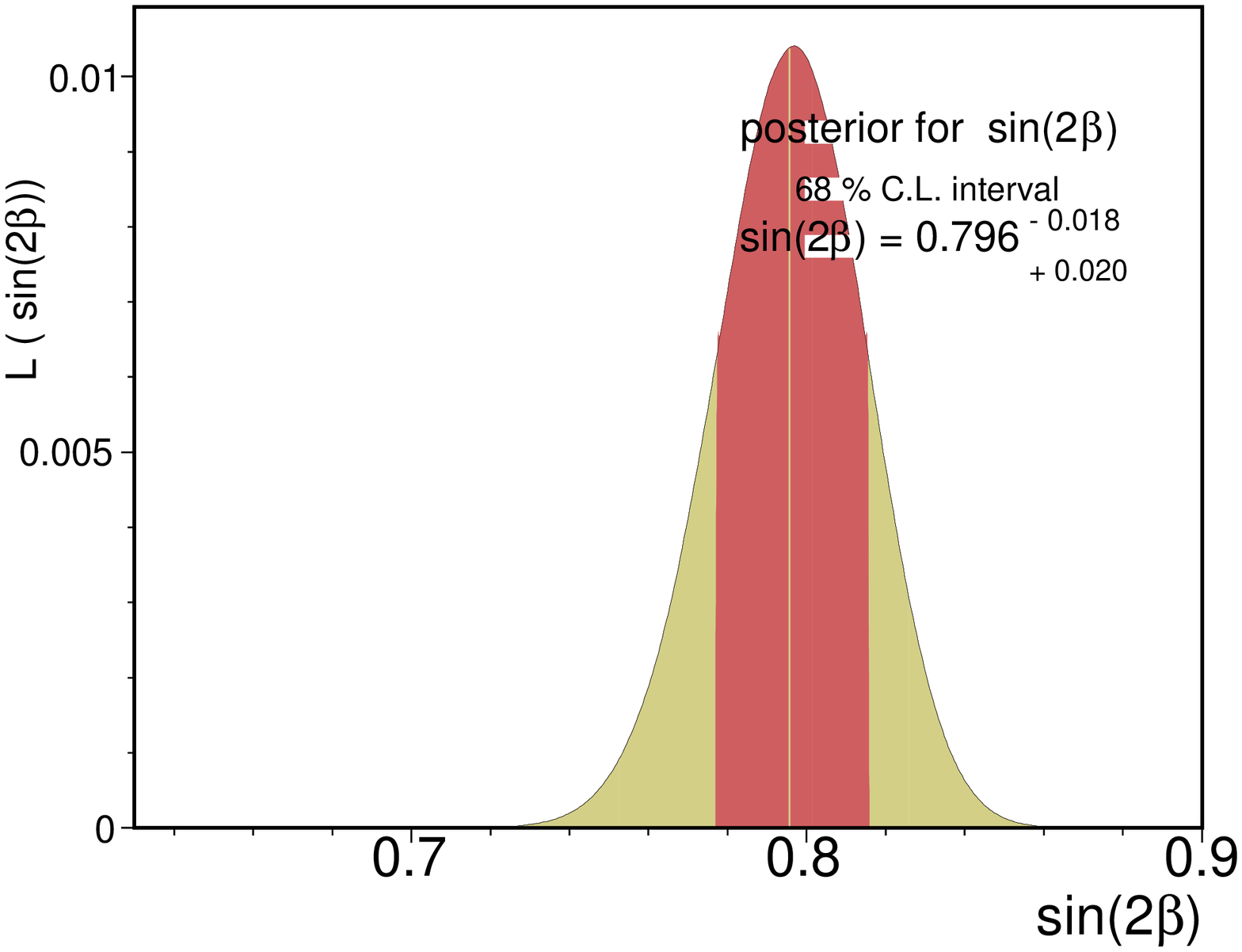}
\vspace*{-0.3cm}
(i)
\end{center}
\end{minipage}
\hfill
\begin{minipage}[t]{5.7cm}
\begin{center}
\includegraphics[width=1\textwidth]{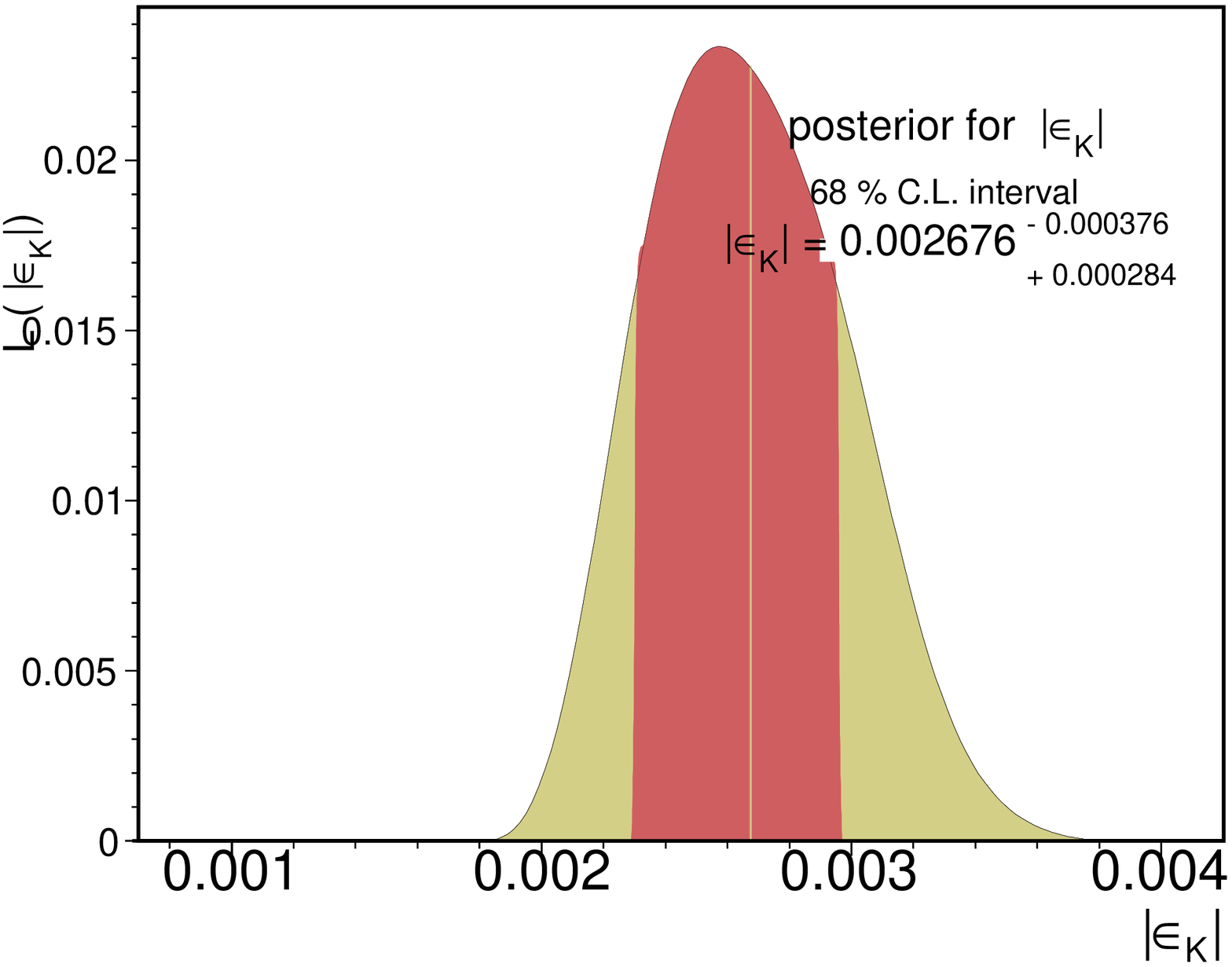}
(ii)
\end{center}
\end{minipage}
\hfill
\vspace*{-0.3cm}
\caption{\small One dimensional projections of (i) $\sin 2\beta$ and (ii) $\ek$ using data of the fit including the $\dms$ measurement with out imposing them as a constraint.}
\label{fig:sin2b_epsk_r1b}
\end{figure}
The output value of $\ek$ for the fit including just $|V_{ub}|/|V_{cb}|$, $\dmd$ and $\dmd/\dms$ is $\ek=(2.230,2.96)\times 10^{-3}$ at 68\%~CL; 
it is consistent with its experimental value of $\ek=(2.11,2.45)\times 10^{-3}$ at 68\%~CL. 
However as we can see by comparing Figure (\ref{fig:sin2b_epsk_r1b} a) to \fig{fig:sin2beta_r0}, 
for both the $\dms$ bound and measurement cases, the value of $\sin 2\beta$ increases if the CP violating constraints $\sin 2\beta$ and $\ek$ are not taken into account:
%both for the bound and the measurement of $\dms$  cases, the value of $\sin 2\beta$ increases if the CP violating constraints $\sin 2\beta$ and $\ek$ are not taken into account:
%
\bea
\sin 2\beta^{\;\rm{CP-conserv.}}=(0.778,0.816) \,,
\label{eq:nocp_s2b} 
\eea
which does leave out the 68\%~CL interval of the experimentally measured value:
\bea
\sin 2\beta^{\;\rm{exp.}}=(0.655,0.719) \,. 
\label{eq:exp_s2b} 
\eea
This inconsistency is already apparent in the fits which include $\sin 2\beta$ and $\ek$ as a constraint, but here we note that the values of Eqs.(\ref{eq:nocp_s2b}) and (\ref{eq:exp_s2b}) are consistent just at 99\%~CL. 
%For the CP-conserving value the 99\%~CL $(0.742,0.856)$ while the experimetal value is $(0.591,0.783)$.
At 99\% CL, the CP-conserving interval is $(0.742,0.856)$ while the experimental one is $(0.591,0.783)$.

One reason for observing this apparent inconsistency may be due to the current set of experimental inputs for the constraints and their errors, especially for $|V_{ub}/V_{cb}|$. A better determination of these semi-leptonic parameters is crucial for checking this inconsistency. Another reason is a potential %that there is a
non-zero phase accounting for processes BSM in $\sin 2\beta$ such that we have $\sin 2\beta=\sin(2\beta^{SM}+2\theta_{B_d})$, as we have discussed in Section (\ref{sec:howtoa_BSM}) and as we will be presenting in Section (\ref{sec:results}).

\clearpage

%%%%%%%%%%%%%%%%%%%%%%%%%%%%%%%%%%%%%%%%%%%%%%%%%%%%%%%%%%%%%%%%%%%%%%%%%
%
\subsection{Determination of $\mathbf{\fsbd}$ and $\mathbf{B_K}$}
%
%%%%%%%%%%%%%%%%%%%%%%%%%%%%%%%%%%%%%%%%%%%%%%%%%%%%%%%%%%%%%%%%%%%%%%%%%
%
The actual measurement of $\dms$ also provides, within the context of the SM, a strong constraint on the non-perturbative QCD parameter $\fsbs$, and through the relation~\eq{dmstheory} on the parameter $\fsbd$, which suffers from large uncertainties. In particular here we can check how its lattice computation,
\bea
\fsbd^{\rm{QCD}}=(0.223\pm 0.033\pm 0.012)~{\rm{GeV^2}},\label{eq:qcd_fsbd}
\eea
 agrees with the current $\dms$ results. 
\begin{figure}[ht]
%\hfill
\begin{minipage}[t]{5.7cm}
{\centering
\includegraphics[width=1\textwidth]{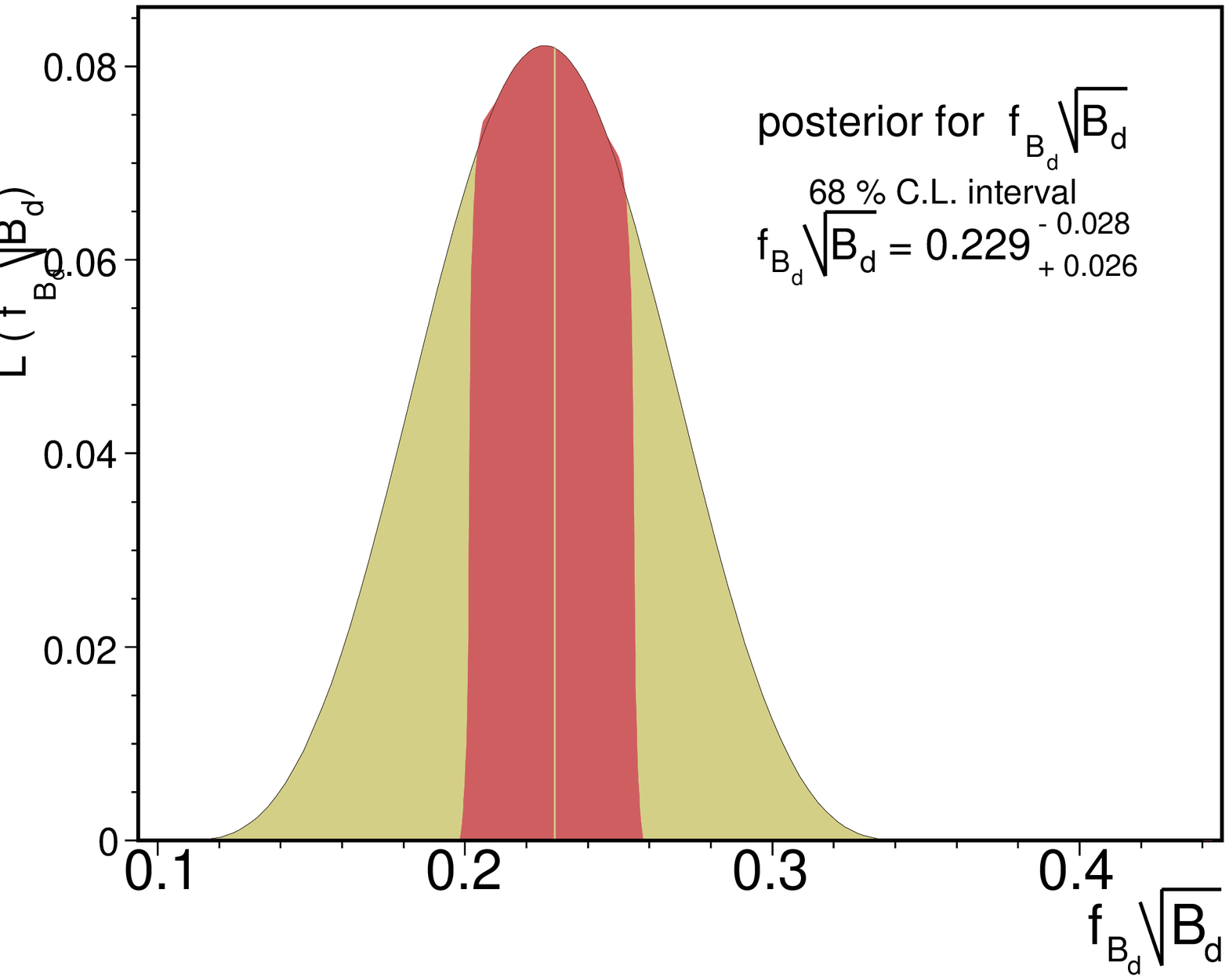}
}
\end{minipage}
\hfill
\begin{minipage}[t]{5.7cm}
{\centering
\includegraphics[width=1\textwidth]{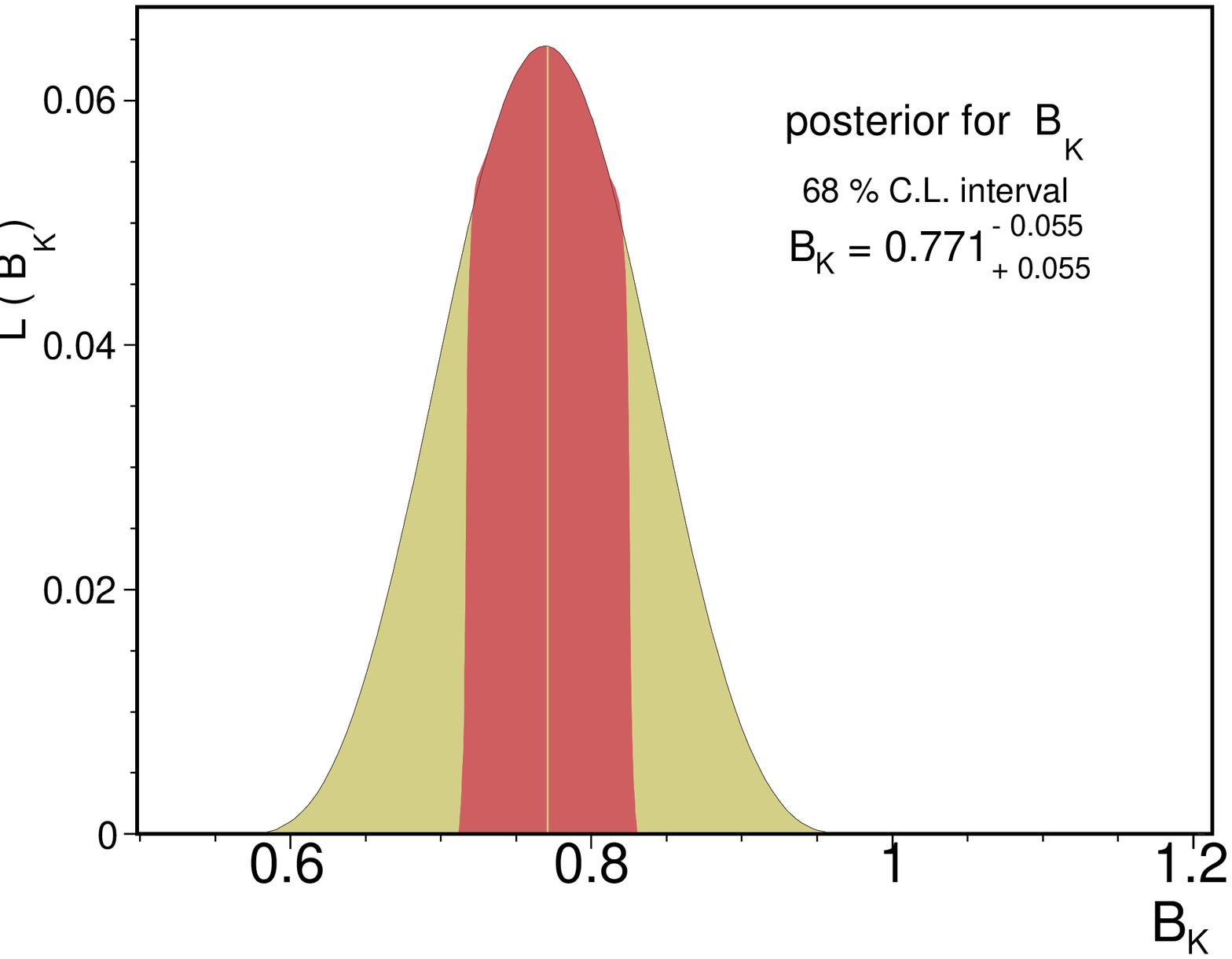}
}
\end{minipage}
\hfill
\vspace*{-0.7cm}
\caption{\small The pdfs for $\fsbd$ [GeV$^2$] and $B_K$ for the bound fit.}
\label{fig:fsbd_BK_r0a}
\end{figure}
\begin{figure}[ht]
%\hfill
\begin{minipage}[t]{5.7cm}
\centering{
\includegraphics[width=1\textwidth]{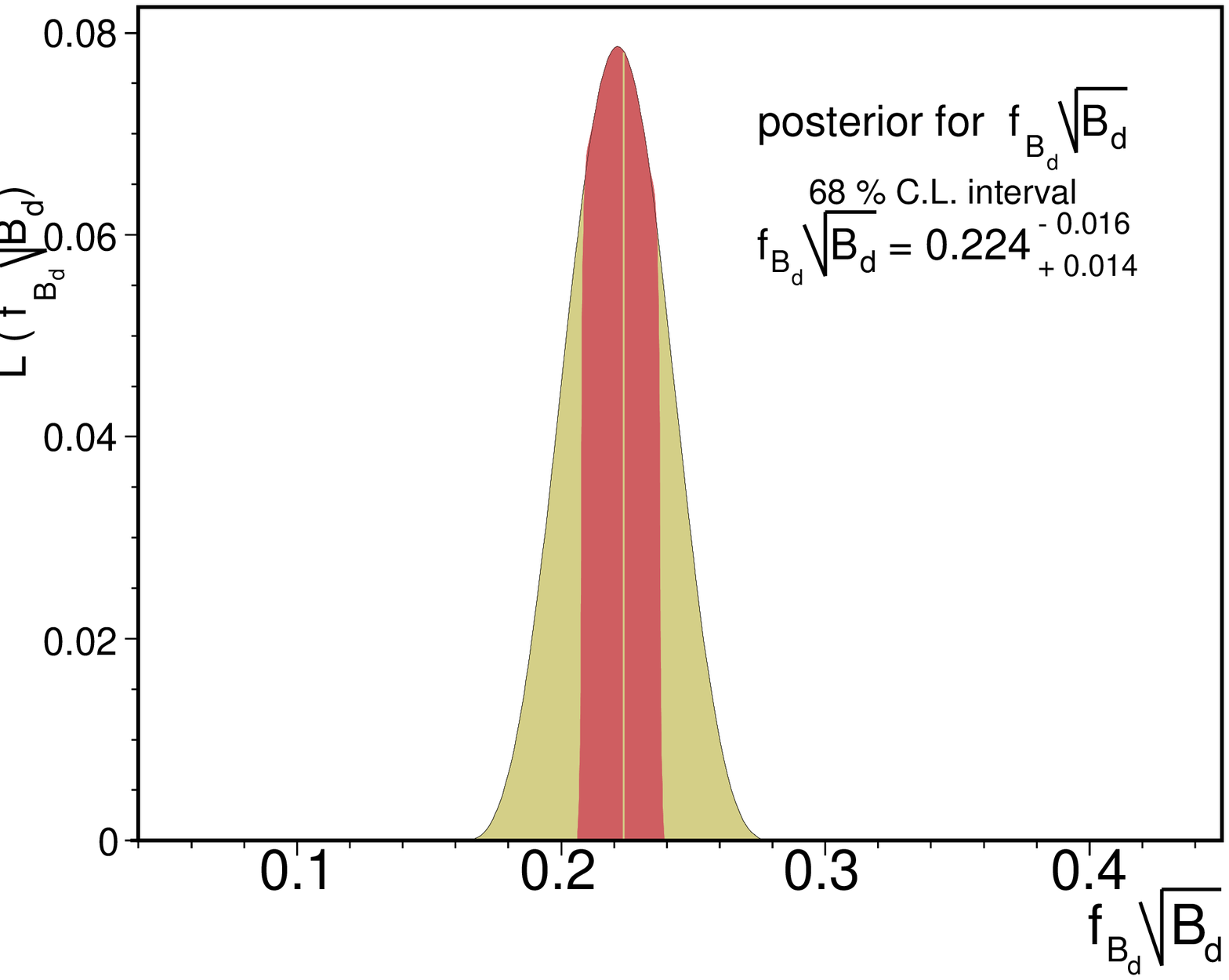}
}
\end{minipage}
\hfill
\begin{minipage}[t]{5.7cm}
\centering{
\includegraphics[width=1\textwidth]{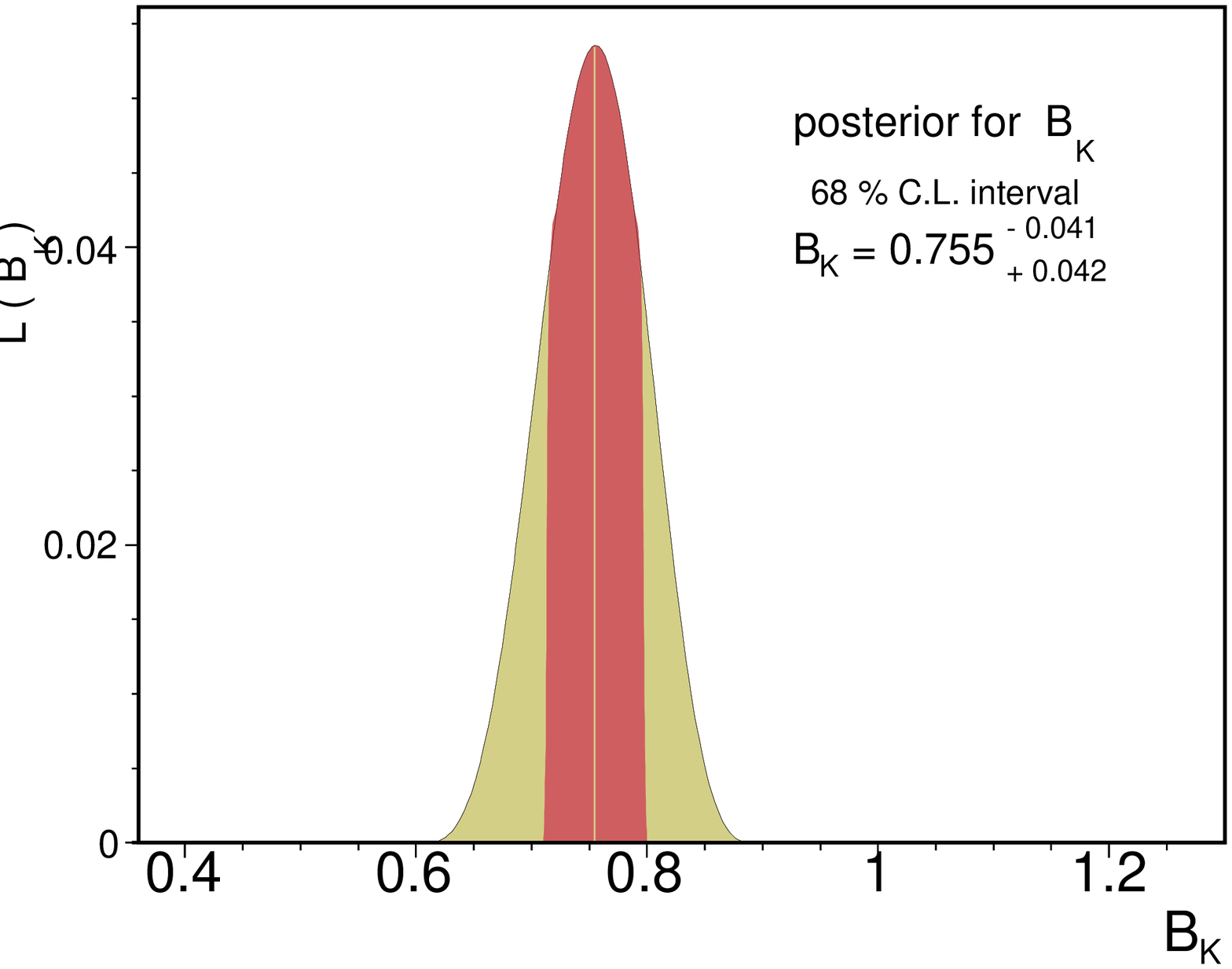}
}
\end{minipage}
\hfill
\vspace*{-0.7cm}
\caption{\small The pdfs for $\fsbd$ [GeV$^2$] and $B_K$ for the fit using the measured $\dms$.}
\label{fig:fsbd_BK_r0b}
\end{figure}

We can read the values of $\fsbd$ from Figures (\ref{fig:fsbd_BK_r0a}) and (\ref{fig:fsbd_BK_r0b}). The values of this parameter (in GeV$^2$) for the overall fit are
\bea
& &\fsbd^{\rm{bound}}       \,=\, 0.229_{-0.028}^{+0.026}\,, \qquad 
   \fsbd^{\rm{meas.}}       \,=\, 0.218_{-0.008}^{+0.007}\,, \nn\\
& &\chi^2_{\fsbd\,(i)}      \,=\, 2.9 \times 10^{-2}\,,   \qquad 
\chi^2_{\fsbd\,(ii)}        \,=\, 8.1 \times 10^{-4}\,,
\eea
hence checking that the QCD Lattice value, \eq{eq:qcd_fsbd}, and the fit value of $\fsbd$ agree remarkably well.

Comparing Figures (\ref{fig:fsbd_BK_r0a}) and (\ref{fig:fsbd_BK_r0b}) we can see that the measured $\Bsmix$ oscillation data has had an important impact on $\fsbd$, by decreasing its value and its uncertainty. This uncertainty is determined to a better accuracy than the QCD determination, \eq{eq:qcd_fsbd}. The impact of $\dms$ on $B_K$  is not as strong as the impact on $\fsbd$ since it is just affected through the overall fit and not through a constraint, $B_K$ only appears in the SM  due to the $|\ek|$ constraint, \eq{eq:ek_sm}. Although the difference, $\delta B_K$, between $B_K$ as obtained in the fit, Figures (\ref{fig:fsbd_BK_r0a}) and (\ref{fig:fsbd_BK_r0b}), and $B_K$ as given by the QCD lattice calculations is smaller in case (ii) of measured $\dms$:
\bea
B^{\rm{QCD}}_K       &=& 0.79 \pm 0.04\,,\nn\\
\chi^2_{B_{K\,(i)}}    &=& 2.26 \times 10^{-1}\,,\nn\\
\chi^2_{B_{K\,(ii)}}    &=& 7.66 \times 10^{-1}\,;
\label{eq:qcd_comp_BK}
\eea 
the uncertainty has increased for the measured $\dms$ case.  
It is interesting to compare the effect on $B_K$ when $|\ek|$ is removed as a constraint, as we can see in \fig{fig:BK_r4b},
\begin{figure}[ht]
%\hfill
\begin{minipage}[t]{5.7cm}
{\centering
\includegraphics[width=1\textwidth]{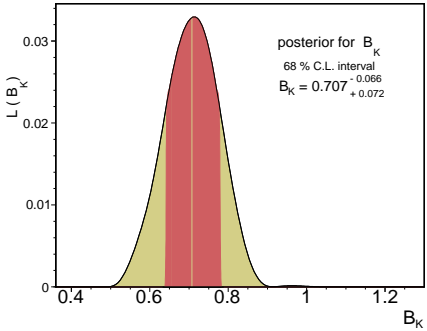}
}
\end{minipage}
\hfill
\begin{minipage}[t]{5.7cm}
{\centering

}
\end{minipage}
\hfill
\vspace*{-0.7cm}
\caption{\small The pdf of $B_K$ when $\ek$ is not imposed as a constraint.}
\label{fig:BK_r4b}
\end{figure}
the value of $B_K$ in this case is
\bea
B_K=0.707_{-0.066}^{+0.072} \,.
\eea
When the $\dms$ bound~\eq{eqn:dms:f05} was presented, we obtained $B_K=0.745^{+0.102}_{-0.072}$ which agreed quite well with its QCD value, \eq{eq:qcd_comp_BK}, although the error was of 12\% in contrast with the 10\% error of the QCD caculation. Now we see that although the error has decreased to 10\% and there is compatibility with the QCD calculation at 68\%~CL, here we have $\chi^2_{B_{K}}=4.3 $.

%
%%%%%%%%%%%%%%%%%%%%%%%%%%%%%%%%%%%%%%%%%%%%%%%%%%%%%%%%%%%%%%%%%%%%%%%%%%%%%%%%
\section{Bounds on processes BSM}
%%%%%%%%%%%%%%%%%%%%%%%%%%%%%%%%%%%%%%%%%%%%%%%%%%%%%%%%%%%%%%%%%%%%%%%%%%%%%%%
%
%
%%%%%%%%%%%%%%%%%%%%%%%%%%%%%%%%%%%%%%%%%%%%%%%%%%%%%%%%%%%%%%%%%%%%%%%%%%%%%%%%
\subsection{Scenario I \label{sbsec:scenario1}}
%%%%%%%%%%%%%%%%%%%%%%%%%%%%%%%%%%%%%%%%%%%%%%%%%%%%%%%%%%%%%%%%%%%%%%%%%%%%%%%
%
%
In this scenario we consider the constraints on models for which the conditions of what has been called Minimal Flavour Violation (MFV) are valid. An effective theory is said to be minimal flavour violating if all higher-dimensional operators constructed from the SM and fields responsible for giving the corresponding structure of Yukawa couplings (e.g. flavon -scalar- fields whose \emph{vev}s determine the effective Yukawa couplings) are invariant under the CP symmetry and the flavour symmetry, determined by the group $G_F$. When this happens the dynamics of flavour violation is completely determined by the structure of the SM Yukawa couplings. If all Yukawa couplings are small except for the top, then the only relevant non-diagonal operators entering in the Hamiltonians of the $\Delta F=2$ transitions  are proportional to $\left(Y_uY^{\dagger}_u\right)_{ij}\sim y^2_t V^*_{ti}V_{tj}$. By looking at \eq{eq:sm_effham} we can see that $V_{td}V^*_{tb}$ and $V_{td}V^*_{ts}$, the main contribution to the mixing in the $\Delta B=2$ and $\Delta S=2$ transitions respectively, are multiplied by the same Inami-Lim function $S(x_t)$ (as expressed in \eq{eq:dmbds_sm} and \eq{eq:ek_sm}) exactly because of the top dominance). Once there is another particle contributing at the same order of the top contributions then the expressions of the dominating diagrams for the processes $\Delta B=2$ and $\Delta S=2$ become different. This is evident in supersymmetric theories when $\tan\beta_s$ is large because the bottom contribution becomes relevant and hence it gives a sizable contribution to the $\Delta B=2$ transitions. Due to the potential difference in these processes, two possibilities are considered:

\begin{enumerate}

\item $S(x_t)\rightarrow S(x_t)+\delta S(x_t)$

\item $\eta_2 S(x_t)\rightarrow \eta_2[S(x_t)+\delta S(x_t)_{\ek}]$,\\ 
      $\eta_B S(x_t)\rightarrow \eta_B[S(x_t)+\delta S(x_t)_{\dmd}]$ 

\end{enumerate}

\subsubsection{Example: supersymmetric MFV scenario in the low \boldmath{$\tan\beta_s$} limit \label{sbsec:susy_mfv}} 

In the context of MSSM models there are four physical phases: $\dckm$, $\theta_A=\rm{arg}$, $\theta_{\mu}=\rm{arg}(\mu)$ (where $A$ is the common trilinear term between left and right s-fermions and a Higgs field and $\mu$ the bilinear term between the two MSSM Higgs fields) and the QCD vacuum parameter $\bar{\theta}_{QCD}$ which however can be conveniently set to be zero.

Experimental upper limits on the electric dipole moments (EDM) of the neutron and electron provide constraints on the phases $\theta_\mu$ and $\theta_A$. In the constrained minimal sypersymmetric standard model (CMSSM) in general the bounds on $\theta_\mu$ are stronger than those to $\theta_A$, where $\theta_\mu$ ranges from $O(10^{-1})$ to $O(10^{-4})$ \cite{Olive:2005ru}. In the minimal SUGRA model $\theta_\mu\sim O(10^3)$ while $\theta_A$ can be $O(1)$ in the small $\theta_\mu$ region \cite{Nihei:1997kh}.

To which extend $\theta_{B_d}$ and $\theta_{B_s}$ are affected by these bounds depend on the values of $\tan\beta_s=v_{u}/v_{d}$, the ratio of the vacuum expectation values of the two Higgs fields in the MSSM. For low-to-moderate values of $\tan\beta_s$ in SUGRA models $\theta_A$ does not shown in the phase of either the matrix elements  $\langle B^0_d| {\mathcal{H}}^{\rm{SUGRA}}_{\rm{eff}} |\bar{B}^0_d \rangle$ and $\langle K^0| {\mathcal{H}}^{\rm{SUGRA}}_{\rm{eff}} |\bar{K}^0 \rangle$, hence 
\bea
{\rm{arg}}\langle B^0_d| {\mathcal{H}}^{\rm{SUGRA}}_{\rm{eff}} |\bar{B}^0_d \rangle 
&=&{\rm{arg}}(V^*_{td}V_{tb}),\nn\\
{\rm{arg}}\langle K^0| {\mathcal{H}}^{\rm{SUGRA}}_{\rm{eff}} |\bar{K}^0 \rangle  
&=&{\rm{arg}}(V^*_{td}V_{ts}),
\eea
thus $\theta_{B_d}$ and $\theta_{B_s}$ can be safely treated as zero in this case. This means that CP violation asymmetries give information just for $\alpha$, $\beta$ and $\gamma$ as defined in the SM and the BSM contribution to $\ek$ is aligned with the  $\bar{t}t$ contribution of the SM ${\rm{arg}}(V^*_{td}V_{ts})$. In this context then the analysis of the UT and CP violation phases $\alpha$, $\beta$ and $\gamma$ can be carried out in a very similar way as in the SM, just by taking into account the contributions to $(V^*_{td}V_{tb})$ and $(V^*_{td}V_{ts})$ without involving additional phases to those in the MSSM models.
For large $\tan\beta_s$ cases it is necessary to take into account new operators whose contributions become relevant for this case. These contributions can be computed for a given theory from the effective Hamiltonian
$H_{\rm{eff}}^{(\Delta B=2)}=\frac{G^2_FM^2_W}{2\pi^2}\sum_{i=1}^{3}C_i(\mu)\mathcal{Q}_i$, 
where $C_i$ are the Wilson coefficients and the operators $\mathcal{Q}_i$ are given by
\bea
{\mathcal{Q}}_1 = \bar{d}^\alpha_L\gamma_{\mu}b^{\alpha}_L\bar{d}^\beta_L\gamma^\mu d^\beta_L,\quad
{\mathcal{Q}}_2 = \bar{d}^\alpha_L b^\alpha_R\bar{d}^\beta_L b^\beta_R,\quad 
{\mathcal{Q}}_3 = \bar{d}^\alpha_\alpha b^\beta_R \bar{d}^\beta_\alpha b^\alpha_R.
\label{eq:DB2_Oi}
\eea
The first two operators are as in the SM, while the supersymmetric contributions to ${\mathcal{Q}}_3$ from which chargino contributions to $C_3(\mu)$ are generically complex relative to the SM contribution and hence can generate a new phase shift in the $\Bdmix$ mixing amplitude. This is quite significant because $C_3(\mu)\propto \left(\frac{m_b}{m_W} \cos\beta \right)^2$. When the EDM constraint in these cases is imposed $\theta_{B_d}$ becomes very small \cite{Baek:1998yn},\cite{Chang:1998uc}.

If we like to accurately test the possible theory for physics beyond the SM, we have to compute all the quantities involved in the box diagrams associated to these mixings. Precisely the NLO QCD parameters in the supersymmetric MFV (MFSV) context required for $\dmd$, $\dms$ and $\ek$ have been calculated under the following simplifications \cite{Krauss:1998mt}:
(i)~the s-quark flavour mixing matrix which diagonalizes the s-quark mass matrix is approximately the same as the $V_{\rm{CKM}}$, apart from the left-right mixing of the top quarks; 
(ii)~the first and second generation s-quarks with the same gauge quantum numbers remain highly degenerate in masses but the third generation (especially $\tilde t$) can be significantly lighter due to RGE of the top Yukawa couplings; (iii)~the phases $\theta_{d,s}$ can be safely set to zero in the entire $\tan\beta_s$ space once the EDM constraint is imposed (considered first in \cite{Ali:1999we}).  Then $\dmd$ can be expressed as
\bea
\dmd=\frac{G^2_FM^2_W}{2\pi^2}m_{B_d}\fsbd {\eta}_B\left[A_{SM}(ii)+ A_{H^{\pm}}(B)+ A_{\chi^\pm}(B)+ A_{\tilde{g}}(B) \right],
\label{eq:dmd_susy}
\eea
where $A_{SM}(B)=S(x_t)|V^*_{tq}V_{tb}|^2$.

The expressions for $A_{H^\pm}(B)$, $A_{\chi^\pm}(B)$ and
$A_{\tilde{g}}(B)$ are obtained from the SUSY box diagrams.
Here, $H^\pm$, $\chi^\pm_j$, $\tilde{t}_a$ and $\tilde{d}_i$ represent,
respectively, the charged Higgs, chargino, s-top and s-down-type
s-quarks. The contribution of the intermediate states involving
neutralinos is small and usually neglected. The expressions for
$A_{H^\pm}(B)$, $A_{\chi^\pm}(B)$ and $A_{\tilde{g}}(B)$ are given
explicitly in -\cite{Branco:1994eb}, \cite{Goto:1998iy}, \cite{Bertolini:1990if}-.

Similarly in this context $|\ek|$ can be expressed by:
\bea
|\ek| = C_{\epsilon}B_K
\left[\mbox{Im}~A_{SM}(K) + \mbox{Im}~A_{H^\pm}(K) + 
\mbox{Im}~A_{\chi^\pm}(K) + \mbox{Im}~A_{\tilde{g}}(K) \right]~,
\label{eq:ek_susy} 
\eea
where, the constant $C_{\epsilon}$ and  $\mbox{Im}~ A_{SM}$ are as given in \eq{eq:ek_sm} and 
the expressions for $\mbox{Im}~A_{H^\pm}(K)$,
$\mbox{Im}~A_{\chi^\pm}(B)$ and $\mbox{Im}~A_{\tilde{g}}(B)$ can be
found in -\cite{Branco:1994eb}, \cite{Goto:1998iy}, \cite{Bertolini:1990if}-.

In the MFSV context~\cite{Ciuchini:1998xy}, apart from the SM degrees of freedom, only
charged Higgs fields, charginos and a light s-top (assumed to be
right-handed) contribute, with all other supersymmetric particles
integrated out. This scenario is effectively implemented in a class of
SUGRA models (both minimal and non-minimal) and gauge-mediated models
\cite{Dine_GMM} in which the first two s-quark generations are heavy and
the contribution from the intermediate gluino - s-quark states is small
\cite{Nihei:1997kh,Goto:1995zk,Goto:1996dh,Goto:1998qv,Goto:1998iy}.

The phenomenological profiles of the unitary triangle and CP phases for the SM and this class of supersymmetric models can thus be meaningfully compared. Given the high precision on the phases $\alpha$, $\beta$ and $\gamma$ expected from experiments at
$B$-factories and hadron colliders, a quantitative comparison of this
kind could provide means of discriminating between the SM and this
class of MSSM's.

In the next subsections we follow the work of Krauss and Soff \cite{Krauss:1998mt} in order to identify the NLO QCD supersymmetric contributions to the $\Delta F=2$ processes.

%%%%%%%%%%%%%%%%%%%%%%%%%%%%%%%%%%%%%%%%%%%%%%%%%%%%%%%%%%%%%%
\subsubsection{ NLO QCD-corrected effective Hamiltonian
for \boldmath{$\Delta B=2$}}
%%%%%%%%%%%%%%%%%%%%%%%%%%%%%%%%%%%%%%%%%%%%%%%%%%%%%%%%%%%%%%
 The NLO QCD-corrected effective Hamiltonian
for $\Delta B=2$ transitions in the minimal flavour violation SUSY
framework can be expressed as follows at the bosonic scale $\mu=O(m^2_{B_d})$\cite{Krauss:1998mt}:
\be
H_{\rm{eff}}^{(\Delta B=2)} 
= \frac{G_F^2}{4 \pi^2} (V_{td}V_{tb}^*)^2 {\eta}_{2, B}(\mu)~ S~ 
{\mathcal{Q}}_{1} (\mu)~,
\label{effHsusy}
\ee
where ${\mathcal{Q}}_1$ is given in \eq{eq:DB2_Oi}, $S$ is a sum of the Inami-Lim functions for the different internal particles, $S=S(x_W,x_H)+\tilde S({x_i,y_a})$
{\footnote{ 
$S(x_W,x_H) = S_{WW}(x_W)+2\,S_{WH}(x_W,x_H)+S_{HH}(x_H)\;$ and \\
$~~~~~~\tilde S(\{x_i,y_a\}) = \sum\limits_{i,j=1}^2
 \sum\limits_{a,b=1}^6 \tilde K_{ij,ab} \tilde S(x_i,x_j,y_a,y_b)$}}, where
\bea
x_{W,H} = \frac{m_t^2}{M_{W,H}^2}\;,\;
x_{d,s,c,b} = \frac{m_{d,s,c,b}^2}{M_W^2}\;,\;
y_a = \frac{\tilde m_{\tilde q_a}^2}{M_W^2}\;,\;
x_i = \frac{\tilde m_{\tilde \chi_i}^2}{M_W^2} \,.
\label{eq:rats_x_y}
\eea 
${\mathcal{Q}}_1(\mu)$ and $\eta_2(B,\mu)$ depend on the scale but once they are run down to the electroweak scale $\mu=M^2_W$ and the matching conditions are performed, then the NLO QCD correction factor $\eta_{2}(B)$ can be expressed as \cite{Krauss:1998mt}:
\bea
\eta_{2}(B) = \alpha_s(m_W)^{\gamma^{(0)}/(2 \beta_{n_f}^{(0)})} \left[
1 + \frac{\alpha_s(m_W)}{4 \pi} \left(\frac{D}{S} + Z_{n_f}\right) \right]~,
\label{eta2s}
\eea
in which $n_f$ is the number of active quark flavours (here $n_f=5$),
$Z_{n_f}= \frac{\gamma_{n_f}^{(1)}}{2 \beta_{n_f}^{(0)}} -
\frac{\gamma^{(0)}}{2 {\beta_{n_f}^{(0)}}^2} \beta_{n_f}^{(1)}$, and $\gamma^{(0)}$ and $\beta_{n_f}^{(0)}$ are the anomalous dimension and beta functions of QCD
{\footnote{\label{ft:and_beta_qcd}
$\gamma^{(0)}       = 6 \frac{N_c-1}{N_c}, ~~~ \beta_{n_f}^{(0)} = 
\frac{11N_c-2 n_f}{3}$ ~,
$\beta_{n_f}^{(1)}  = \frac{34}{3}N_c^2 - \frac{10}{3} N_c n_f - 2 C_fn_f$~,
$\gamma_{n_f}^{(1)} = \frac{N_c-1}{2N_c} \left[-21 +\frac{57}{N_c} - 
\frac{19}{3}N_c + \frac{4}{3} n_f \right]$.
}} with $N_c=3$ and $C_F=4/3$.
 The function $D$ comes about from identifying the QCD correction factor
once the Wilson coefficient is computed (in the naive dimensional regularization scheme using $\overline{MS}$) and is a function of \eq{eq:rats_x_y}, 
$D=D(x_W,x_H,x_{\mu_0})+\tilde{D}(\{x_i,y_a\},x_{\mu_0})$. The factor $[1+\frac{\alpha_s(MW)}{4\pi}\left(\frac{D}{S}+Z_{n_f}\right)]$ is quite stable against variations of the functions $D$ and $S$ and has been found to be \cite{Krauss:1998mt} and \cite{Ali:1999we}, around $0.89$ and hence $\eta_2(B)$ it is found to be $\eta_2(B)=0.51$ in the $\overline{MS}$ scheme.
 
The Hamiltonian given above for $\Bdmix$ mixing
leads to the mass difference
\bea
\dmd = \frac{G_F^2}{6 \pi^2} (V_{td}V_{tb}^*)^2 \eta_{2}(B)~ S ~
\fbd^2 B_{B_d}~,
\label{eq:dmdsusy}
\eea
and analogously for $\dms$. Since in the MFV supersymmetric context, the QCD correction factors are identical for $\dmd$ and $\dms$ then $\dmd$ and $\dms$ have the same enhancement, with respect to their values in the SM. Hence the ratio $\dms/\dmd$ is the same as in the SM.
%
%%%%%%%%%%%%%%%%%%%%%%%%%%%%%%%%%%%%%%%%%%%%%%%%%%%%%%%%%%%%%%
%
\subsubsection{NLO QCD-corrected effective Hamiltonian
for \boldmath{$\Delta S=2$}}
%
%%%%%%%%%%%%%%%%%%%%%%%%%%%%%%%%%%%%%%%%%%%%%%%%%%%%%%%%%%%%%%
%
The NLO QCD-corrected Hamiltonian for $\Delta S=2$ transitions in the
MFSV framework has also been
obtained in Ref.~\cite{Krauss:1998mt}. From this, the result for $|\ek|$ can
be written as:
\bea
|\ek| &=&C_{\epsilon}B_K
A^2 \lambda^6 \bar{\eta} [-\eta_1x_c + A^2 \lambda^4\left(1-\bar{\rho}\right)\eta_2(K)~S~  +\eta_3S(x_c,x_t)],\nn
\label{eps2}
\eea
where the NLO QCD correction factor is \cite{Krauss:1998mt}:
\bea
\eta_{2}(K) &=& \alpha_s(m_c)^{\gamma^{(0)}/(2\beta_3^{(0)})}
\left(\frac{\alpha_s(m_b)}{\alpha_s(m_c)}\right)^{\gamma^{(0)}/(2\beta_4^{(0)})}
\left(\frac{\alpha_s(M_W)}{\alpha_s(m_b)}\right)^{\gamma^{(0)}/(2\beta_5^{(0)})} 
\nn\\
&& \left[ 1 + \frac{\alpha_s(m_c)}{4 \pi} (Z_3 - Z_4) +
\frac{\alpha_s(m_b)}{4 \pi} (Z_4 - Z_5) +
\frac{\alpha_s(M_W)}{4 \pi} (\frac{D}{S}+ Z_5)\right],\label{eta2k}
\eea
again, as it happens with $\eta_2(K)$, the factor involving  the functions $Z_i$, $D$ and $S$, is quite stable and has been calculated~\cite{Krauss:1998mt, Ali:1999we} to give $0.84$, consequently $\eta_2(K)$ is estimated to give $\eta_2(K)=0.5 3$ in the $\overline{MS}$ scheme.

\subsubsection{Effects of MVSV for \boldmath{$\dmd$} and \boldmath{$\ek$}}
Here we summarize the effects of MVSV for the transitions $\Delta B=2$ and $\Delta S=2$ which with respect to the SM may be characterized by the shifts
\bea
\label{eq:susySxtchg}
\eta_B S(x_t)\quad &\rightarrow& \quad \eta_2(B)~ S,\nn\\
\eta_2 S(x_t)\quad &\rightarrow& \quad \eta_2(K)~ S,
\eea
where $\eta_2(B)$, $\eta_2(K)$ and $S$ are functions of $(m_{\chi_2^\pm},m_{\tilde{t}_2}, m_{H^\pm},\tan\beta_s)$. In Section (\ref{sbsec:outputs}) we are going to use the approximations of \cite{Ali:1999we} and \cite{Krauss:1998mt} for $\eta_2(B)$ and $\eta_2(K)$
\bea
\eta_2(B)&=&0.51,\nn\\
\eta_2(K)&=&0.53
\label{eq:eta2_B_K_mfvs}
\eea
and compute explicitly $S$ for different supersymmetric inputs.

%
%%%%%%%%%%%%%%%%%%%%%%%%%%%%%%%%%%%%%%%%%%%%%%%%%%%%%%%%%%%%%%%%%%%%%%%%%%%%%%%
\subsection{Scenario II \label{sbsec:scnii}}
In this context we can test models for which the BSM contribution to $\ek$ it is considerably larger than the corresponding contribution to $\dmd$ and $\dms$ or vice-versa. Hence we can expect variations of the top-W box diagram function, $S(x_t)$, different for $\dmd$ (or $\dms$) and for $\ek$.

An example of this possibility occurs in models with a horizontal symmetry where often the CKM matrix is no longer the only source of CP violation because there are off diagonal elements in the soft squared matrices as well as in the trilinear terms. In this respect these models depart from the condition of MFV since the dynamics of flavour violation is no longer completely determined from the structure of the SM Yukawa couplings. However if the models behave in the low energy limit as a supersymmetric extension of the SM, i.e. 
all fields other than the SM ones 
%all other fields not being the SM fields 
and their supersymmetric partners have been integrated out at the scales where the supersymmetry is broken; then we can describe $\dmd$ and $\ek$ as in \eq{eq:dmd_susy} and \eq{eq:ek_susy} respectively where we can have sizable contributions for example from the gauginos.

In \cite{Ross:2004qn} we have constructed and analyzed the predictions of a supersymmetric model with an underlying $SU(3)$ horizontal symmetry and with spontaneous CP violation. A full numerical analysis of this type of models, including the exact diagonalization of the soft mass matrices, needs to be performed, but since in the context of the Mass Insertion approximations (MI) {\footnote{Defined by $\delta^q_{AB}=\frac{V^{q\dagger}_A \tilde{M}^2_{\tilde{q}} V^{q}_B }{\tilde{m}_{\tilde{q}}}$, where $A,B=L,R$ refers to the chirality of the super-partners of the internal lines of the relevant $\Delta F$ processes and $V^q_A$ are the matrices diagonalizing the Yukawa couplings, e.g. $Y^d_{\rm{diag}}=V^{d\dagger}_L Y^d V^d_R$}} the $\Delta F=2$ Hamiltonian can be expressed \cite{Gabbiani:1996hi} in terms of $\delta^d_{LL}$, $\delta^d_{RR}$, $\delta^d_{LR}$ and  $\delta^d_{RL}$, we have compared the experimental bounds with the predictions of the model. The model puts the following limits
\bea
\sqrt{{\rm{Im}}(\delta^d_{RR})_{12}} \leq 6.8 \times 10^{-4} \sin \Phi_1\nn\\
\sqrt{[{\rm{Im}} (\delta^d_{RR})_{12} (\delta^d_{LL})_{12}]} \leq 2 \times 10^{-4} \sin \Phi_1,
\eea
where $\Phi_1$ is a phase of the model, preferred to be small. These limits can be compared with the experimental limits, obtained in the kaon sector, of 
$\sqrt{{\rm{Im}}(\delta^d_{RR})_{12}} \leq 3.2 \times 10^{-3}$ and
$\sqrt{[{\rm{Im}} (\delta^d_{RR})_{12} (\delta^d_{LL})_{12}]} \leq 2.2 \times 10^{-4}$, which correspond to average s-quark masses of $\tilde{m}_{\tilde{q}}= 500$ GeV. These can be scaled as $\tilde{m}_{\tilde{q}}{\rm{GeV}}/ 500$, so we can see that for large values of $\tilde{m}_{\tilde{q}}$, we can have a large contribution to $\ek$. However as we will see for the current results on possible contributions of non SM physics to $\ek$, these are rather small, limiting the possibilities of a big contribution to $\ek$ from these kind of theories. In the $B$ sector the contributions can be negligible, so this is an example in which $\ek$ can receive big contributions, while $\dmd$ is left practically unchanged.

%%%%%%%%%%%%%%%%%%%%%%%%%%%%%%%%%%%%%%%%%%%%%%%%%%%%%%%%%%%%%%%%%%%%%%%%%%%%%%%
%

\subsection{Results \label{sec:results}}
\subsubsection{Procedure to extract information on bounds BSM}

The procedure to obtain the pdfs for $r_{B_d}$, $\theta_{B_d}$ and $r_{\ek}$ follows from the very simple observation that we can write these parameters as
\bea
r_{B_d} &=& \frac{\Delta m^{\rm{exp.}}_{\dmd}}{q_{\dmd} S(x_t)},\quad 
\sin(2\beta^{\rm{SM}}+2\theta_{B_d})=\sin 2\beta^{\rm{exp.}}, \nn\\
r_{\ek} &=& \frac{\Delta \ek^{\rm{exp.}} }{q_{\ek,1} + q_{\ek,2} S(x_t)},
\eea
where $q_\dmd$, $q_{1,\ek}$ and $q_{2,\ek}$ are functions of the Wolfenstein parameters (and of the correspondent $\Delta B=2$ and $\Delta S=2$ transition amplitudes). 
We can extract the pdf for $r_{B_d}$ and $r_{\ek}$ as
$p(r)\propto \int p(\rb,\eb) |\partial \rb /\partial r| d\eb$, 
exactly as in the case of the parameters of the Classic Fit, \eq{eq:lik_parx}.
The most conservative way to obtain $r_{B_d}$, $\theta_{B_d}$ and $\ek$ is to use only the constraint $|V_{ub}|/|V_{cb}|$ and experimental constraints on $\alpha$ and $\gamma$, such as 
those associated with 
the decays $B\to \rho \rho,\ \rho\pi$ and $\rho \pi$ for $\alpha$ and $B\to D K$ for $\gamma$, which are $\Delta F=1$ processes. However we use the constraint $\dms/\dmd$ for two reasons: the first one  is to study the impact of the $\dms$ measurement and the second one to probe the models for which this ratio is left invariant, such as the MFV case. In order to present the impact on the parameters measuring processes BSM we perform the following sets of fits.
\begin{enumerate}

%\item Constraints $|V_{ub}|/|V_{cb}|$  and $\dms/\dmd$ to obtain $r_{B_d}$, $\theta_{B_d}$ and $r_{\ek}$,

\item Constraints $|V_{ub}|/|V_{cb}|$, $\sin 2\beta$, $\dms/\dmd$ and $\dmd$ to obtain $r_{\ek}$,

\item  Constraints $|V_{ub}|/|V_{cb}|$, $\dms/\dmd$ and $\ek$ to obtain $r_{\dmd}$ and $\theta_{B_d}$,

\item  Constraints $|V_{ub}|/|V_{cb}|$, $\sin 2\beta$, $\dms/\dmd$  to obtain $r_{\ek}$, $r_{\dmd}$ and $\theta_{B_d}$.

\end{enumerate}
% 
%In the (iii)-th case, 
In case (iii), 
$\theta_{B_d}$ should be consistent with zero at some CL since we are imposing the $\sin 2\beta$ constraint.
From these fits we can extract then information for bounds on the different examples presented in the previous section.
\subsubsection{Outputs\label{sbsec:outputs}}
%
%{\bf{(i)}} Results using the constraints $\mathbf{|V_{ub}|/|V_{cb}|}$ and $\mathbf{\dms/\dmd}$ to obtain $r_{B_d}$, $\theta_{B_d}$ and $r_{\ek}$.\\

\noindent {\bf{(i) Results using the constraints $\mathbf{|V_{ub}|/|V_{cb}|}$, $\mathbf{\sin 2\beta}$, $\mathbf{\dms/\dmd}$ and $\mathbf{\dmd}$}}\\
\begin{figure}[ht]
%\hfill
\begin{minipage}[t]{5.7cm}
\begin{center}
\includegraphics[width=1\textwidth]{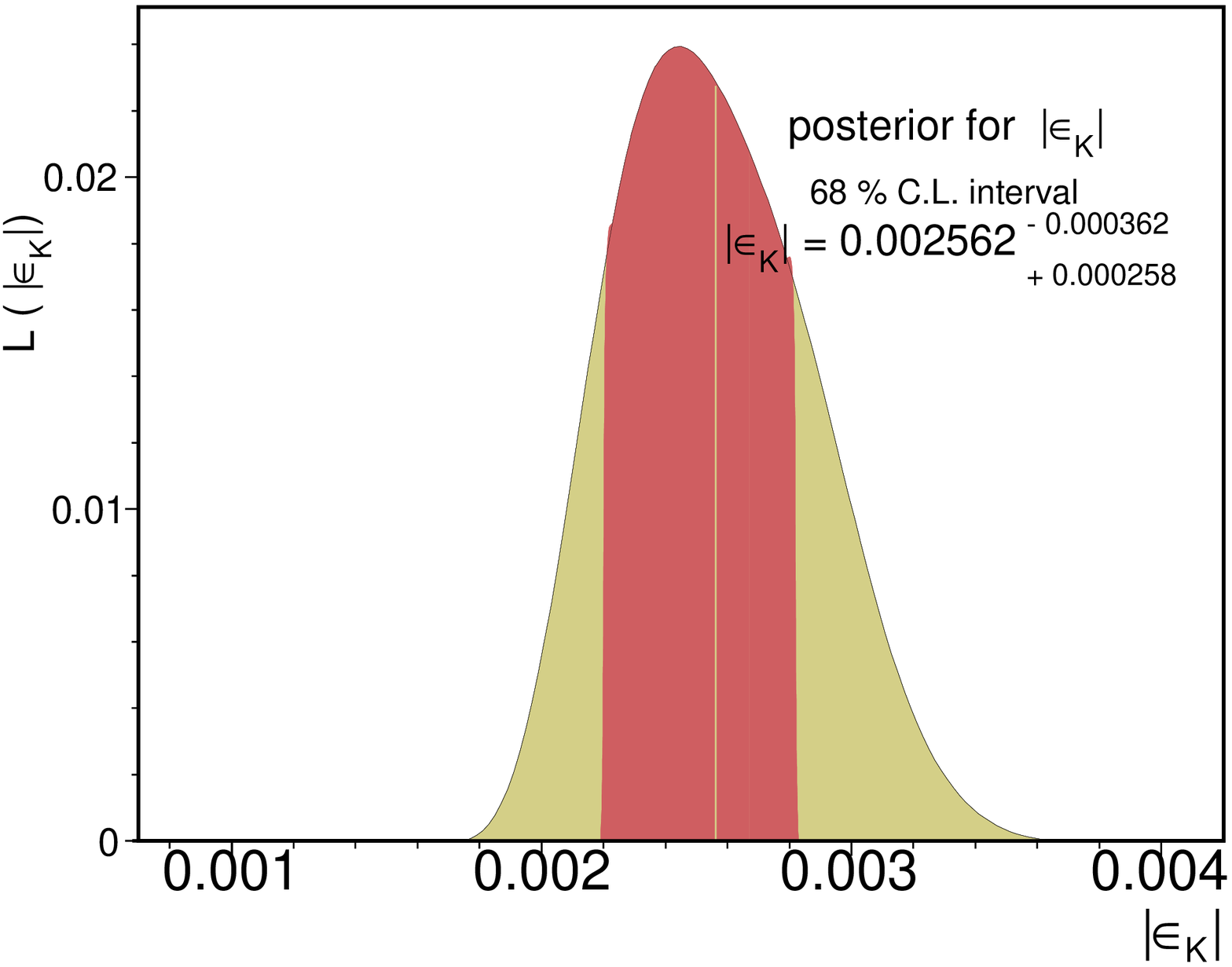}\label{fig:epsk_r4b}
(i)
\end{center}
\end{minipage}
\hfill
\begin{minipage}[t]{5.7cm}
\begin{center}
\includegraphics[width=1\textwidth]{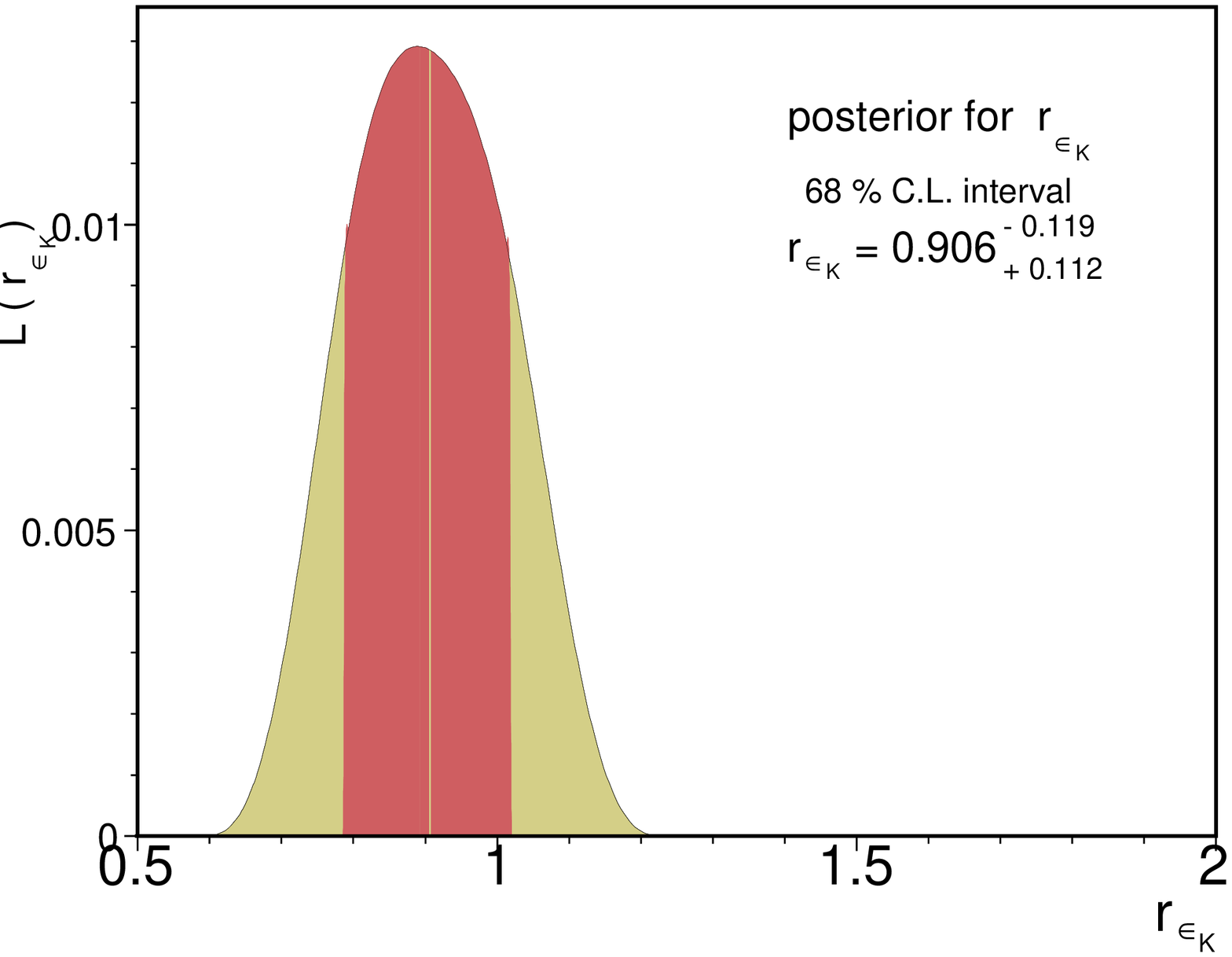}\label{fig:reK_r4b}
(b)
\end{center}
\end{minipage}
\hfill
\end{figure}
\begin{figure}[ht]
%\hfill
\begin{minipage}[t]{5.7cm}
\begin{center}
\includegraphics[width=1\textwidth]{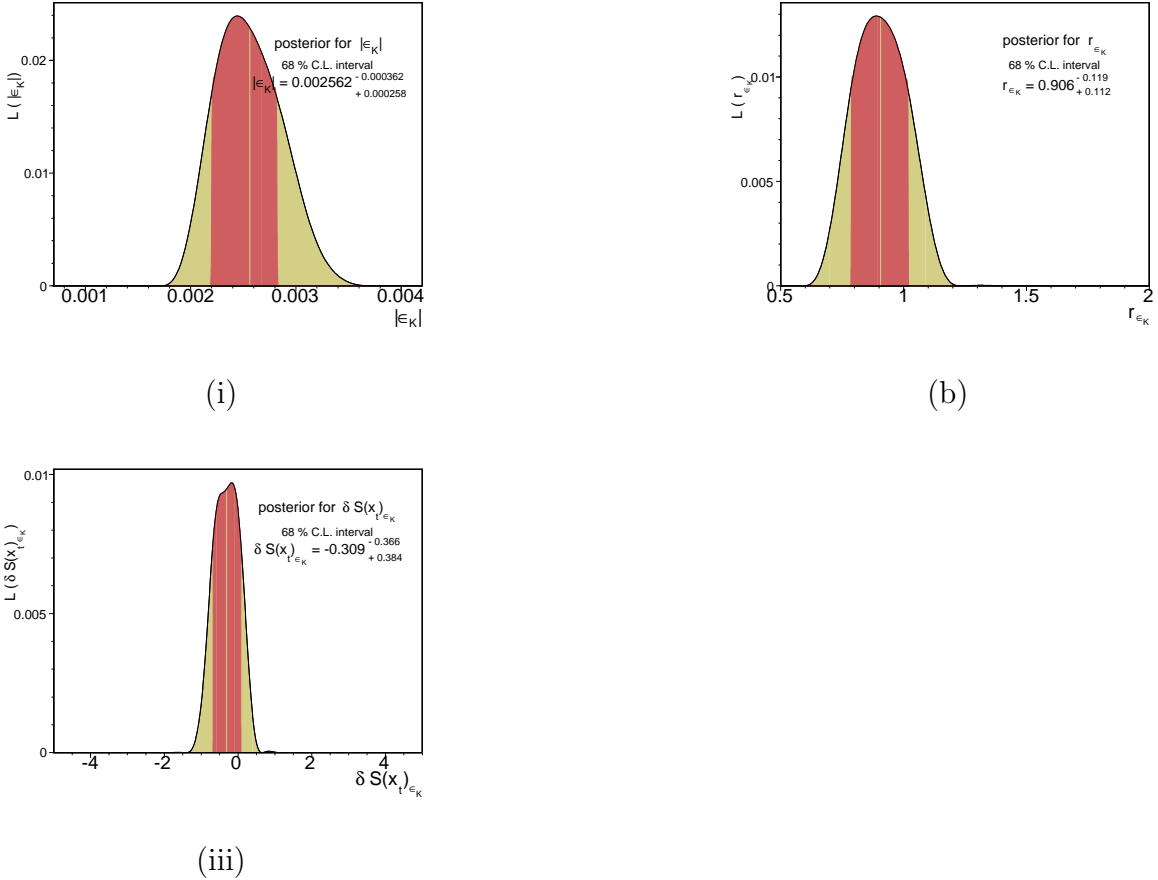}\label{fig:delepsk_r4b}
(iii)
\end{center}
\end{minipage}
\hfill
\begin{minipage}[t]{5.7cm}

\end{minipage}
\hfill
\caption{\small One dimensional probabilities for (i) $\ek$, (ii) $r_{\ek}$ and (iii) $\delta S(x_t)_{\ek}$ when $\ek$ is not imposed as a constraint.}
\end{figure}

The results of this fit are relevant to probe the models for which $\dmd$ is expected to be left invariant while a change in $\ek$ may happen, as it is the case of the $SU(3)$ flavour models \cite{Ross:2004qn}, as pointed out in Section (\ref{sbsec:scnii}). First we can check that the experimental value of $\ek$ lies within a third of the 68\%~CL of the output value of $\ek$: $2.56^{-0.36}_{+0.26}\times 10^{-3}$. Then we see that $r_{\ek}$ is compatible with unity:
\bea
r_{\ek}=0.906_{-0.119}^{+0.112}
\label{eq:rek}
\eea
although the relative uncertainty in this parameter is about $13\%$. % of its value. 
Taking into account that the relative uncertainty in the hadronic parameter $B_K$ is of about $ 12\% $, since $B_K=0.707\pm 0.0066\pm 0.072$, we see then  that the uncertainty in the determination of $r_{\ek}$ is due to the uncertainty on $B_K$, since indeed we can only determine with a good accuracy $r_{\ek}B_K$. However it is important to remark that the value of $r_{\ek}$ does not cluster exaclty around 1, which leaves interesting possibilites of processes BSM in this case. For the example of $SU(3)$ model presented in Section (\ref{sbsec:scnii}) we see that if this model is to satify the present determination of $r_{\ek}$ then the s-quark masses should not exceed $500$ GeV and the phase $\Phi_1$ should be quite small.

In the general context of MFV D'Ambrosio \emph{et al.}~\cite{D'Ambrosio:2002ex} have classified the operators for which we can expect contributions BSM. In particular for the $\Delta F=2$ operators, ${\mathcal{O}}_n$ the effective Hamiltonian can be written as
\bea
\mathcal{H}_{\rm{eff}}=\frac{1}{\Lambda^2} \sum_n a_n {\mathcal{O}}_n + h.c. \quad \rightarrow\quad
\frac{G_F\alpha}{2\sqrt{2\pi}\sin^2\theta_{W}}V^*_{ti}V_{tj}\sum_n C_n {\mathcal{Q}}_n +h.c.,
\label{eq:ham_lambda}
\eea
where $\Lambda$ is the scale at which the operators become relevant and $a$ its effective coefficient. The term after the arrow is the standard notation for the Hamiltonian using the Wilson coefficients $C_n$ and the QCD operators ${\mathcal{Q}}_n$. The Hamiltonian in \eq{eq:ham_lambda} can be normalized such that the electroweak contribution is of order one, i.e. for the SM $a_{\rm{SM}}=1$ and the scale $\Lambda_{\rm{SM}}=\Lambda_o$ can be defined such that $\Lambda_o=y_t \sin^2\theta_W M_W/\alpha\approx 2.4$~TeV. Hence we consider the ratios $r_{B_d}$ and $r_{\ek}$, defined through \eq{eq:gen_ek_dms},  %we are 
getting a measure of $(\Lambda_o/\Lambda)^2$. In particular 
\bea
\delta S(x_t)=\frac{4a}{a_{\rm{SM}}}\left(\frac{\Lambda_o}{\Lambda}\right)^2,
\label{eq:bound_lambda}
\eea
such that we can put bounds on the scale $\Lambda$ at which a process BSM could take place, by considering $a=1$. For this case, taking into account the full 68\%~CL range, the bounds for $\Lambda$ are $\Lambda>5.84$ TeV for a negative $\delta S(x_t)$ contribution with respect to the SM contribution and $\Lambda> 17.52$ TeV for a positive contribution.
\vspace*{0.5cm}

\noindent {\bf{(ii) Results using the constraints $\mathbf{|V_{ub}|/|V_{cb}|}$, $\mathbf{\dms/\dmd}$ and $\mathbf{\ek}$}}\\

\begin{figure}[ht]
%\hfill
\begin{minipage}[t]{5.7cm}
\vspace{-5cm}
\begin{center}
\includegraphics[width=1\textwidth]{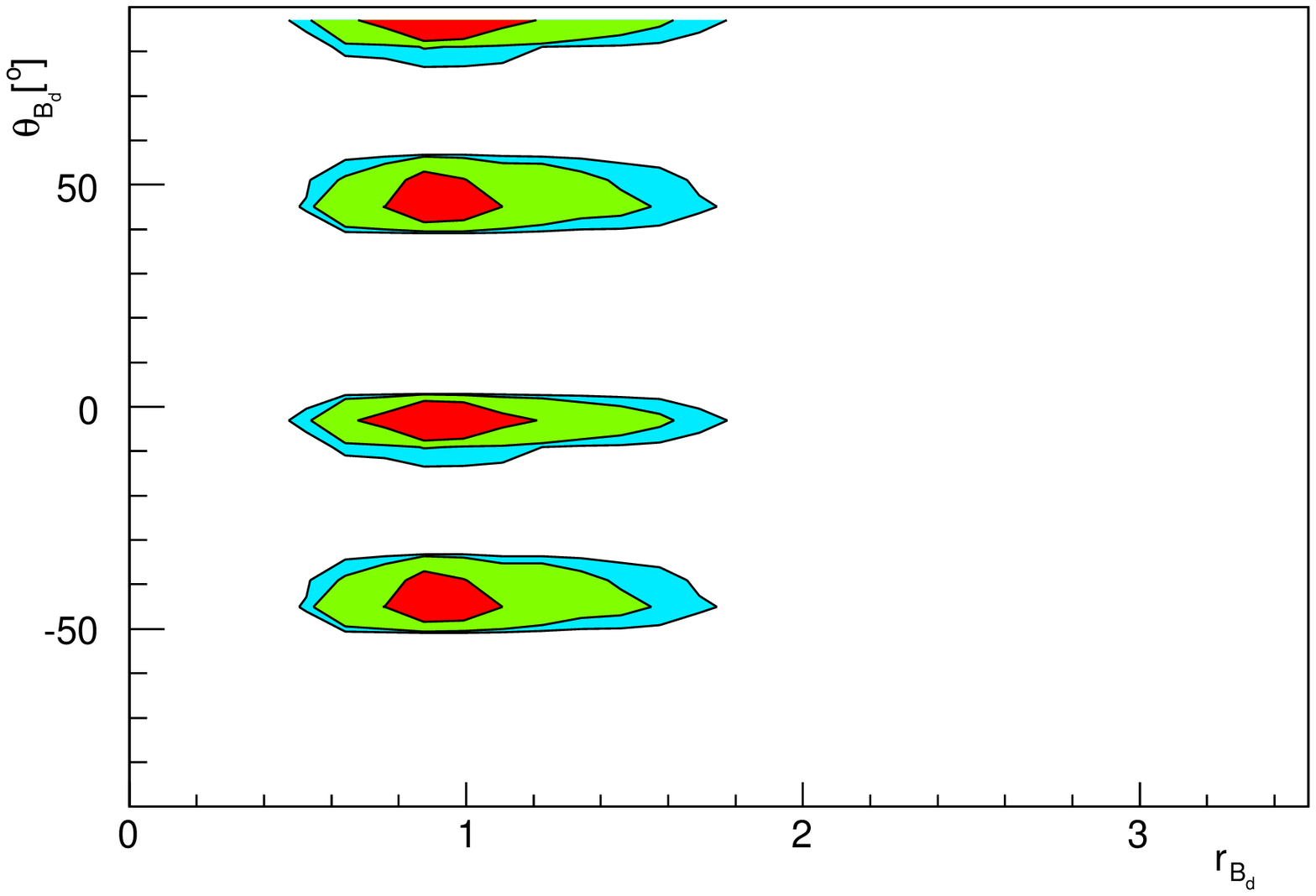}\label{fig:rBd_thetaBd_r5b}
(i)
\end{center}
\end{minipage}
\hfill
\begin{minipage}[t]{5.7cm}
\begin{center}
\includegraphics[width=1\textwidth]{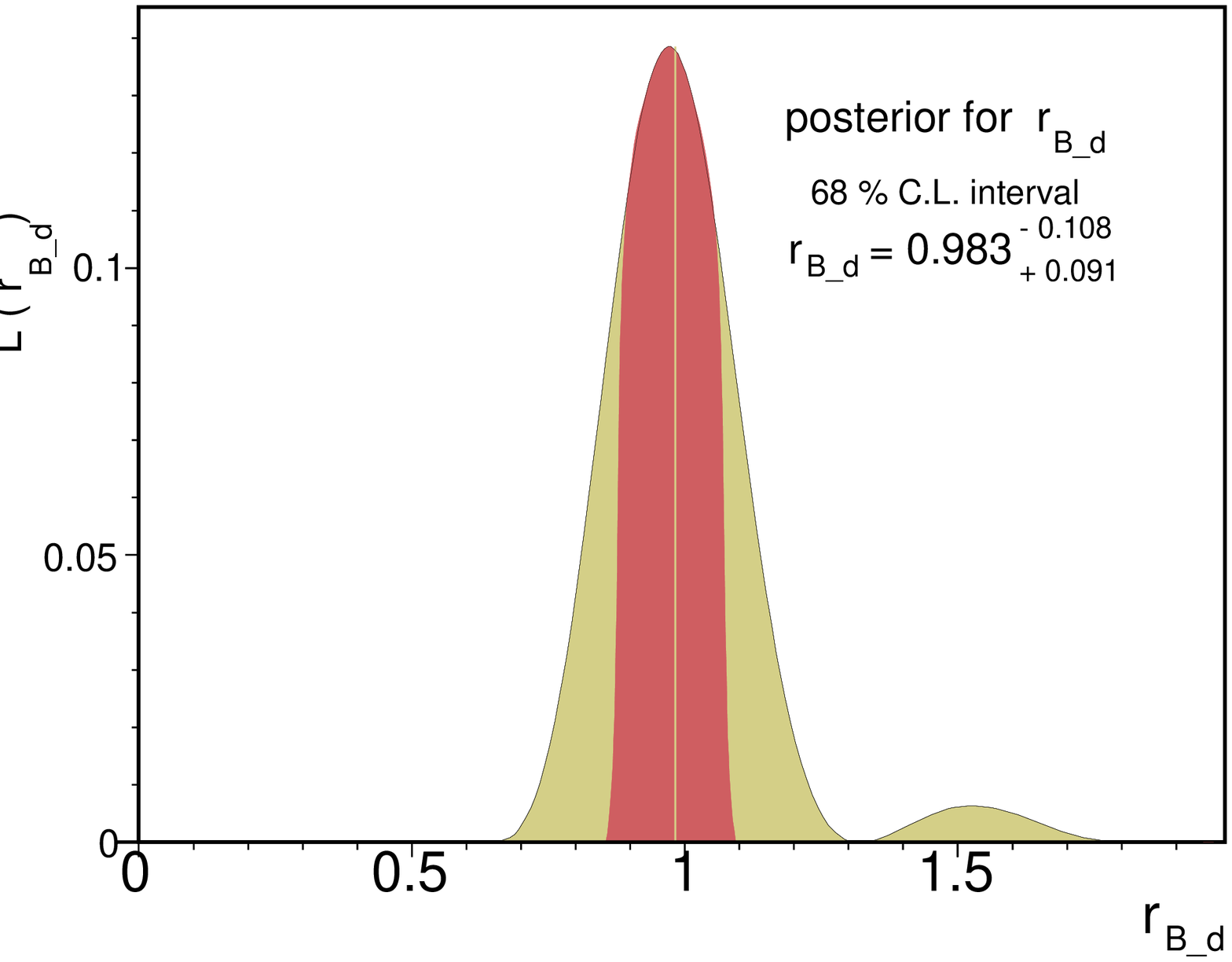}\label{fig:rBd_r5b}
(ii)
\end{center}
\end{minipage}
\hfill
\caption{\small The two dimensional probability (i) for $r_{Bd}$ and $\theta_{B_d}$ and the one dimensional probability (ii) for  $r_{Bd}$. As we can see this solution agrees remarkably well with the SM, for which $r_{Bd}=1$ and $\theta_{B_d}=0$, although there could be a phase different from zero, giving new contributions to the phase $\beta$ and $\alpha$.}
\end{figure}
Here it is important to compare the output value of the fit for $\dmd$ to its experimental counterpart and to the output of the complete fit to check that this fit is consistent with the experimental value.
The output values  of $\dmd$ for this fit and the complete fit are respectively
\bea
\dmd=(0.488-0.294+0.291){\rm{ps}}^{-1},\quad \dmd=(0.476-0.196+0.183){\rm{ps}}^{-1},\nn
\eea
which are to be compared to the experimental measurement $\dmd=(0.506\pm 0.005)~\ips$. 
%so we 
We can see that the output value of $\dmd$ agrees remarkably well with its experimental counterpart, consequently the output value for $r_{B_d}$, 
\bea
r_{B_d}=0.983-0.108+0.091,
\eea
is still consistent with 1. This value and its relative uncertainty ($10\%$) however have decreased with respect to the value used when considering the last bound on $\dms$, which gave $r_{B_d}=1.095-0.315+0.173$ with a relative uncertainty of about 23\%, leaving some possibilities for processes BSM. We can check this also by determining the phase $\theta_{B_d}$, as defined in \eq{eq:bsm_cont_b_a},
\bea
\theta_{B_d}=(-3.35-3.06+4.03)[{\rm{^o}}],
\label{eq:thetaBd_sii}
\eea
which is compatible with the SM at $68\%$, although we could have an interesting possibility of having a contribution to the angles $\beta$ and $\alpha$, through the relation \eq{eq:bsm_cont_b_a}, while no contribution BSM to $\dmd$.
\vspace*{0.5cm}

%%%%%%%%%%%%%%%%%%%%%%%%%%%%%%%%%%%%%%%%%%%%%%%%%%%%%%%%%%%%%%%%%%%%%%%%%%%%%%
%
\noindent {\bf{(iii) Results using the constraints $\mathbf{|V_{ub}|/|V_{cb}|}$, $\mathbf{\sin 2\beta}$ and $\mathbf{\dms/\dmd}$}}\\

%
%%%%%%%%%%%%%%%%%%%%%%%%%%%%%%%%%%%%%%%%%%%%%%%%%%%%%%%%%%%%%%%%%%%%%%%%%%%%%%
%

The results of this fit can be used to test the hypothesis of MFV violation and specifically the supersymmetric case, with the extra assumptions mentioned in Section (\ref{sbsec:scenario1}). In these cases it is assumed that $\dmd$ and $|\ek|$ will have the same change in $S(x_t)$, the box diagram function of the top-W boson interaction, thus in general we can compare $r_{\ek}$ and $r_{B_d}$ as defined in \eq{eq:ratio_bsm_sm}. We quote here the values using the last $\dms$ bound and the current $\dms$ measurement (the pdfs of these can be found in figures (\ref{fig:r3b_Kaon} ii )  and (\ref{fig:r3b_Bd} ii), which show important differences. 
\bea
r_{\ek}=1.211-0.195+0.280, &\quad&  r_{B_d}=1.258-0.422+0.354,\nn\\
r_{\ek}=0.904-0.121+0.110, &\quad&  r_{B_d}=0.933-0.097+0.102,
\eea
where the first line corresponds to the last bound on $\dms$ and the second to its current measurement. Using the last bound on $\dms$, $\delta S(x_t)_{\ek}$ and  $\delta S(x_t)_{\dmd}$ agreed pretty well with the MFV scenario:
\bea
\delta S(x_t)_{\ek}= 0.684-0.565+0.280,\quad
\delta S(x_t)_{\dmd}=0.587-0.950+0.846,
\label{eq:delta_Sxt_iii}
\eea
but the uncertainties were quite big. With the current $\dms$ measurement, although there is %region of 
overlap for the values of $S(x_t)$ extracted from $\ek$ and $\dmd$, 
\bea
\delta S(x_t)_{\ek} = -0.373-0.375 + 0.343,\quad
\delta S(x_t)_{\dmd}= -0.082-0.720 + 0.359
\eea
the MFV assumption appears to be weakened.
\begin{figure}[ht]
%\hfill
\begin{minipage}[t]{5.7cm}
\begin{center}
\includegraphics[width=1\textwidth]{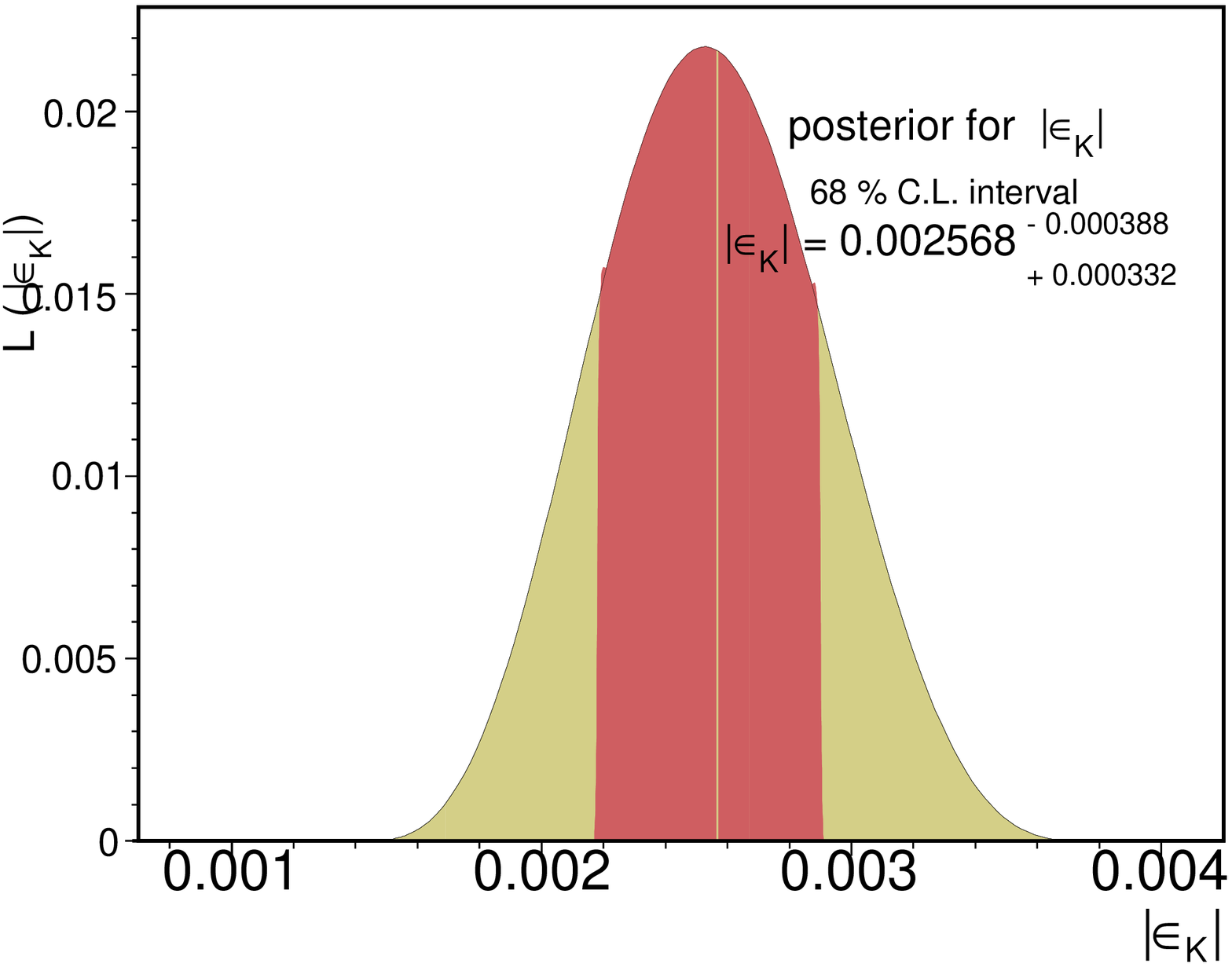}\label{fig:epsk_r3b}
(i)
\end{center}
\end{minipage}
\hfill
\begin{minipage}[t]{5.7cm}
\begin{center}
\includegraphics[width=1\textwidth]{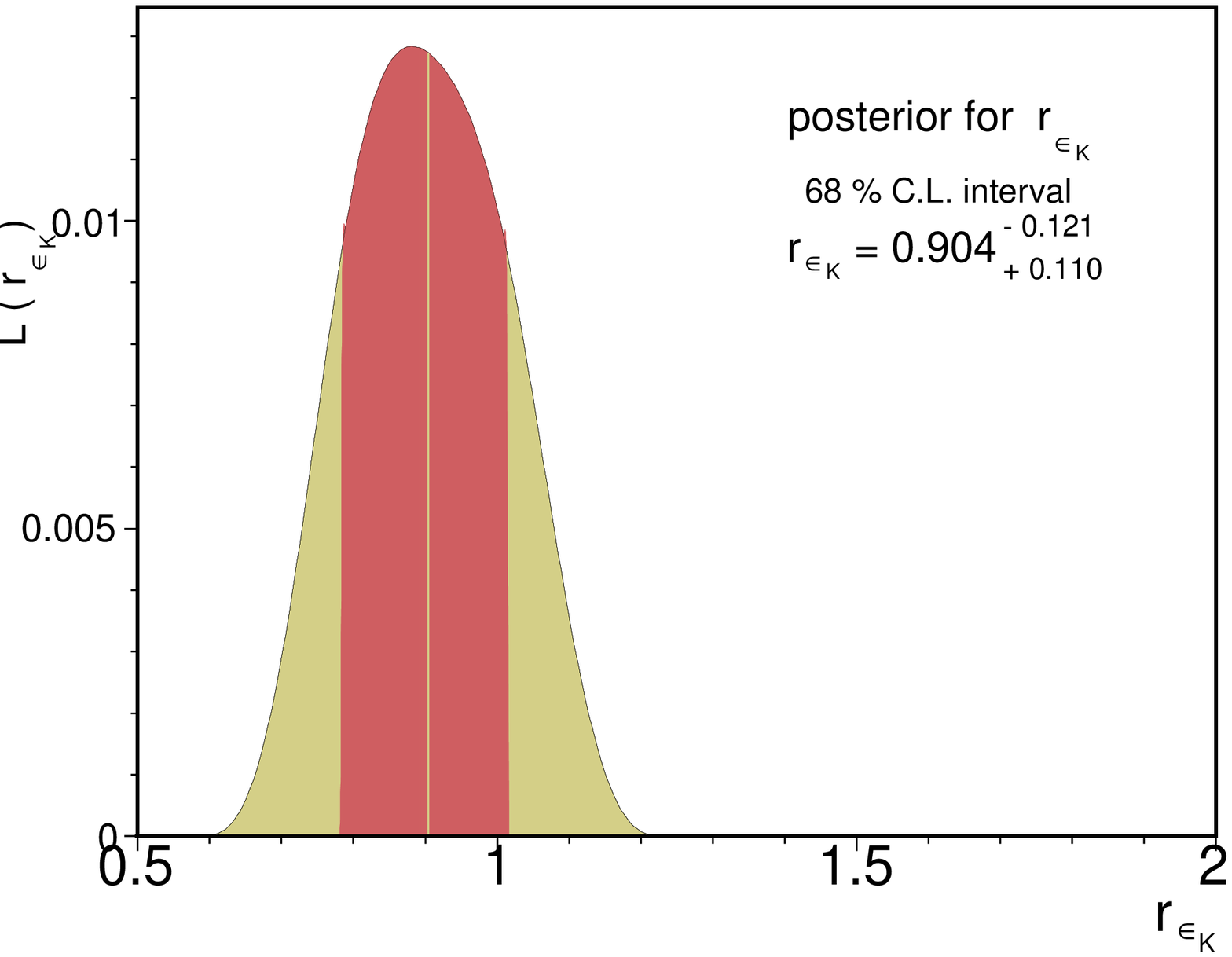}\label{fig:reK_r3b}
(ii)
\end{center}
\end{minipage}
\hfill
\begin{minipage}[t]{5.7cm}
%\vspace*{-5cm}
\begin{center}
\includegraphics[width=1\textwidth]{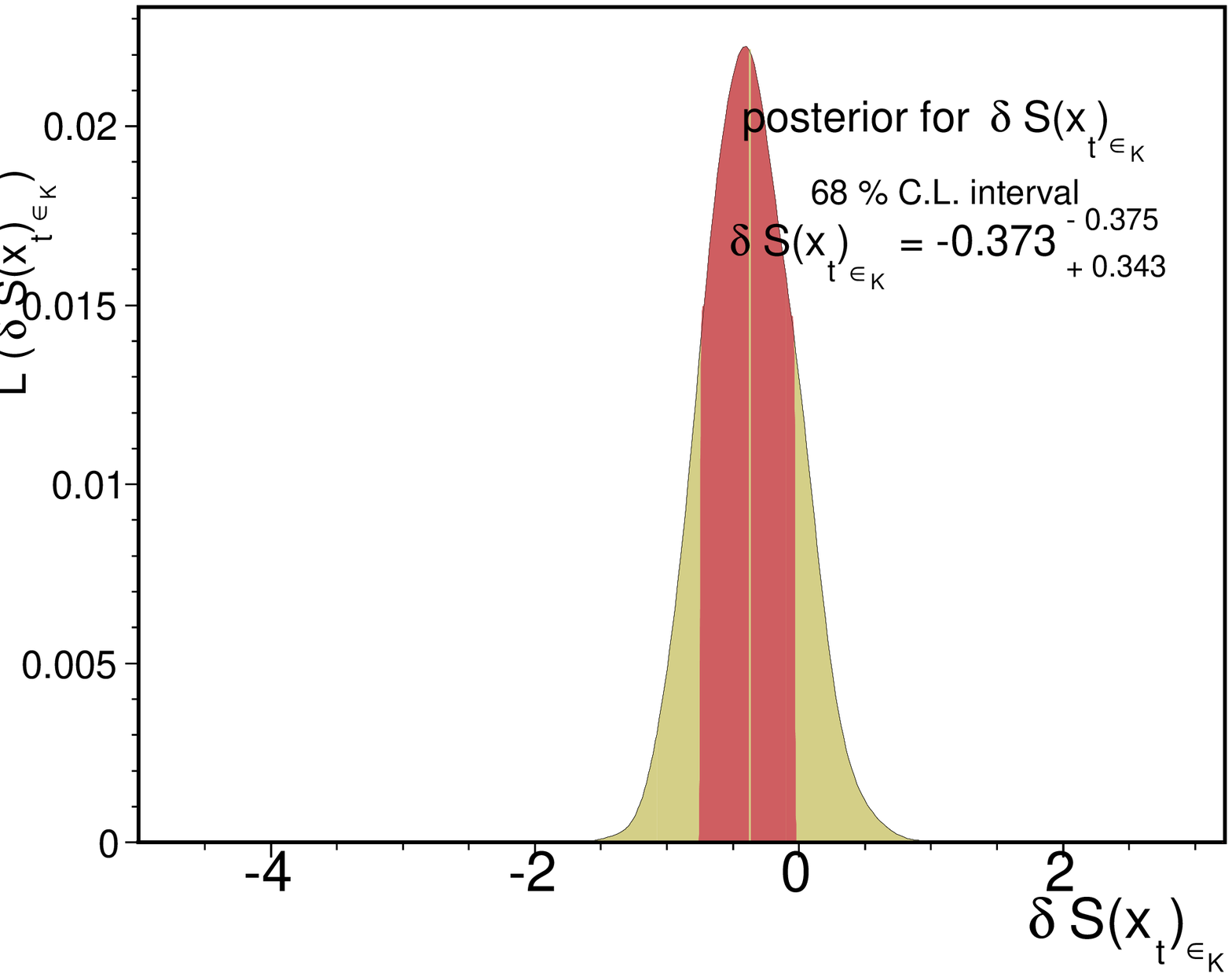}\label{fig:delepsk_r3b}
(iii)
\end{center}
\end{minipage}
\hfill
\begin{minipage}[t]{5.7cm}
{\centering
%...+mt[0];
}
\end{minipage}
\hfill
\caption{\small One dimensional probabilities for (i) $\ek$, (ii) $r_{\ek}$ and (iii) $\delta S(x_t)_{\ek}$ when $\ek$ and $\dmd$ are not imposed as constraints for the fit.}
\label{fig:r3b_Kaon}
\end{figure}

\begin{figure}[ht]
%\hfill
\begin{minipage}[t]{5.7cm}
\vspace*{-5cm}
{\begin{center}
\includegraphics[width=1\textwidth]{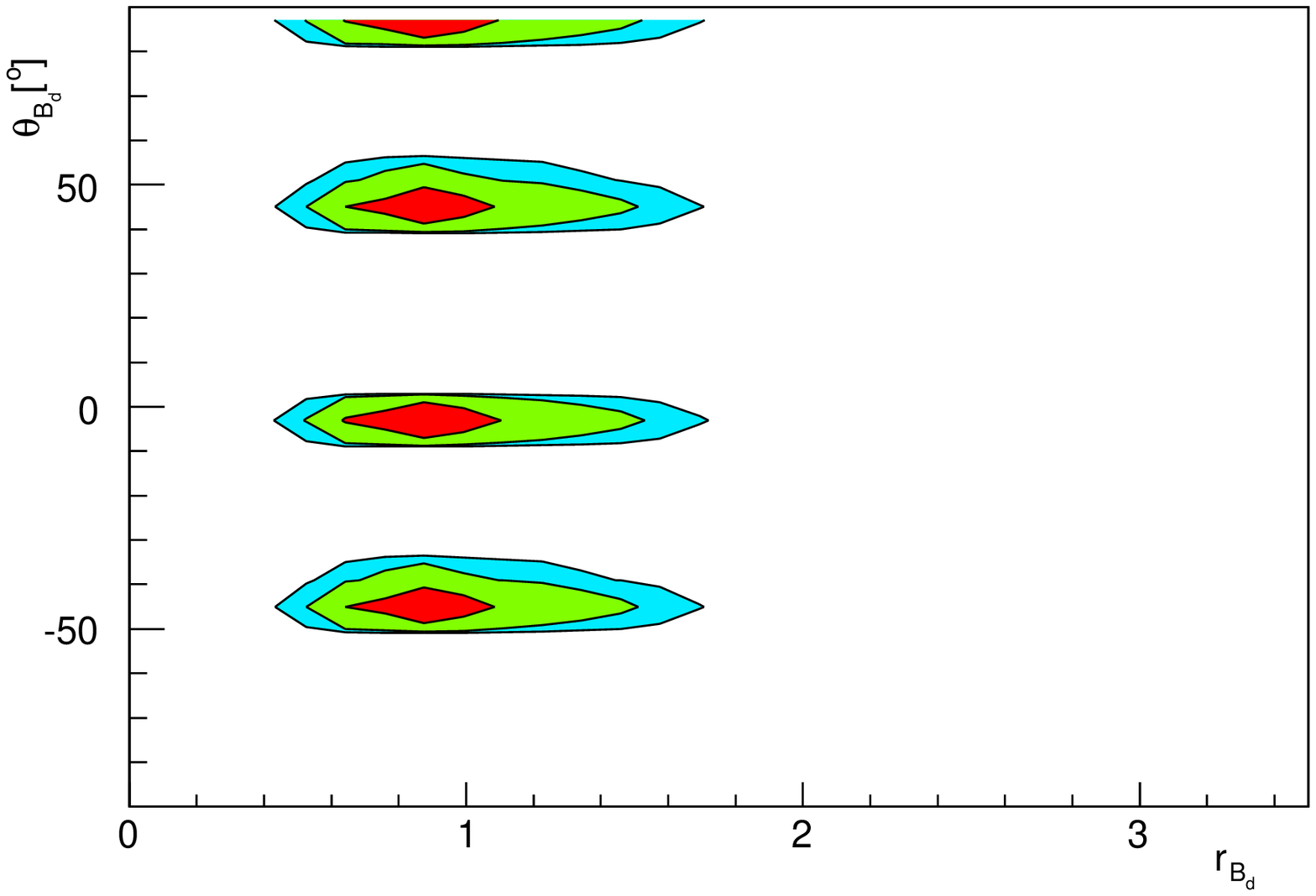}\label{fig:rBd_thetaBd_2d_r3b}
\vspace*{0.4cm}
(i)
\end{center}
}
\end{minipage}
\hfill
\begin{minipage}[t]{5.7cm}
{\begin{center}
\includegraphics[width=1\textwidth]{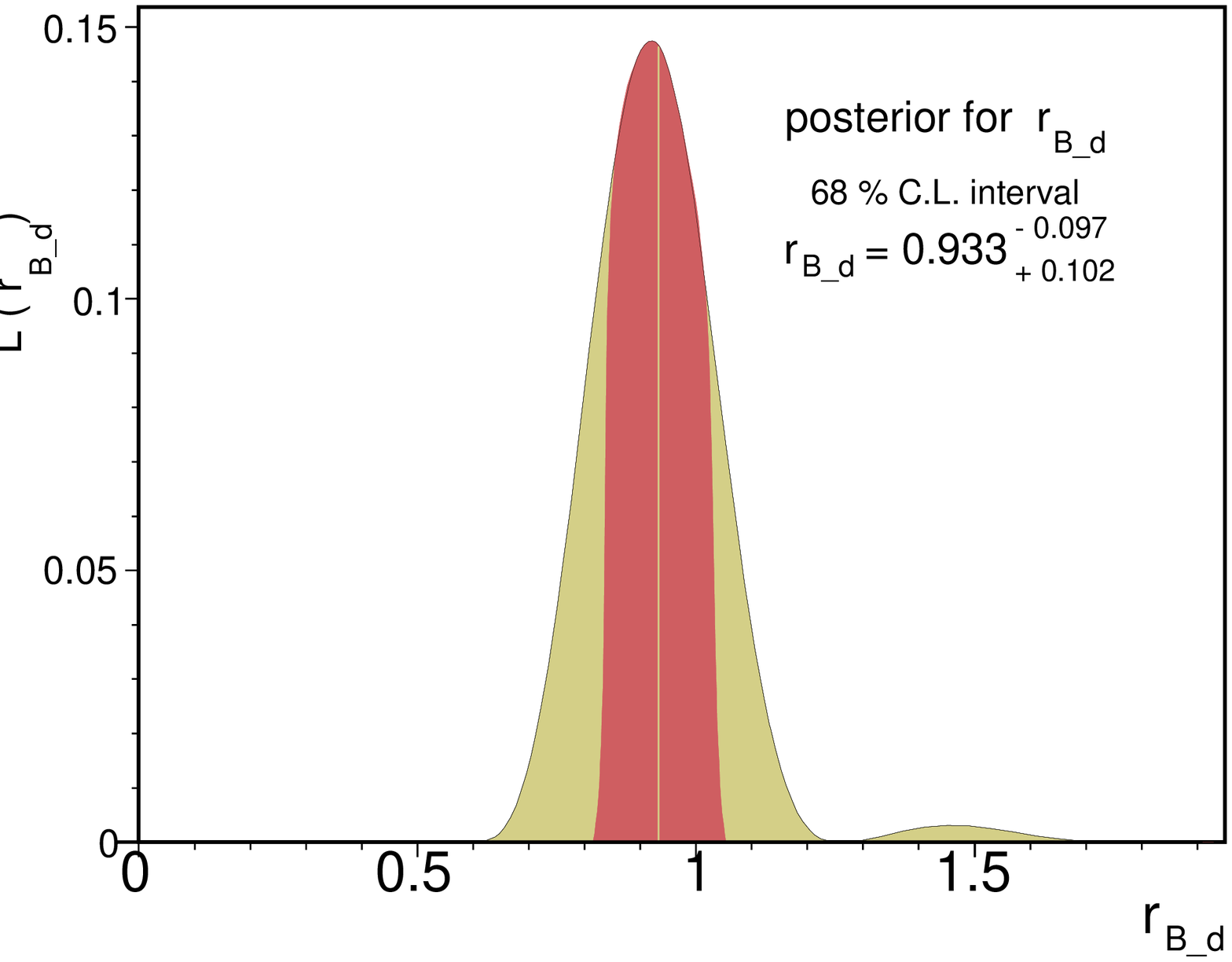}\label{fig:rBdcp_r3b}
\vspace*{-0.2cm}
(ii)
\end{center}
}
\end{minipage}
\hfill
\begin{minipage}[t]{5.7cm}
{\begin{center}
\includegraphics[width=1\textwidth]{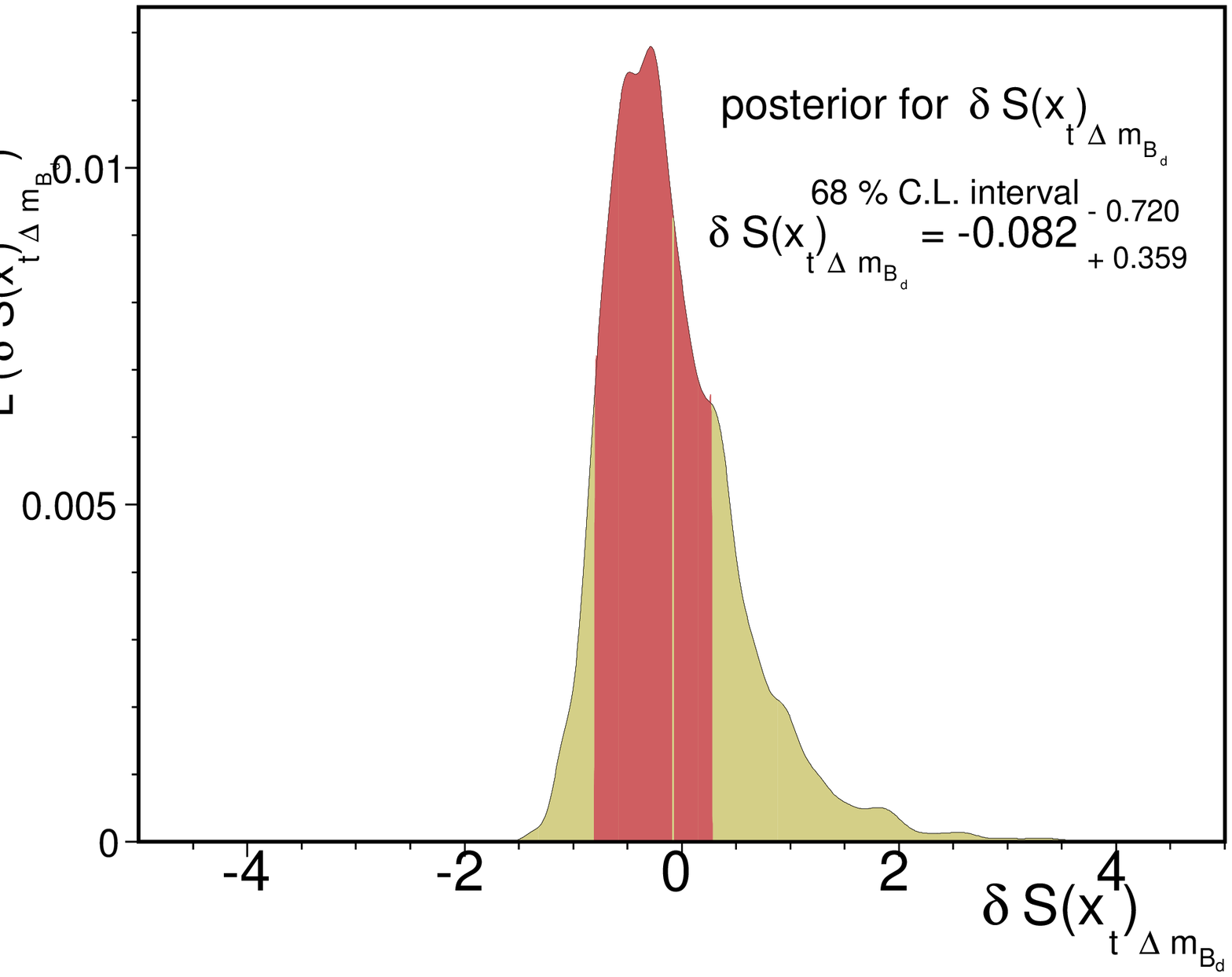}\label{fig:deldmd_r3b}
\vspace*{-0.2cm}
(iii)
\end{center}
}
\end{minipage}
%\vspace*{-0.5cm}
\caption{\small In figure (i) we have presented the contour levels of $r_{B_d}$ versus $\theta_{B_d}$ at 99\%, 95\% and 68\%. In (ii) the one dimensional pdf of $r_{B_d}$ and in (iii) the one dimensional pdf of the solution for $\theta_{B_d}$ compatible with the SM, whose 68\%CL is $(-5.99,-2.17)[^o]$ with mean at $-4.03^o$ and another solution whose 68\%~CL is $(43.01,46.21)[^o]$ about $44.64^o$.}
\label{fig:r3b_Bd}
\end{figure}

According to \eq{eq:bound_lambda} the limits on $\lambda$ from \eq{eq:delta_Sxt_iii} are for $\ek$: $\Lambda>4.6$ TeV. In the 68\%~CL range in this case $\delta S(x_t)_{\ek}$ is a positive contribution. For $\dmd$: $\Lambda>8 $ TeV for a negative contribution with respect to the SM and $\Lambda > 4 $ TeV for a positive contribution.

For the case of the supersymmetric MFV we can say more, since we have the change of \eq{eq:susySxtchg}, thus we can determine to which values of $\eta_2(B)$, $\eta_2(K)$ and  $S$ the parameters determined from the fits correspond to
\bea
S_{\dmd} &=& r_{B_d}\frac{\eta_B}{\eta_2(B)}S(x_t)\quad \quad \quad \quad ~ =2.290^{+0.323}_{-0.296}\nn\\
S_{\ek}  &=& \frac{\eta_{tt}}{\eta_2(K)}[S(x_t)+ \delta S(x_t)_{\ek}]= 2.229^{+0.343}_{-0.354}
\label{eq:S_exp-det_meas}
\eea
where we have used  $S(x_t)=2.2764-0.06466+0.0652$, as given by the experimental information on $M_W$ and $m_t$.
Here we would like to mention the great improvement with respect to the values obtained using the bound~\eq{eqn:dms:f05} %of $\dms > 16.6 {\rm{ps}}^{-1}$
\bea
S_{\dmd} &=& 3.0006^{-1.0065}_{+0.8444},\nn\\
S_{\ek}  &=& 3.2062^{-0.6819}_{+1.0800} ,
\label{eq:S_exp-det_bound}
\eea
For $S_{\dmd}$ there is an improvement in the relative uncertainty from $33\%$ to $14\%$ and for $S_{\ek}$ from $31\%$ to $16\%$.
The values of $\eta_{tt}$ and $\eta_B$ can be read from Table (\ref{tbl:fixedp}). The values of $\eta_2(B)$ and $\eta_2(K)$ are given in Section (\ref{sbsec:susy_mfv}) in \eq{eq:eta2_B_K_mfvs}.

We can determine values of $S$, with a given input of representative supersymmetric extra-constraints to the MFV hypothesis. In Table (\ref{tbl:susy_inputs_mfv}) we present some of these examples, which need to satisfy the conditions
\bea
150~\rm{GeV}   &\leq& m_{\tilde{t}_R}\leq 300~\rm{GeV},\quad
m_{\tilde{t}_R} \leq m_{\tilde{t}_L}\leq 600~\rm{GeV},\nn \\
m_{\tilde{t}_L} &\leq &m_{\tilde{q}}\leq 900~\rm{GeV},\nn \\
100~\rm{GeV}   &\leq& m_{\tilde{\chi}_2},\quad   m_{\tilde{\chi}_1}\leq
400~\rm{GeV},
\eea
for $m_{\tilde{q}}=m_{\tilde{u}_{1,2,4,5}}$. We can see that larger values of $S$ correspond to small values of $\tan\beta_s$ and a  heavy supersymmetric spectra, except for the charginos, which can be somewhat lighter than the rest of the involved s-particles. 

All examples here presented, Table (\ref{tbl:susy_inputs_mfv}), are compatible with the results of \eq{eq:S_exp-det_bound}, at 68\%~CL, specially those with $\tan\beta_s=8$. This is quite interesting because the values using the last bound on $\dms$, \eq{eq:S_exp-det_bound}, had a limited compatibility, from the the examples presented in Table (\ref{tbl:susy_inputs_mfv}) the ones for $\tan\beta_s=5$ are  in agreement with the determination of $S$ from the observed values of $r_{B_d}$, $S(x_t)$ and $\delta S(x_t)_{\ek}$ at 68\%~CL, while the last solution for  $\tan\beta_s=8$ was in agreement 
just at 99\%~CL. 
%at just the 99\%~CL. 

More constraints on the determination of the UT and a better determination of the QCD parameters will reduce the uncertainties in the determination of $S$ from \eq{eq:S_exp-det_meas} and hence will help to possibly rule out some possibilities within the supersymmetric MFV scenario.

%%%%%%%%%%%%%%%%%%%%%%%%%%%%%%%%%%%%%%%%%%%%%%%%%%%%%%%%%%%%%%%%%%%%%%%%%%
%
\begin{table}[!ht]
%\vspace*{-0.75cm}
\begin{center}
\begin{tabular}{|l l l  l l l  l c|}
\hline
$m_H$ & $m_{\tilde{q}}$ & $m_{\tilde{t}_L}$ &  $m_{\tilde{t}_R}$ & 
$m_{\tilde{\chi_1}}$ & $m_{\tilde{\chi_2}}$ &  $\tan\beta_s$ & S\\
\hline
$115$ & $880$ & $400$ &  $150$ & $170$ & $189$ &  $5$  & $2.56 \pm 0.08$\\
$175$ & $880$ & $400$ &  $150$ & $170$ & $189$ &  $5$  & $2.52^{-0.08}_{+0.06}$\\
$115$ & $880$ & $400$ &  $150$ & $300$ & $380$ &  $5$  & $2.46 \pm 0.07$\\
$115$ & $700$ & $350$ &  $160$ & $170$ & $189$ &  $5$  & $2.69 \pm 0.09$\\
$200$ & $700$ & $350$ &  $160$ & $170$ & $189$ &  $5$  & $2.63 \pm 0.09$\\
$115$ & $700$ & $350$ &  $160$ & $170$ & $189$ &  $8$  & $2.58 \pm 0.09$\\
$115$ & $700$ & $350$ &  $160$ & $300$ & $380$ &  $8$  & $2.35 \pm 0.07$\\
\hline
\end{tabular}
\end{center}
\caption{\small \small{Examples of input parameters to determine $S$, all the masses are expressed in GeV.}}
\label{tbl:susy_inputs_mfv}
\end{table}
%
%%%%%%%%%%%%%%%%%%%%%%%%%%%%%%%%%%%%%%%%%%%%%%%%%%%%%%%%%%%%%%%%%%%%%%%%%%
\section{Conclusions}

The goal of this note was to study the impact of the present limit on $\dms$ on the determination of the unitary triangle (UT) parameters within the context of the SM and also to put bounds for certain classes of processes BSM for which the current UT analysis it is useful.

 The first version of this note used the bound on $\dms>14.5 {\rm{ps}}^{-1}$ and the bound of $\dms>16.6 {\rm{ps}}^{-1}$ in order to begin to study the impact of $\dms$ on the SM and other models. These bounds were set by combining all the $\dms$ information available at the end of 2005, including the one reported by the CDF and  D\O\ collaborations prior and after Tevatron run II, respectively. The first measurement of $\dms=17.33\pm 0.4$, which has an uncertainty of $2\%$ its value, of course makes a stronger impact on the determination of the CKM parameters and possible BSM scenarios, being limited only by uncertainties from hadronic computations.

Using as an input the bound  $\dms>16.6 {\rm{ps}}^{-1}$ we obtained an output value of  $\dms=19.96^{-1.71}_{+1.01}$ which agreed remarkably well with its output value with out using $\dms$ as a constraint: $\dms=21.08^{-2.83}_{+2.73}$. However we can see that the measurement of $\dms=17.33\pm 0.4$ differs from this estimation, although it is not incompatible.

We tested  the consistency between the CP conserving processes - $|V_{ub}|/|V_{cb}|$, $\dmd$ and $\dmd/\dms$ -- and  the CP violating ones -- $\ek$ and $\sin2\beta$. With the last bound on $\dms$ and the used set of experimental information there was an slight disagreement between the resulting value for $\sin 2\beta$ considering just CP conserving processes, which in this case renders a value of $\sin 2\beta^{\;\rm{CP-conserv.}}=(0.751,0.841)$, and the experimental reported value of $\sin 2\beta^{\;\rm{exp.}}=(0.623,0.751)$, both intervals at 95\%~CL. With the  measurement of $\dms=17.33\pm 0.4$ the compatibility it is just at 99\%~CL since for this case we have $\sin 2\beta^{\;\rm{CP-conserv.}}=(0.742,0.856)$. 

 An interesting possibility is that there is a beyond the SM (BSM) contribution to $\sin 2\beta$ which it may be expressed as $\sin 2\beta= \sin( 2\beta^{\rm{SM}}+\theta_{B_d})$. We have checked that this case it is indeed possible by taking into account the results for this phase. From the parameterization $r_{B_d}e^{2\theta_{B_d}}=\langle B^0_{d,s}|{\mathcal{H}}^{\rm{Total}}_{\rm{eff}} |\bar{B}^0_{d,s}\rangle / B_{d,s}|{\mathcal{H}}^{\rm{SM}}_{\rm{eff}} |\bar{B}^0_{d,s}\rangle$ and the results of \eq{eq:thetaBd_sii}, we obtain $\theta_{B_d}=(-3.35-3.06+4.03)[^{\rm{o}}]$ and $r_{B_d}=0.983-0.108+0.091$, both at the 68\%~CL.

Theories satisfying the minimal flavour violating (MFV) conditions, for which $|V_{ub}|/|V_{cb}|$ and $\dmd/\dms$, used for the unitary triangle fits in the SM remain in the same form, can currently be tested at the same level of precision at which the analyses in the SM are carried out. In particular a supersymmetric version of the MFV scenario where the first two generations of s-quarks are taken to be heavy and degenerate, allowing just the contribution of the third generation, $\tilde{t}_R$ and $\tilde{t}_L$ can be tested in more detail since the change at the QCD NLO is known \cite{Krauss:1998mt}. We have tested this possibility for some samples of supersymmetric spectra, finding an agreement with most part of the supersymmetric examples considered, specially for $\tan\beta_s$ near to $5$ (in this scenario $\tan\beta_s$ cannot be large).

The first measurement of $\dms=17.33\pm 0.4$ and its relatively small uncertainty ($2\%$) has provided a good estimation of the region of compatibility of the MFV supersymmetric scenario, a better determination $\dms$ and $|V_{ub}|$ and $|V_{cb}|$ will help to narrow the uncertainties in all the parameters related to it in the UT analysis, specifically the parameters related to observations BSM and hence helping us to disentangle more possibilities in this respect. After the first version of this work was presented some related works, specially considering the impact of $\dms$ on the MFV scenario or other BSM scenarios appeared in the literature \cite{relatedworksbsm} and the unitary triangle updates after the first measurement of $\dms$ \cite{relatedworksut}.

%
%
%%%%%%%%%%%%%%%%%%%%%%%%%%%%%%%%%%%%%%%%%%%%%%%%%%%%%%%%%%%%%%%%%%%%%%%%%%

%
%%%%%%%%%%%%%%%%%%%%%%%%%%%%%%%%%%%%%%%%%%%%%%%%%%%%%%%%%%%%%%%%%%%%%%%%%%%%%%%%%%%%%%
%
\section*{Acknowledgments}
L. Velasco-Sevilla would like to thank F. Krauss for the clarification of the notation of his work in hep-ph/9807238, M. Voloshin for discussions and also take here the opportunity to deeply thank A. Stocchi for help understanding the bayesian approach when developing the code for the determination of the CKM parameters. L. Velasco-Sevilla is supported in part by the DOE grant DE-FG02-94ER.

\clearpage
%
%%%%%%%%%%%%%%%%%%%%%%%%%%%%%%%%%%%%%%%%%%%%%%%%%%%%%%%%%%%%%%%%%%%%%%%%%%%%%%%%%%%%%
%%%%%%%%%%%%%%%%%%%%%%%%%%%%%%%%%%%%%%%%%%%%%%%%%%%%%%%%%%%%%%%%%%%%%%%%%%%%%%%%%%%%%%
%
\appendix

\section{Experimental Inputs \label{sec:app_exp_inp}}
%%%%%%%%%%%%%%%%%%%%%%%%%%%%%%%%%%%%%%%%%%%%%%%%%%%%%%%%%%%%%%%%%%%
%
\begin{table}[!ht]
%\vspace*{-0.75cm}
\begin{center}
\begin{tabular}{|l l l|}
\hline
\multicolumn{2}{c}{{Fixed Parameters}}\\ 
\hline
Parameter      & Value                                     & Reference       \\ 
\hline
&  & \\
$G_{F}$        & $1.16639\times 10^{-5}{\rm GeV}^{-2}$     & \cite{Eidelman:2004wy} \\ 
$M_{W}$        & $(80.425\pm 0.038){\rm GeV}$              &  ''  \\
$f_{K}$        & $(0.1598\pm 0.0015){\rm GeV}$             &  ``  \\
$m_{K}$        & $(0.49765\pm 0.00002){\rm GeV}$           &  ``  \\
$\Delta m_{K}$ & $(3.4606\pm 0.006)\times 10^{-15}{\rm GeV}$&  ``  \\
$|\epsilon _{K}|$ & $(2.280\pm 0.017)\times 10^{-3}$       &  ``  \\
$\eta _{tt}$ & $(0.574\pm 0.004)$                          &  \cite{etas}\\
$m_{B_{d}}$    & $(5.2794\pm 0.0005){\rm GeV}$             &  \cite{Eidelman:2004wy}  \\
$\eta _{B}$    & $0.55\pm 0.007$                           &   ``        \\
$m_{B_{s}}$    & $(5.375\pm 0.0024){\rm GeV}$              &   ``        \\
\hline
\end{tabular}
\end{center}
\caption{\small \small{Input values of \emph{fixed} parameters.}}
\label{tbl:fixedp}
\end{table}
%
%
%
%%%%%%%%%%%%%%%%%%%%%%%%%%%%%%%%%%%%%%%%%%%%%%%%%%%%%%%%%%%%%%%%%%%
%
\begin{table}[!h]
\begin{center}
\begin{tabular}{|l l l l|} 
\hline
\multicolumn{3}{c}{Fitted Parameters}\\ 
\hline 
Parameter            & Value $\pm$ Gaussian errors      & Flat errors            &Referen.\\
\hline
&  & &\\
$|V_{cb}|({\rm{incl.}})$      & $(41.6\pm 0.7 )\times 10^{-3}$   &                      & \cite{Alexander:2005cx} \\
$|V_{cb}|({\rm{excl.}})$      & $(41.3\pm 1.0 )\times 10^{-3}$   & $1.8\times 10^{-3}$  &  `` \\
$|V_{ub}|({\rm{incl.}})$      & $(43.9\pm 2.0 )\times 10^{-4}$   & $2.7\times 10^{-4}$  &  `` \\
$|V_{ub}|({\rm{excl.}})$      & $(38.0\pm 2.7 )\times 10^{-4}$   & $4.7\times 10^{-4}$  &  `` \\
$|V_{us}|$             & $0.2258\pm 0.0014$             &        &                         `` \\

%$|\frac{V_{ub}}{V_{cb}}|^{{\rm av}}$  & $0.1017 \pm 0.0029$             &  $\pm 0.0058$ 
% &\cite{Eidelman:2004wy} \\
$B_K$               & $0.79\pm 0.04$                  & $\pm 0.09$ 
                                                         &\cite{Eidelman:2004wy} \\
$\overline{m}_c        $       & $(1.3\pm 0.1) {\rm GeV}$        &        &'' \\
$\overline{m}_t        $       & $(161.5\pm 3.0){\rm GeV}$       &        &\cite{topcdf:2005sum}*\\
$\eta_{cc}        $ & $1.38\pm 0.53 $                 &          & \cite{Buras:1990fn}\\
$\eta_{ct}        $ & $0.47\pm 0.04$                  &          & \cite{etas}\\
%$\eta_{tt}        $ & $0.5765\pm 0.0065$              &        &\\
$\sin 2\beta$       & $0.687\pm 0.032$                &          & \cite{Eidelman:2004wy} \\
$\dmd$              & $(0.506\pm 0.005){\rm ps}^{-1}$ &          & \cite{hfag:2005sum} \\
$f_{B_s}\sqrt{B_{B_s}}$ & $(0.276\pm 0.038){\rm GeV}$ &          & \cite{Bona:2005vz}\\
%$f_{B_d}\sqrt{B_{B_d}}$ & $(0.223\pm 0.033){\rm GeV}$ & $\pm 0.012{\rm GeV}$ &\cite{talkLell02}\\
$\xi$               & $1.21\pm 0.04$                  & $\pm0 .06$    & \cite{Battaglia:2003in}\\
\hline
$\dms$              & $(17.33\pm 0.4)\ {\rm ps}^{-1}$ &  &\\
\hline
\end{tabular}
\end{center}
\caption{\small Input values of the \emph{fitted} parameters. $*$ Taken from \cite{topcdf:2005sum} and calculated at the pole.}
\label{tbl:fittedp}
\end{table}
\section{Tables of outputs}

%%%%%%%%%%%%%%%%%%%%%%%%%%%%%%%%%%%%%%%%%%%%%%%%%%%%%%%%%%%%%%%%%%%
%
\begin{table}[h]
\vspace{-0.5cm}
\begin{center}
\begin{tabular}{|l l l l| } 
\hline
\multicolumn{3}{c}{Outputs}\\ 
\hline 
Parameter     &$\!\!\dms>16.6$ ps $^{-1}\!\!\!$                             &  No $\dms$               & $\!\!\dms=(17.33\pm 0.4)$ ps $^{-1}\!\!\!$                              \\
\hline
$\rb$         & $0.212^{-0.032}_{+0.034}$                & $0.238^{-0.032}_{+0.034}$                 & $0.200^{-0.021}_{+0.022}$                 \\
$\eb$         & $0.358^{-0.026}_{+0.028}$                & $0.345^{-0.024}_{+0.025}$                 & $0.363^{-0.015}_{+0.015}$                 \\
$\dms[{\rm ps}^{-1}]$ & $19.18^{-1.84}_{+1.35}$          & $21.08^{-2.83}_{+2.73}$                   & $17.42^{-0.38}_{+0.27}$      \\
%$|V_{ub}|$    & $(3.  \pm 0.1 ) \times 10^{-3}$          & $(3.0\pm ..)\times 10^{-3}$               & $(3.\pm ) \times 10^{-3}$           \\
%$|V_{cb}|$    & $( \pm   )\times 10^{-2}$          & $(4. \pm ..)\times 10^{-2}$            & $(4.\pm  )\times 10^{-2}$           \\
$|V_{ub}|/|V_{cb}|$   & $0.0097^{-0.005}_{+0.005}$       & $0.0098^{-0.005}_{+0.005}$                & $0.0097^{-0.006}_{+0.005}$ \\

$\dmd[{\rm ps}^{-1}]$ & $0.504^{-0.124}_{+0.057}$        & $0.483^{-0.125}_{+0.125}$                 & $0.476^{-0.196}_{+0.183}$  \\
%$|\ek|$       &  $..... $                                & $(...)\times 10^{-3}$                     & $(2.291^{-0.011}_{+0.013})\times 10^{-3}$    \\
$\sin 2\beta$ & $0.755^{-0.019}_{+0.023}$                & $0.755^{-0.024}_{+0.023}$                  & $0.765^{-0.019}_{+0.028}$                \\
$\alpha[^o]$  & $98.90^{-4.29}_{+4.60}$                  & $102.98^{-6.47}_{+7.35}$              & $97.49^{-6.83}_{+6.99}[^o]$                  \\
$\gamma[^o]$  & $59.39^{-3.85}_{+5.07}$                  & $55.32^{-5.70}_{+5.39}$               & $60.97^{-4.04}_{+4.64}$                   \\
$R_b$         & $0.419^{-0.008}_{+0.013}$                & $0.421^{-0.016}_{+0.020}$                 & $0.416^{-0.014}_{+0.019}$                   \\
$R_t$         & $0.865^{-0.028}_{+0.036}$                & $0.836^{-0.039}_{+0.035}$                 & $0.877^{-0.022}_{+0.022}$ \\
$B_K$         & $0.764^{-0.046}_{+0.044}$                & $0.786^{-0.04}_{+0.047} $                      & $0.755^{-0.041}_{+0.042}$ \\
$f_{B_d}\sqrt{B_{B_d}}[{\rm GeV}]$ & $0.226^{-0.016}_{+0.014}$ &  $0.234^{-0.017}_{+0.016}$               & $0.224^{-0.016}_{+0.014}$ \\
\hline
\hline
\end{tabular}
\end{center}
\caption{\small Output values of the \emph{fitted} parameters. The three columns correspond respectively to the complete {\it Classic Fit} using the bound $\dms>16.6$ ps $^{-1}$, without using any information on $\dms$ and finally using the measurement of $\dms=(17.33\pm 0.4)$ ps $^{-1}$.}
\label{tbl:completfit}
\end{table}
%

%
%%%%%%%%%%%%%%%%%%%%%%%%%%%%%%%%%%%%%%%%%%%%%%%%%%%%%%%%%%%%%%%%%%%
\begin{table}[!h]
\vspace{-0.6cm}
\begin{center}
\begin{tabular}{|l l l l| } 
\hline
\multicolumn{3}{c}{Outputs}\\ 
\hline 
Parameter     & (i)                                      &  (ii)                                     & (iii)                                     \\
\hline
$\rb$         & $0.185^{-0.023}_{+0.030}$                & $0.209^{-0.020}_{+0.023}$                 & $0.184^{-0.023}_{+0.030}$                 \\
$\eb$         & $0.375^{-0.020}_{+0.018}$                & $0.379^{-0.017}_{+0.013}$                 & $0.375^{-0.018}_{+0.017}$                 \\
$\dms[{\rm ps}^{-1}]$ & $17.37^{-0.38}_{+0.37}$          & $17.48^{-0.39}_{+0.41}$   
        & $17.36^{-0.42}_{+0.38}$      \\
$|V_{ub}/V_{cb}|$    & $0.097^{0.002}_{0.003} $          & $0.101^{0.003}_{0.003}$            & $0.097^0.002_{0.003}$           \\
$\dmd[{\rm ps}^{-1}]$ & $0.505^{-0.178}_{+0.158}$        & $0.488^{-0.294}_{+0.291}$  
        & $0.506^{-0.297}_{+0.294}$  \\
$|\ek|$       & $(2.56^{-0.36}_{+0.23})\times 10^{-3}$   & $(2.30^{-0.06}_{+0.05})\times 10^{-3}$  & $(2.57^{-0.39}_{+0.33})\times 10^{-3}$    \\
$\sin 2\beta$ & $0.766^{-0.017}_{+0.018}$                & $0.789^{-0.020}_{+0.023}$                  & $0.766^{-0.017}_{+0.018}$                \\
$\alpha[^o]$  & $94.79^{-4.19}_{+4.67}$                  & $97.36^{-2.82}_{+3.01}$              & $94.98^{-4.38}_{+4.50}[^o]$                  \\
$\gamma[^o]$  & $63.61^{-5.10}_{+3.57}$                  & $60.99^{-3.27}_{+3.04}$               & $63.65^{-4.91}_{+3.03}$                   \\
$R_b$         & $0.420^{-0.012}_{+0.021}$                & $0.434^{-0.012}_{+0.015}$                 & $0.420^{-0.012}_{+0.015}$                   \\
$R_t$         & $0.873^{-0.054}_{+0.048}$                & $0.876^{-0.025}_{+0.019}$                 & $0.897^{-0.038}_{+0.022}$ \\
$B_K$         & $0.707^{-0.066}_{+0.072}$                & $0.743^{0.036}_{0.030}$                            & $0.706^{-0.070}_{+0.073}$ \\
$f_{B_d}\sqrt{B_{B_d}}[{\rm GeV}]$ & $0.219^{-0.014}_{+0.011}$ &   $0.225^{-0.008}_{+0.007}$                                 & $0.219^{-0.014}_{+0.011}$ \\
\hline
\hline
\end{tabular}
\end{center}
\caption{\small Output values of the \emph{fitted} parameters. The three columns correspond to the three cases presented in Section (\ref{sbsec:outputs}): (i)Using the constraints $|V_{ub}|/|V_{cb}|$, $\sin 2\beta$, $\dms/\dmd$ and $\dmd$; (ii)  $|V_{ub}|/|V_{cb}|$, $\dms/\dmd$ and $\ek$ and (iii) $|V_{ub}|/|V_{cb}|$, $\sin 2\beta$ and $\dms/\dmd$.}
\label{tbl:outputs_i_iii_cases} 
\end{table}

\end{document}